%% file: cygnus2009Whitepaper.tex
\pdfoutput=1 

\documentclass{ws-ijmpa}
\input sgml.sty

\usepackage{graphicx}
\usepackage{epsfig}
\usepackage{dcolumn}
\usepackage{bm}

\newcommand{\be}{\begin{equation}}
\newcommand{\ee}{\end{equation}}
\newcommand{\bea}{\begin{eqnarray}}
\newcommand{\eea}{\end{eqnarray}}
\newcommand{\cstwo}{CS$_2$}
\newcommand{\cffour}{CF$_4$}

\newcommand{\mmm}{m$^3$}
\newcommand{\simgt}{\lower.5ex\hbox{$\; \buildrel > \over \sim \;$}}
\newcommand{\simlt}{\lower.5ex\hbox{$\; \buildrel < \over \sim \;$}}

\newcommand{\he}{$^4$He }
\newcommand{\hee}{$^4$He}
\newcommand{\hid}{$^1$H }

\newcommand{\het}{$^3$He }
\newcommand{\hett}{$^3$He}
\newcommand{\fl}{$^{19}$F }

\newcommand{\alu}{$^{27}$Al }
\newcommand{\iso}{C$_4$H$_{10}$ }

\newcommand{\bostonuniversity}{$^{1}$}
\newcommand{\brandeisuniversity}{$^{2}$}
\newcommand{\ucla}{$^{3}$}
\newcommand{\ceaSaclay}{$^{4}$}
\newcommand{\edinburghuniversity}{$^{5}$}
\newcommand{\harvarduniversity}{$^{6}$}
\newcommand{\huelvauniversity}{$^{7}$}
\newcommand{\ioanninauniversity}{$^{8}$}
\newcommand{\kekTsukuba}{$^{9}$}
\newcommand{\kyotouniversity}{$^{10}$}
\newcommand{\lpscGrenoble}{$^{11}$}
\newcommand{\zaragozauniversity}{$^{12}$}
\newcommand{\lawrenceBerkeleyNationalLab}{$^{13}$}
\newcommand{\mitCambridge}{$^{14}$}
\newcommand{\nagasakiInstitute}{$^{15}$}
\newcommand{\nagoyauniversity}{$^{16}$}
\newcommand{\unmAlbuquerque}{$^{17}$}
\newcommand{\newYorkUniversity}{$^{19}$}
\newcommand{\universityOfNottingham}{$^{19}$}
\newcommand{\occidentalCollege}{$^{20}$}
\newcommand{\universityOfPennsylvania}{$^{21}$}
\newcommand{\perimeterInstitute}{$^{22}$}
\newcommand{\sagaUniversity}{$^{23}$}
\newcommand{\jaxaJapan}{$^{24}$}
\newcommand{\universityOfSheffield}{$^{25}$}
\newcommand{\templeUniversity}{$^{26}$}
\newcommand{\kamiokaObservatory}{$^{27}$}
\newcommand{\universityOfUtah}{$^{28}$}
\newcommand{\universityOfWarwick}{$^{29}$}
\newcommand{\universityOfWaterloo}{$^{30}$}
\newcommand{\authorSpace}{\ }

\begin{document}

\markboth{Battat et al.}
{Directional Dark Matter Detection}

%
\catchline{}{}{}{}{}
%


\title{The case for a directional dark matter detector and the status of current experimental efforts}

\input{authors.tex}


\date{\today}

\maketitle


\begin{abstract}
We present the case for a dark matter detector with directional sensitivity.  This document was developed at the 2009 CYGNUS workshop on directional dark matter detection, and contains contributions from theorists and experimental groups in the field.  We describe the need for a dark matter detector with directional sensitivity; each directional dark matter experiment presents their project's status; and we close with a feasibility study for scaling up to a one ton directional detector, which would cost around \$150M.
\keywords{Dark matter; directional detection}
\end{abstract}

\ccode{PACS numbers}

\input{sectionTheory.tex}
\input{sectionExperiment.tex}
\input{sectionOutlook.tex}

\input{sectionAcknowledgements.tex}


\bibliographystyle{ws-ijmpa}
\bibliography{journal-abbreviations,dmrefs}

\end{document}

%% file: authors.tex

\author{
S.~AHLEN,\bostonuniversity\authorSpace 
N.~AFSHORDI,\perimeterInstitute$^{,}$\universityOfWaterloo\authorSpace
J.~B.~R.~BATTAT,\mitCambridge\authorSpace
J.~BILLARD,\lpscGrenoble\authorSpace
N.~BOZORGNIA,\ucla\authorSpace
S.~BURGOS,\occidentalCollege\authorSpace 
T.~CALDWELL,\mitCambridge${,}$\universityOfPennsylvania\authorSpace
J.~M.~CARMONA,\zaragozauniversity\authorSpace
S.~CEBRIAN,\zaragozauniversity\authorSpace 
P.~COLAS,\ceaSaclay\authorSpace
% ...
}

\address{
\bostonuniversity Boston University, Boston, MA  02215, USA\\
\brandeisuniversity Brandeis University, Waltham, MA 02453, USA\\
\ucla University of California Los Angeles, Los Angeles, CA 90095, USA\\
\ceaSaclay CEA Saclay, C\'edex, France\\
\edinburghuniversity University of Edinburgh, Edinburgh, EH9 3JZ, UK\\
\harvarduniversity Harvard University, Cambridge, MA, 02140, USA\\
\huelvauniversity Universidad de Huelva, Campus El Carmen, 21071 Huelva, Spain\\
\ioanninauniversity University of Ioannina, Ioannina, Gr 451 10, Greece\\
\kekTsukuba\authorSpace Institute of Particle and Nuclear Studies, KEK, Tsukuba, Japan\\
\kyotouniversity Kyoto University Kitashirakawa-oiwakecho, Sakyo-ku, Kyoto, 606-8502, Japan\\
\lpscGrenoble Laboratoire de Physique Subatomique et de Cosmologie, Universit\'e Joseph Fourier Grenoble 1, CNRS/IN2P3, Institut National Polytechnique de Grenoble, 53, rue de Martyrs, 38026 Grenoble, France\\
\zaragozauniversity Laboratorio de Fisica Nuclear y Astroparticulas, Universidad de Zaragoza, 50009 Zaragoza, Spain, and Laboratorio Subterr\'aneo de Canfranc, Huesca, Spain \\
\lawrenceBerkeleyNationalLab Lawrence Berkeley National Laboratory, Berkeley, CA 94720, USA\\
\mitCambridge Massachusetts Institute of Technology, Cambridge, MA 02139, USA\\
\nagasakiInstitute Nagasaki Institute of Applied Science, Nagasaki, Japan\\
\nagoyauniversity Fundamental Particle Physics Laboratory, Nagoya University, Japan\\
\unmAlbuquerque University of New Mexico, Albuquerque, NM, 87131, USA\\
\newYorkUniversity New York University, New York, NY 10003, USA\\
\universityOfNottingham University of Nottingham, University Park, Nottingham, NG7 2RD, UK\\
\occidentalCollege Occidental College, Los Angeles, CA 90041, USA\\
\universityOfPennsylvania University of Pennsylvania, Philadelphia, PA 19104, USA\\
\perimeterInstitute Perimeter Institute for Theoretical Physics, Waterloo, ON N2L 2Y5, Canada\\
\sagaUniversity Saga University, Saga, Japan\\
\jaxaJapan Scientific Balloon Laboratory, ISAS, JAXA, Yoshinodai 3-1-1, Sagamihara, Kanagawa, 229-8510, Japan\\
\universityOfSheffield University of Sheffield, Sheffield S3 7RH, UK\\
\templeUniversity Temple University, Barton Hall, Philadelphia, PA 19122, USA\\
\kamiokaObservatory Kamioka Observatory, ICRR, The University of Tokyo, Higashi-Mozumi, Kamioka cho, Hida 506-1205 Japan\\
\universityOfUtah University of Utah, Salt Lake City, UT 84112, USA\\
\universityOfWarwick University of Warwick, Coventry, CV4 7AL, UK\\
and \\
\universityOfWaterloo University of Waterloo, Waterloo, Ontario, N2L 3G1 Canada\\
\email{\mitCambridge jbattat@mit.edu}
}

%% file: sectionTheory.tex
\section{Theoretical motivation}

Diverse astrophysical observations demonstrate that the majority of
the matter in the Universe is in the form of non-baryonic cold dark
matter.\cite{BertonePhysRept2005} Understanding the nature of the dark
matter is one of the major outstanding problems of both astrophysics
and particle physics.

The Weakly Interacting Massive Particle (WIMP) is a generically good
dark matter candidate. A stable, weakly-interacting particle that was
in thermal equilibrium in the early Universe will have roughly the
right present-day density to comprise the dark matter.  Furthermore,
well-motivated extensions of the standard model of particle physics
provide us with concrete WIMP candidates.\cite{JungmanPhysRept1996}
Supersymmetric models, in which every standard model particle has a
supersymmetric partner, are motivated by the gauge hierarchy problem,
the unification of coupling constants, and string theory. In these
models, there is usually a conserved quantum number, R-parity
(required to avoid proton decay), which renders the lightest
supersymmetric particle (LSP) stable. In many cases the LSP is the
lightest neutralino (a mixture of the supersymmetric partners of the
photon, the Z and the Higgs) which is a good WIMP candidate. There has
also been heightened interest recently in Universal Extra Dimension
models, where the Lightest Kaluza-Klein particle is a WIMP
candidate. A successful direct detection campaign will not only
confirm the existence of dark matter, but will also probe high
energy particle physics.

WIMPs can be detected in three ways: at particle
colliders,\cite{KaneModPhysLett2008} indirectly
(astrophysically),\cite{JungmanPhysRept1996} or directly in the
laboratory.\cite{Goodman1985} The production and detection of
WIMP-like particles at the LHC would be extremely exciting. As well as
demonstrating that such particles exist in nature, it would provide
information about their properties (mass and interactions). It would
not, however, demonstrate that these particles are the dark matter in
the Universe. In particular it would not show that the particles
produced are stable on cosmological timescales. Astrophysical,
indirect detection experiments attempt to detect the products of WIMP
annihilation, e.g. high-energy gamma-rays, anti-matter and neutrinos,
within the Milky Way and beyond.  The detailed signals depend on the
dark matter distribution, on details of the annihilation process, and
on the propagation of charged particles in the Milky Way's magnetic
field. As demonstrated by recent studies of the PAMELA positron
excess,\cite{PAMELA2009} a WIMP signal will need to be distinguished
from astrophysical backgrounds from, for instance, pulsars and
supernova remnants.  Furthermore, the measurements themselves are very
challenging, and at present, several indirect detection experiments,
including ATIC and FERMI,\cite{ATIC2008,FERMILAT2009} are in conflict
with each other.  This work focuses on the third option: direct
detection, more specifically, on dark matter detectors that have
sensitivity to the direction of arrival of dark matter particles.

\subsection{Direct detection}
Direct detection experiments\cite{Gaitskell2004} aim to detect WIMPs
via their elastic scattering off of target nuclei in the laboratory.
Specifically, they look for the energy deposited in a detector by a
nuclear recoil from individual scattering events. Since WIMPs have a
very small cross section with matter, these events are rare and the
energy of the nuclear recoils is relatively small, $\sim 10-100$~keV.
Nonetheless, current experiments have already achieved the sensitivity
required to rule out regions of supersymmetric parameter space
consistent with all other observational and experimental
constraints.\cite{CDMS2009,AnglePRL2008} Neutrons, for instance from
cosmic-ray muons or local radioactivity, can also produce nuclear
recoils, which, on an event-by-event basis, cannot be distinguished
from WIMP-induced nuclear recoils. Direct detection experiments
minimize neutron backgrounds by operating deep underground, by using
radiopure components, and by shielding the detectors appropriately.
An unambiguous detection of WIMPs requires a smoking gun signal to
demonstrate that the observed recoils are indeed due to WIMPs rather
than neutrons or other backgrounds.

There are three potential WIMP signals, namely the time, direction,
and target nucleus dependence of the energy spectrum of the
recoils. Due to the kinematics of elastic scattering, the shape of the
energy spectrum depends on the mass of the target nuclei, and, for
spin-independent interactions (where the WIMP interacts coherently
with the nucleus), the normalisation of the spectrum is proportional
to the square of the mass number. In principle, the consistency of
energy spectra measured in two or more experiments with different
target nuclei could demonstrate the WIMP origin of the dark matter
interaction candidate events.\cite{LewinAndSmith1996} This is often
referred to as the materials signal. In practice, this would require
the detection of a large number of events with both targets (in order
to measure the energy spectra), the operation of experiments in
similar background environments, and accurate calculations of the
nuclear form factors.

\begin{figure}
\centering
\includegraphics[width=0.75\textwidth]{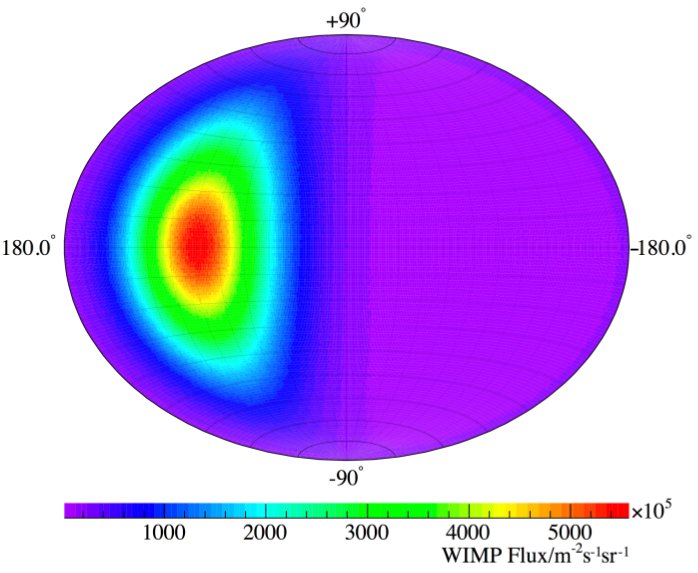}
\caption{\label{fig:theoryFig1} Hammer-Aitoff projection of the WIMP
  flux in Galactic coordinates.  A WIMP mass of 100~GeV has been
  assumed (from Ref.~\protect\refcite{MorganPhysRevD2005}).  }\end{figure}

\begin{figure}
\centering
\includegraphics[width=0.45\textwidth]{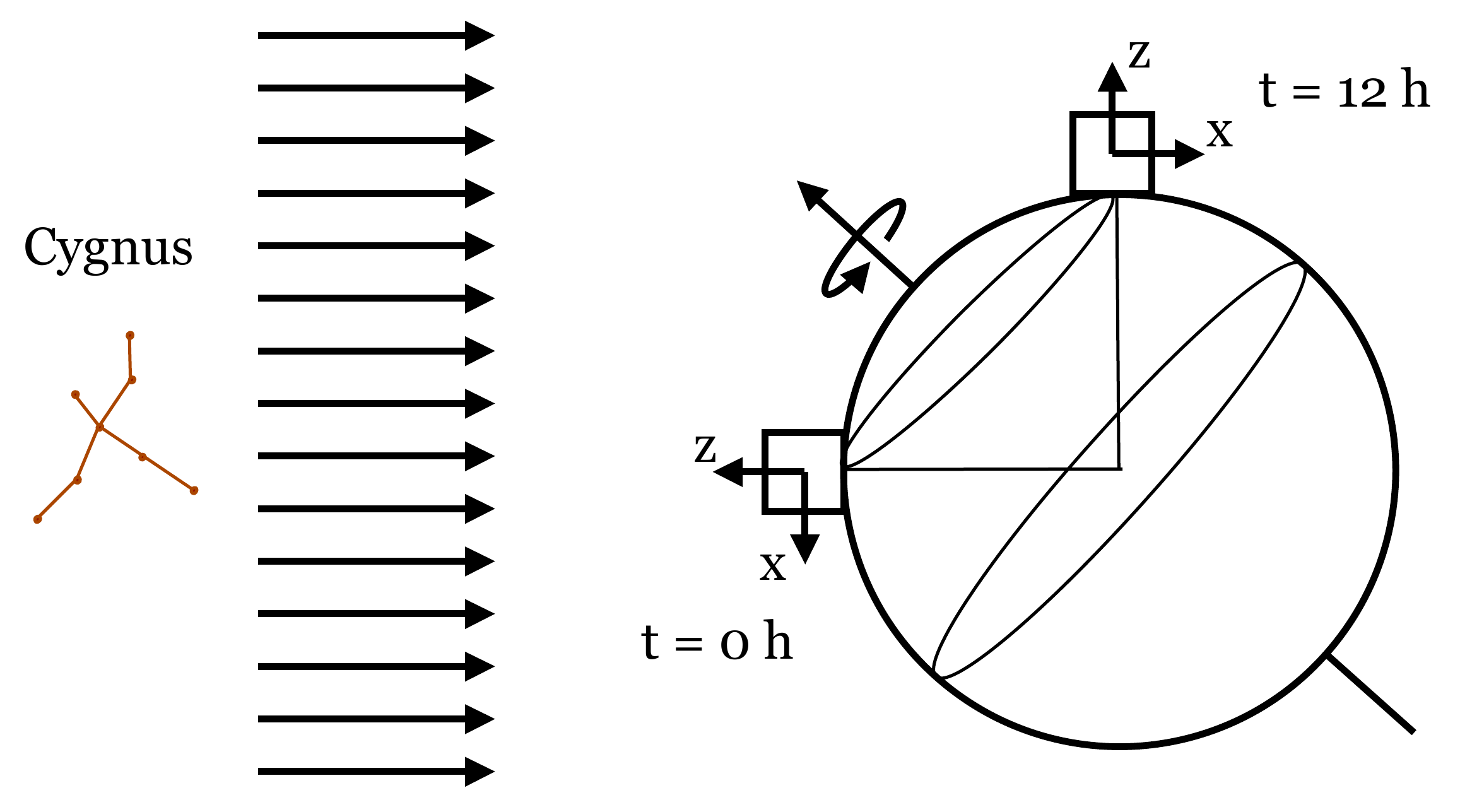}
\hfill
\includegraphics[width=0.45\textwidth]{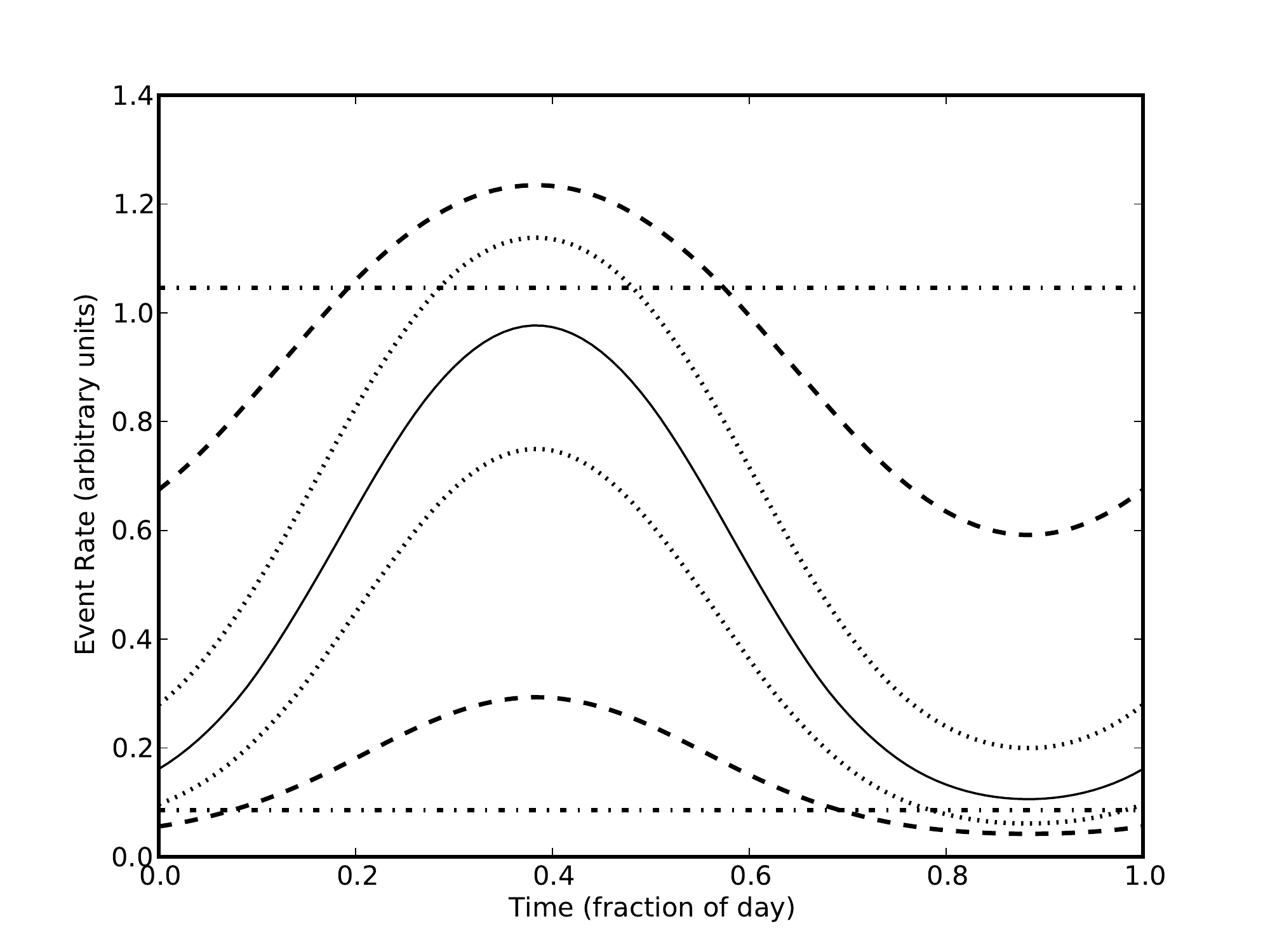}
\caption{\label{fig:theoryFig2} (left) The daily rotation of the Earth
  introduces a modulation in recoil angle, as measured in the
  laboratory frame.  (right) Magnitude of this daily modulation for
  seven lab-fixed directions, specified as angles with respect to the
  Earth's equatorial plane.  The solid line corresponds to zero
  degrees, and the dotted, dashed, and dash-dot lines correspond to
  $\pm 18^\circ$, $\pm 54^\circ$ and $\pm 90^\circ$, with negative
  angles falling above the zero degree line and positive angles below.
  The $\pm90^\circ$ directions are co-aligned with the Earth's
  rotation axis and therefore exhibit no daily modulation.  This
  calculation assumes a WIMP mass of 100~GeV and CS$_2$ target gas.
  (from Ref.~\protect\refcite{Vergados2007}).
}\end{figure}

\subsection{Temporal modulation}
The Earth's yearly orbit about the Sun produces an annual modulation
of the nuclear recoil energy spectrum.\cite{Drukier1986,Freese1988}
The Earth's net speed with respect to the Galactic rest frame is
largest in summer (when the Earth's orbital vector is aligned with the
Sun's orbital vector).  This boosts the WIMP speed distribution in the
laboratory frame towards higher speeds and hence leads to a larger
number of high energy recoils (and a deficit of low energy recoils).
Because the Earth's orbital speed is small compared to the Sun's
speed with respect to the Galactic rest frame, the amplitude of the
annual modulation is small, of order a few percent. Therefore its
detection requires the stable operation of a large detector mass over
a long period of time.  Furthermore, the details of the annual
modulation signal, in particular its phase and amplitude, are somewhat
dependent on the ultra-local (sub-milliparsec) WIMP velocity
distribution.  The DAMA/LIBRA collaboration has measured an annual
modulation in their event rate.\cite{DAMALIBRA2008} Even with ample
statistics and even though the modulation amplitude, frequency, and
phase are consistent with the expectations of Galactic dark matter,
the DAMA/LIBRA measurement has not been widely interpreted by the
physics community as evidence for dark matter, in part because of the
possibility that an annual modulation of background, rather than WIMP
flux, could fake the signal.  It is clear that the wide acceptance of
a dark matter detection claim requires a less ambiguous signature.

\subsection{Direction modulation}
\label{sec:theoryDirectionModulation}
The Earth's motion with respect to the Galactic rest frame also
produces a direction dependence in the recoil
spectrum.\cite{Spergel1988} The peak WIMP flux comes from the
direction of solar motion, which happens to point towards the
constellation Cygnus (see e.g.  Fig.~\ref{fig:theoryFig1}).  Assuming
a smooth WIMP distribution, the recoil rate is then peaked in the
opposite direction.  In the laboratory frame, this direction varies
over the course of the sidereal day due to the Earth's rotation (see
Fig.~\ref{fig:theoryFig2}), thereby providing a robust signature of
the Galactic origin of a WIMP signal.  As shown in
Fig.~\ref{fig:theoryFig2}, the number of recoils along a particular
direction in the laboratory frame will change over the course of the
day.  The amplitude of this modulation depends on the relative
orientation between the lab-fixed direction and the spin axis of the
Earth, with no modulation along directions parallel to the Earth's
spin axis.  No known background can mimic this signal.  The lab-frame
recoil directions can be rotated into galactic coordinates to produce
a 2D skymap of recoil directions. A directional detection experiment
can then look for anisotropies in the galactic skymap of nuclear
recoils.

The expected directional signal is far larger than the annual rate
modulation.  For a simplified, but representative WIMP halo model (the
isothermal sphere), the event rate in the forward direction is roughly
an order of magnitude larger than that in the backward direction.  In
this halo model, a detector capable of measuring the nuclear recoil
momentum vector (the axis and direction of the recoil, also called the
``head-tail'' of the track) in 3-dimensions, with good angular
resolution, could distinguish a WIMP signal from isotropic background
with only O(10) events.\cite{Copi1999,MorganPhysRevD2005} Directional
detection therefore provides the best opportunity to unambiguously
demonstrate the Galactic origin of a nuclear recoil signal.

\subsection{Non-spherical halo models}
Signal calculations for direct detection experiments (both directional
and non-directional) are usually based on the isothermal sphere halo
model, containing a smooth WIMP distribution with an isotropic
Maxwellian speed distribution, truncated at the galactic escape
velocity.\cite{LewinAndSmith1996} Plausible astrophysical and particle
physics deviations from this baseline model can produce other
interesting signals in directional experiments.

This standard halo model is unlikely to be a good approximation to the
real Milky Way halo. The details of the directional signal depend on
the ultra-local WIMP distribution which is not currently well-known.
Numerical simulations have finite resolution and can only probe the
dark matter distribution on kiloparsec scales.  For reference, the
Earth-Sun distance is 5~$\mu$pc, and in one year, the Solar System
moves approximately 0.2~milliparsec along its galactocentric orbit.
It has been argued that on the sub-milliparsec scales probed by direct
detection experiments, the dark matter distribution may not be
completely smooth.\cite{Stiff2003} In the extreme case of a velocity
distribution consisting of a small number of discrete peaks, the
recoil spectrum would still be highly anisotropic, albeit not peaked
in the direction opposite to Cygnus, and possibly time dependent on a
1--10 year timescale. It is, however, extremely unlikely that the WIMP
induced recoil spectrum would be completely isotropic (this would
require a dark matter halo with bulk co-rotation). The dependence of
the directional recoil spectrum on the WIMP velocity distribution in
fact provides an opportunity. With a larger sample of events, a
directional detector would be able to do WIMP astronomy,
reconstructing the ultra-local WIMP velocity distribution and hence
shedding light on the dynamics of the Milky Way halo by finding
streams clustered in momentum space.  This would be highly
complementary to the mapping of the full six-dimensional phase-space
distribution, (positions and velocities), of the stellar component of
the Milky Way by astronomical surveys (such as SDSS, and in the
future, GAIA).

\begin{figure}
\centering
\includegraphics[width=0.48\textwidth]{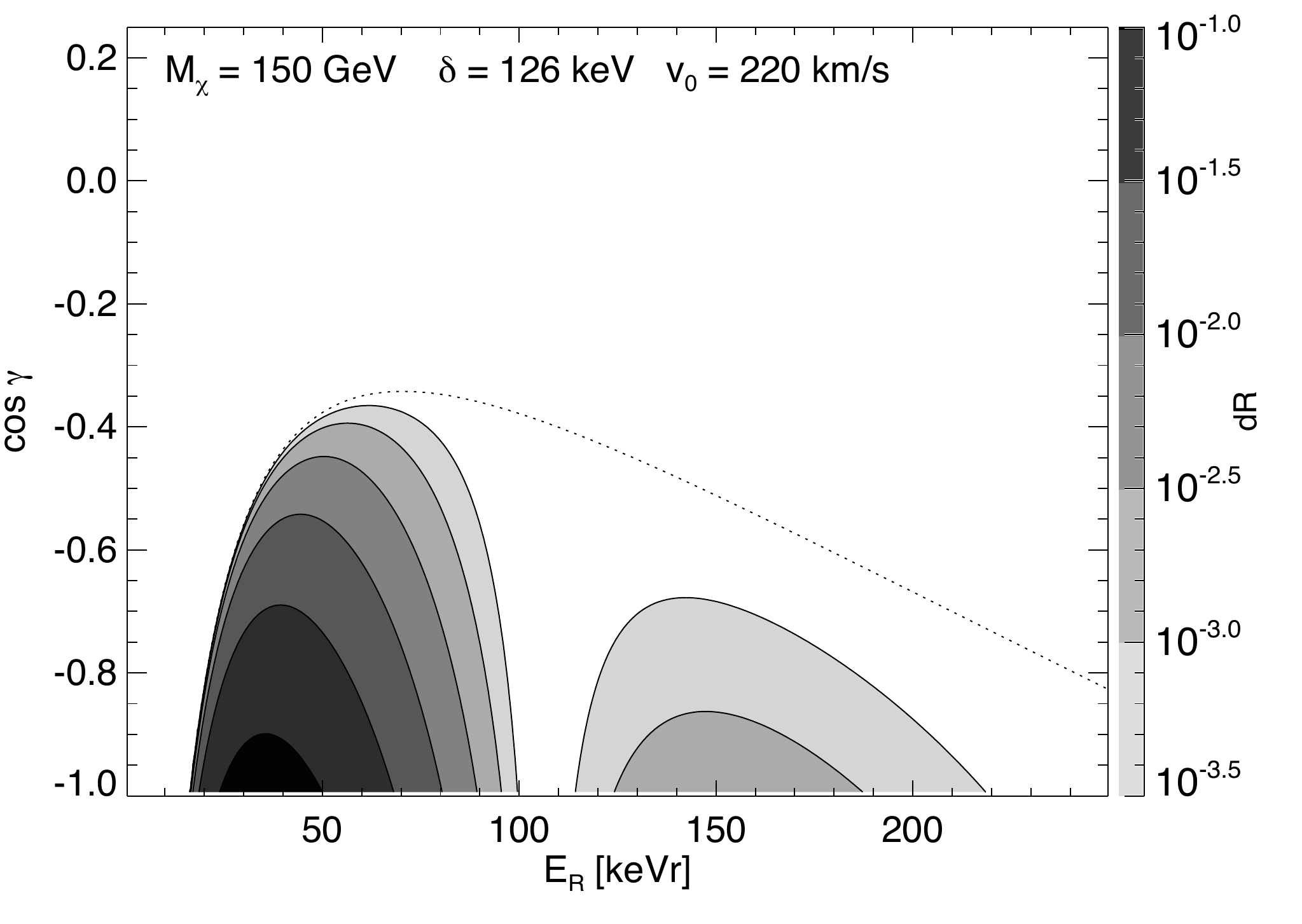}
\includegraphics[width=0.48\textwidth]{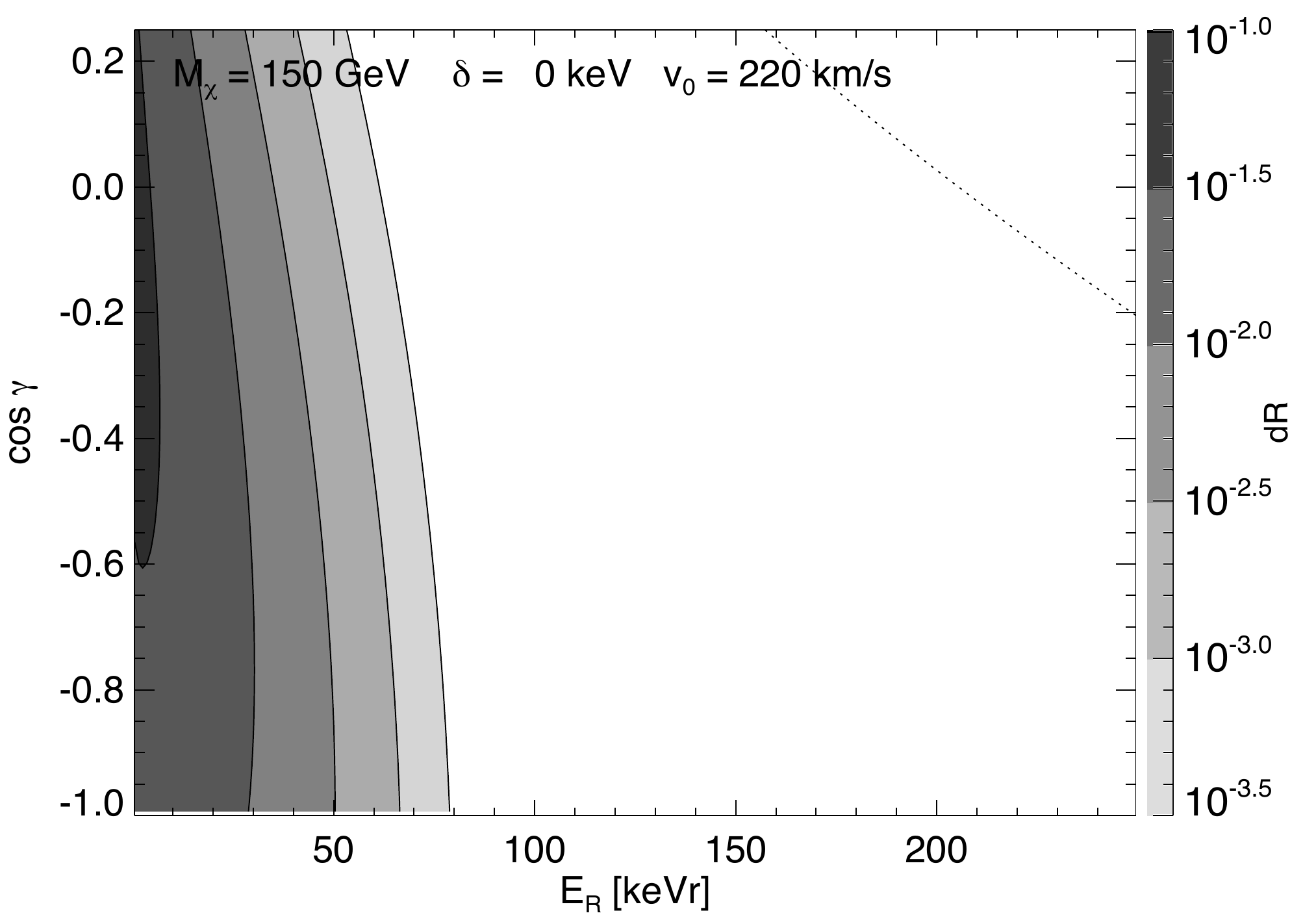}
\caption{\label{fig:theoryFig3} Differential rates $dR/(dE_R d cos
  \gamma)$ for the benchmark models given in
  Ref.~\protect\refcite{Finkbeiner2009} with $v_{esc}=500$~km/s, for
  (a) inelastic and (b) elastic scattering. In each case, the
  differential rate is normalized so that the total rate is
  unity. Outside the region indicated by the dashed line, scattering
  events are kinematically forbidden.  Note that inelastic scattering
  preserves directional information better than elastic scattering.}
\end{figure}

\subsection{Inelastic scattering}
\label{sec:theoryIDM}
There has been interest recently in inelastic WIMP scattering (the iDM
scenario). This happens if the WIMP effectively has a ground state and
an excited state separated by $\sim$100~keV, which can be the case in
models with a new dark sector force, composite dark matter or mirror
dark matter.\cite{TuckerSmith2001} This scenario is compatible with
all current direct detection experiments, including DAMA. Inelastic
scattering is qualitatively different from elastic scattering. The
event rate at low energies is suppressed, due to the energy required
to excite the higher energy state.  Furthermore, the nuclear recoils
are more correlated with the initial direction of the incident WIMP,
thereby producing a stronger directional signal than in the case of
elastic scattering (see Fig.~\ref{fig:theoryFig3} and
Ref.~\refcite{Finkbeiner2009}).  A directional experiment with a heavy
target (required for sufficient kinetic energy in the center of
momentum frame) could provide a crucial test of the iDM scenario even
with a modest exposure ($\sim10^3$~kg-day).\cite{Finkbeiner2009}

In summary, the direction dependence of the direct detection rate
provides a powerful tool for distinguishing WIMP-induced nuclear
recoils from background events, and can unambiguously demonstrate the
Galactic origin of a dark matter signal. A detector capable of
measuring the nuclear recoil vectors in 3-dimensions could detect the
directional signal with only of order 10 events. With a large exposure
it would be possible to do ``WIMP astronomy'' and measure the
ultra-local WIMP velocity distribution. Directional detection can also
test the inelastic scattering scenario.

%% file: sectionExperiment.tex
\section{Status of experimental efforts}
Directional dark matter experiments face several common challenges.
Most importantly, the detector must accurately reconstruct low-energy
(tens of keV) recoil tracks in its active volume.  In diffuse gas
detectors, these track lengths typically only extend a few
millimeters, which necessitates a readout system with high spatial
resolution.  Of particular interest is the head-tail, or vector
direction, of the recoil (see
Section~\ref{sec:theoryDirectionModulation}).  By leveraging the
non-uniform energy loss of recoiling nuclei as a function of recoil
distance ($dE/dx$) experiments can measure the head-tail of a recoil.

Several groups have well-established research programs devoted to
directional dark matter detection.  Various target materials and
readout techniques have been employed and evaluated.
Sections~\ref{sec:DRIFT}-\ref{sec:emulsions} present the current
status of, and recent results from five directional dark matter
detection experiments (DRIFT, DMTPC, NEWAGE, MIMAC and Nuclear
Emulsions).  For convenience, Table~\ref{tab:experimentSummary}
summarizes these experiments.  In addition,
Sections~\ref{sec:pixelChips} and \ref{sec:zaragoza} describe the
status of R\&D efforts on novel detector readout schemes: silicon
pixel chips and micromegas.

Even as these experiments work toward developing ton-scale detectors
with directional sensitivity, there is near-term science that can be
achieved with more modest target masses.  For example, an exposure of
0.1~kg-year with \cffour~gas (equivalent to three months of live time
with a one cubic meter detector filled to 75~Torr) would improve
current constraints on the spin-dependent cross section by a factor of
$\sim$50 over current limits.  In addition, in the iDM scenario (see
Section~\ref{sec:theoryIDM}), an exposure of $\sim$3~kg-yr with a
heavy target (e.g. Xenon) and directionality could either rule out or
support the DAMA/LIBRA signal under the inelastic dark matter
scenario.

\begin{figure}
\centering
\begin{tabular}{cc}
\includegraphics[height=0.3\textheight]{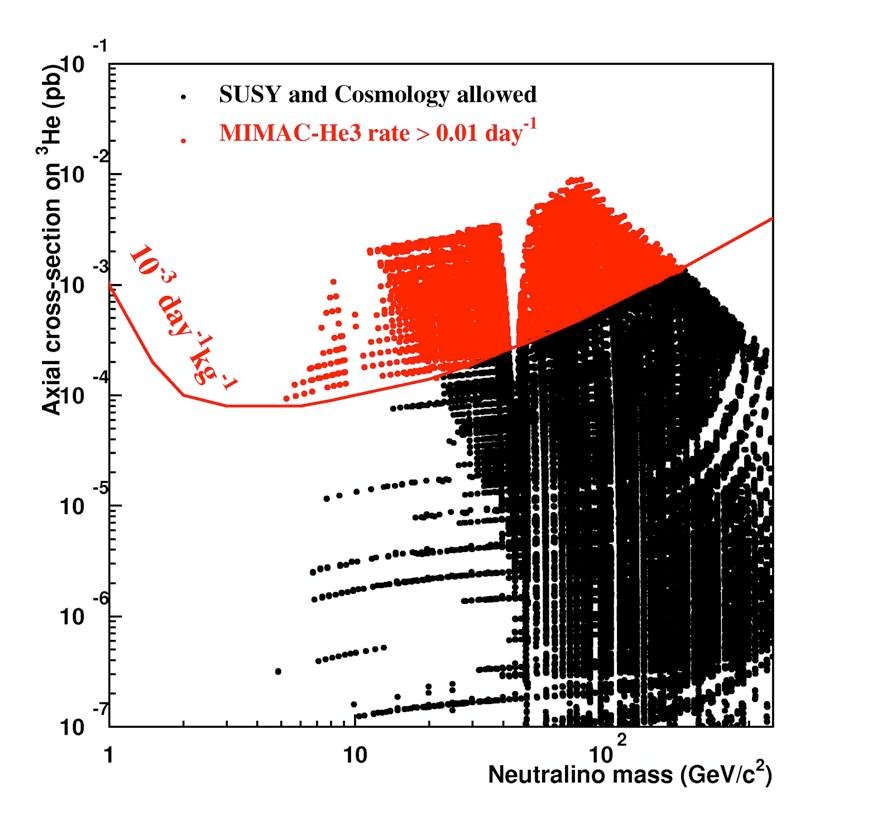}&
\includegraphics[width=0.3\textheight]{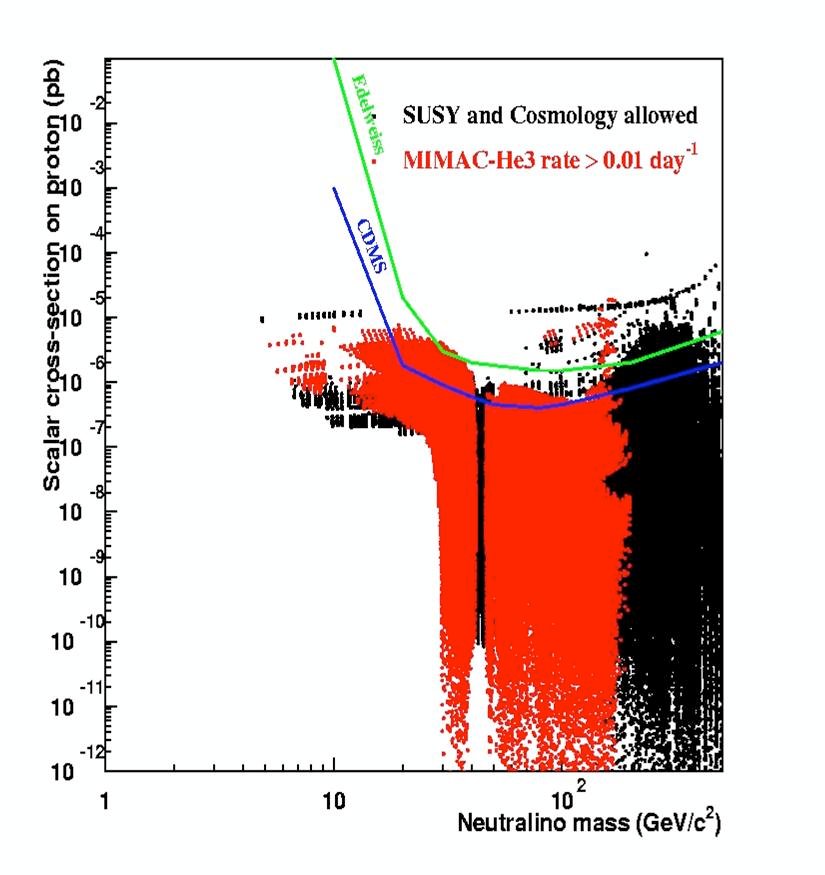}
\end{tabular}
\caption{SUSY non-minimal models, calculated with the DarkSusy
  code.\protect\cite{GondoloJCAP2004} (left) Axial (spin-dependent)
  cross section on nucleon as a function of the neutralino mass.
  (right) Scalar (spin-independent) cross section of the same
  models. In red the models giving an axial cross section higher than
  a projected exclusion curve for an experiment with 10~kg of
  \hett. These models are compared with the exlusion plots of scalar
  experiments (not updated). There are models, in red on the right,
  which require a level of sensitivity that will be very difficult to
  achieve with scalar
  experiments.\protect\cite{MayetPLB2002,MoulinPLB2005}}
\label{fig:mimac1}
\end{figure}

Finally, it has been shown\cite{MayetPLB2002,MoulinPLB2005} that WIMP
spin-independent and spin-dependent cross sections with nucleons can
be uncorrelated: meaning that a given supersymmetric dark matter
candidate may have a relatively large spin-dependent nuclear cross
section, but a very weak spin-independent cross section.  Therefore
even modest constraints on the spin-dependent cross section can rule
out SUSY models that will remain out of reach of traditional dark
matter direct detection experiments.  Fig.~\ref{fig:mimac1}
demonstrates this by showing the spin-dependent (left) and
spin-independent (right) cross sections for a class of SUSY models.
The curve on the left plot shows the hypothetical sensitivity of a
directional experiment that uses \het as a target gas.  The points
above the curve could be ruled out by such an experiment.  The
spin-independent plot shows that models ruled out by the \hett-based
experiment have a broad range of spin-independent cross sections,
extending below $10^{-12}$~pb, well below the sensitivity of the next
generation of traditional direct detection experiments.  Although
these plots are made for \het gas, similar results follow for other
targets with high spin-dependent sensitivity (e.g. fluorine).

\begin{table}
\tbl{Summary of existing directional dark matter detection experiments.  TPC stands for Time Projection Chamber, NITPC stands for negative-ion TPC, and SI and SD refer to spin-independent and spin-dependent WIMP-nucleon interactions.  The column labeled ``head-tail'' specifies whether the experiment has successfully demonstrated head-tail sensitivity.  The last column lists the active volume for each experiment.}
{\begin{tabular}{llllllr} \toprule
Collaboration & Technology & Target        & Interactions & Head-tail & Readout                & V (m$^3$) \\ \colrule
DRIFT         & NITPC      & CS$_2$, CS$_2$-CF$_4$  & SI/SD     & yes       & MWPC 2D + timing       & 1                   \\
DMTPC         & TPC        & CF$_4$                 & SI/SD     & yes       & Optical (CCD) 2D       & 0.01                \\
NEWAGE        & TPC        & CF$_4$                 & SI/SD     & no        & $\mu$PIC 2D + timing   & 0.03                \\
MIMAC         & TPC        & $^3$He/CF$_4$          & SI/SD     & yes       & Micromegas 2D + timing & 0.00013 \\
Emulsions     & emulsions  & AgBr                   & SI/SD     & no        & Microscope 3D          & N/A \\
\botrule
\end{tabular} \label{tab:experimentSummary}}
\end{table}

\input{sectionDRIFT.tex}     
\input{sectionDMTPC.tex}     
\input{sectionNEWAGE.tex}    
\input{sectionMIMAC.tex}     
\input{sectionEmulsions.tex} 
\input{sectionVahsen.tex}    
\input{sectionZaragoza.tex}  

%% file: sectionDRIFT.tex
\section{DRIFT -- Scale-up tests with low background and head-tail discrimination}
\label{sec:DRIFT}
The Directional Recoil Information From Tracks (DRIFT) dark matter
collaboration at Boulby has, since 2001, pioneered construction and
operation underground of low background directional TPCs at the
1~\mmm~scale with Multi-wire Proportional Counter (MWPC) readout using
negative ion (NITPC) CS$_2$ gas to suppress diffusion without magnetic
fields.\cite{MartoffNIMA2000,OhnukiNIMA2001} The NITPC concept, as
demonstrated first in DRIFT I, allows larger drift distances
($>50$~cm) than is feasible with conventional gases like \cffour,
thereby reducing the required readout area and hence
cost.\cite{AlnerNIMA2004,MorganNIMA2003} Operation with 1~m$^2$ MWPC
readout planes allows the study of realistic size detectors
underground with near-conventional technology.

In DRIFT, the ionization generated from recoil events (mainly S
recoils) goes to create tracks of CS$_2$ negative ions.  Under the
influence of an applied electric field, the negative ions drift to one
of the two back-to-back MWPC planes for readout.  The MWPCs include
two orthogonal layers (x and y) of 512 20~micron wires with 2~mm
spacing.  Wires are grouped to reduce the number of readout channels.
Reconstruction is feasible in 3D using timing information for the z
direction (perpendicular to the x-y plane of wires).  Additional R\&D
is underway to allow absolute z positioning, though some z positioning
is feasible already through pulse shape analysis.  Calibration is
undertaken typically every 6 hours using internal $^{55}$Fe sources
(one for each MWPC) that are shielded by an automated shutter system when not in use.

\begin{figure}
\centering
\includegraphics[width=0.7\textwidth]{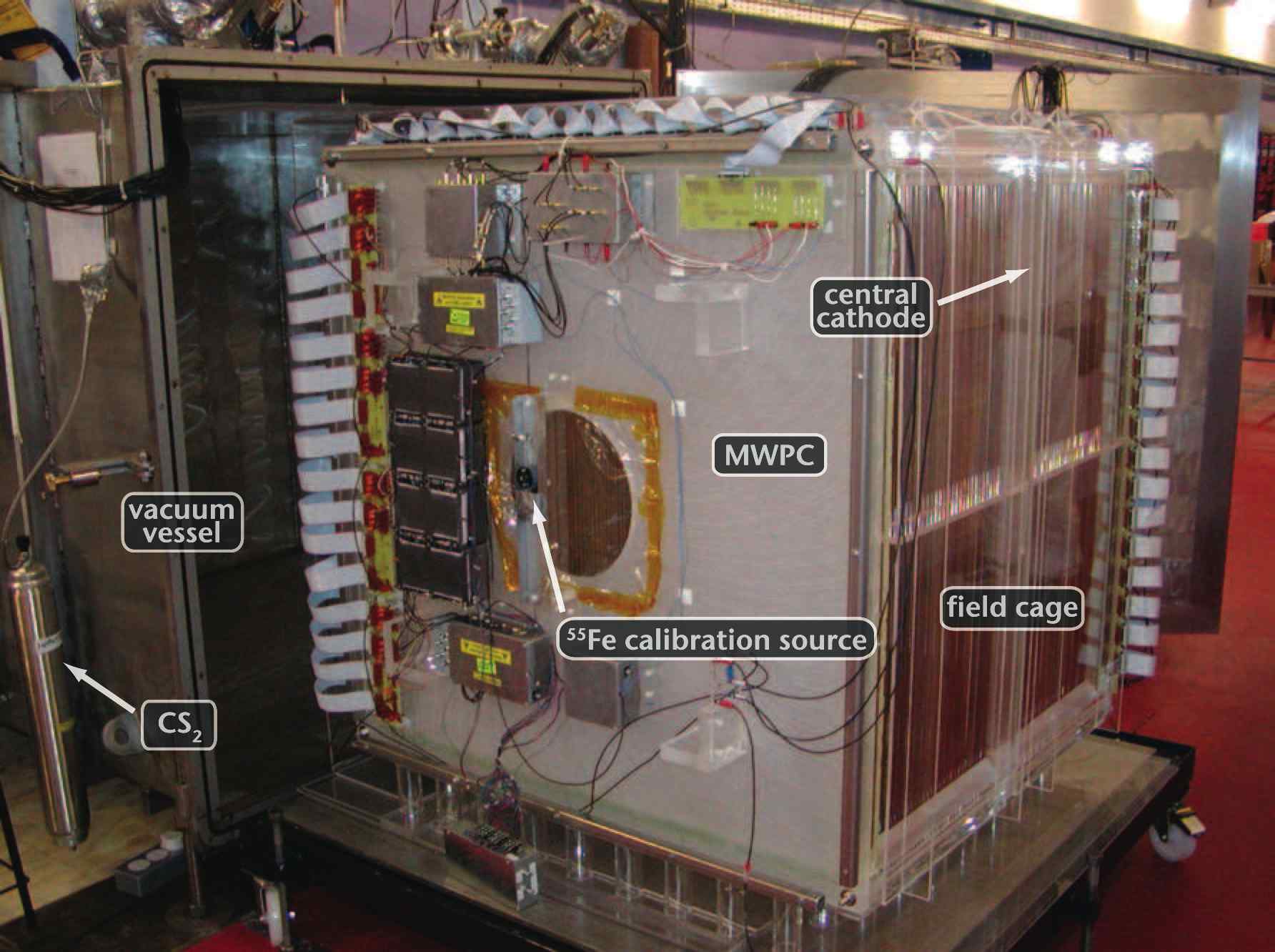}
\caption{\label{fig:driftFig1}DRIFT IIb detector at Boulby mine.  Two
  back-to-back TPCs each with a 50~cm drift distance, share a common
  vertical central cathode.  Readout is done with two 3-layer MWPCs
  with 2~mm wire spacing.  Operation is with negative ion \cstwo~gas
  at 40~Torr (170~g target mass).\protect\cite{AlnerNIMA2005}}
\end{figure}

The Boulby program, particularly with the second generation 1~m$^3$
scale DRIFT IIa-d experiments since 2006 (see Fig.~\ref{fig:driftFig1}
and Ref.~\refcite{AlnerNIMA2005}), has recently made progress on the
practical understanding of all background types for directional TPCs
operated underground, on scale-up issues such as safety and
backgrounds, and on directional sensitivity, for instance
demonstrating for the first time sensitivity to recoil direction sense
(head-tail discrimination) at low energy (47~keV S
recoil).\cite{BurgosAstroPart2009}

\begin{figure}
\centering
\includegraphics[width=0.8\textwidth]{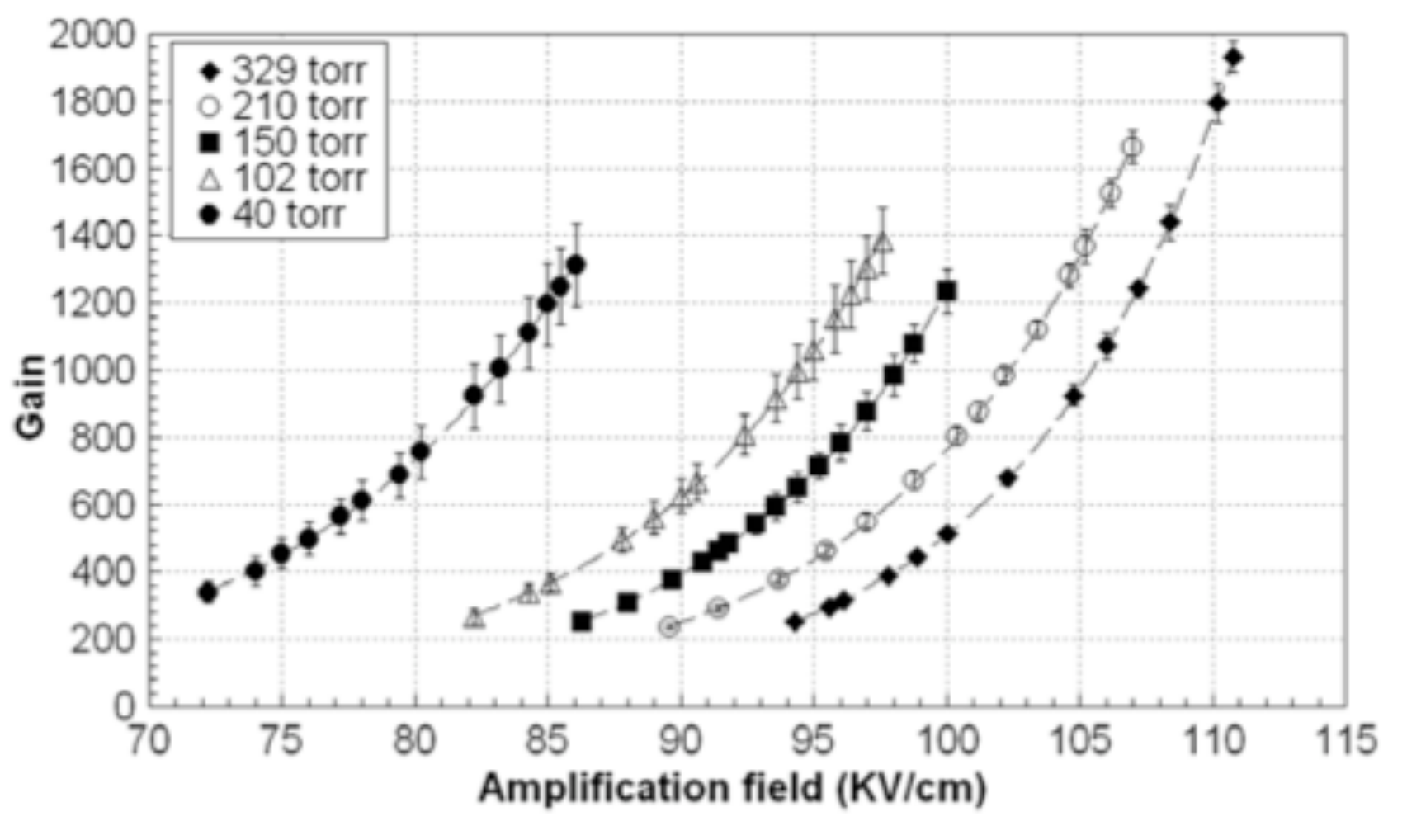}
\caption{Example gain curves for the operation of a 36~x~36~mm bulk
  micromegas device with negative ion \cstwo~gas.  From
  Ref.~\protect\refcite{LightfootAstroPart2007}.}
\label{fig:driftFig2}
\end{figure}

DRIFT recognizes that micromegas/strip readout can offer improvements
in resolution which could, for instance, allow for higher pressure
operation.\cite{MartoffNIMA2005} Initial small-scale studies have
already confirmed that micromegas technology can work with negative
ion CS$_2$ gas (see
Fig.~\ref{fig:driftFig2}).\cite{MiyamotoNIMA2004,LightfootAstroPart2007}
The parallel progress made by MIMAC/CAST (Grenoble, Darmstadt) in
developing larger area full x-y readout devices in the laboratory with
\cffour, including the necessary advanced electronics, is therefore
very well matched to the DRIFT activity, complementing the practical
experience of using the \mmm-scale low pressure TPCs of DRIFT
underground, and offering a collaborative route to demonstrating
feasibility of micromegas readout at this larger volume.  It builds
also on the long-standing Sheffield/Darmstadt collaboration to show
that DRIFT could be used for axion detection (see
Section~\ref{DRIFT:axions}).\cite{MorganAstroPart2005} Relevant
advances from DRIFT are outlined in the following sections (DRIFT is a
UK/US collaboration).

\subsection{Background rejection and event fiducialisation}
\begin{figure}
\centering
\includegraphics[width=0.31\textwidth]{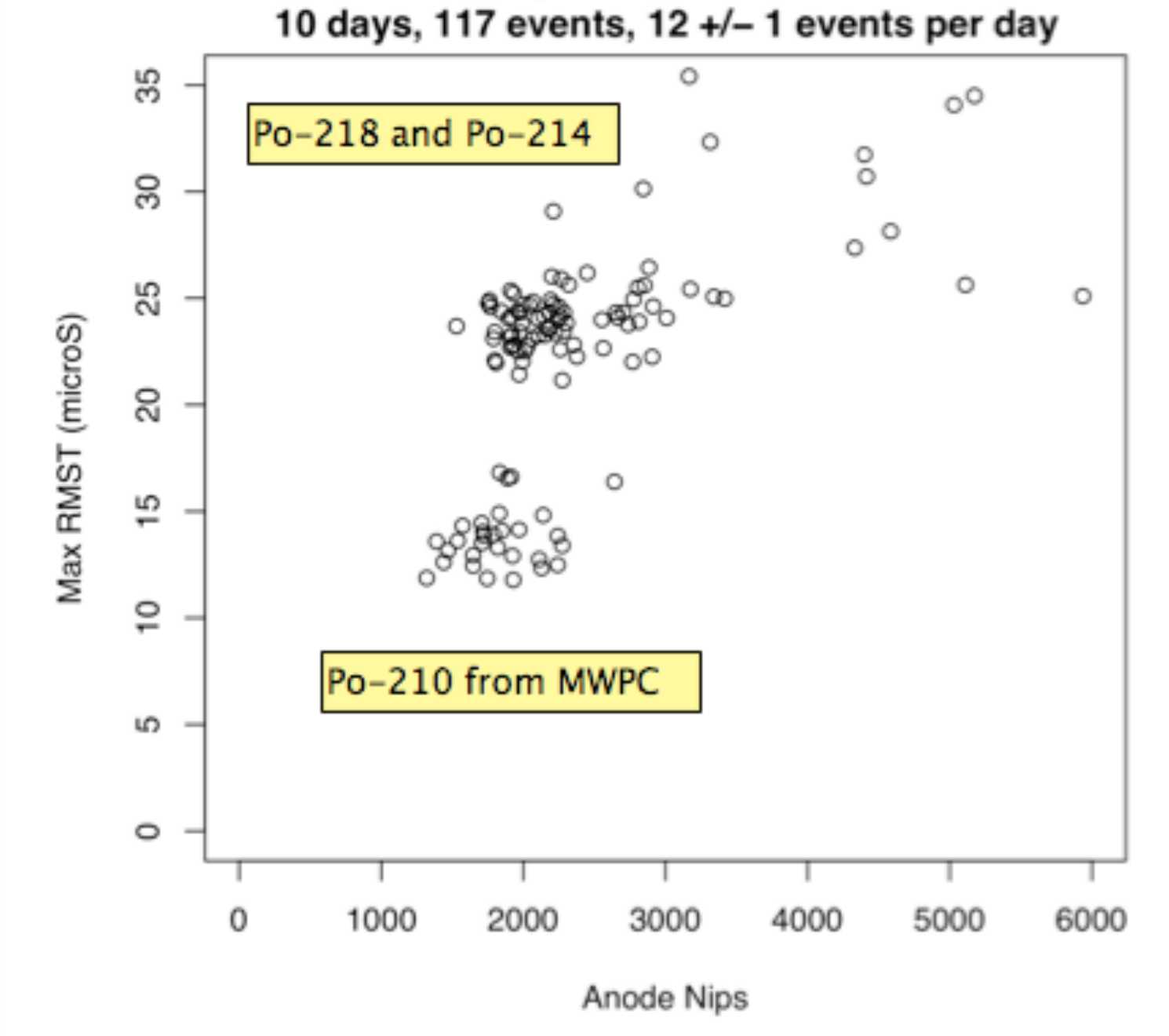}
\includegraphics[width=0.31\textwidth]{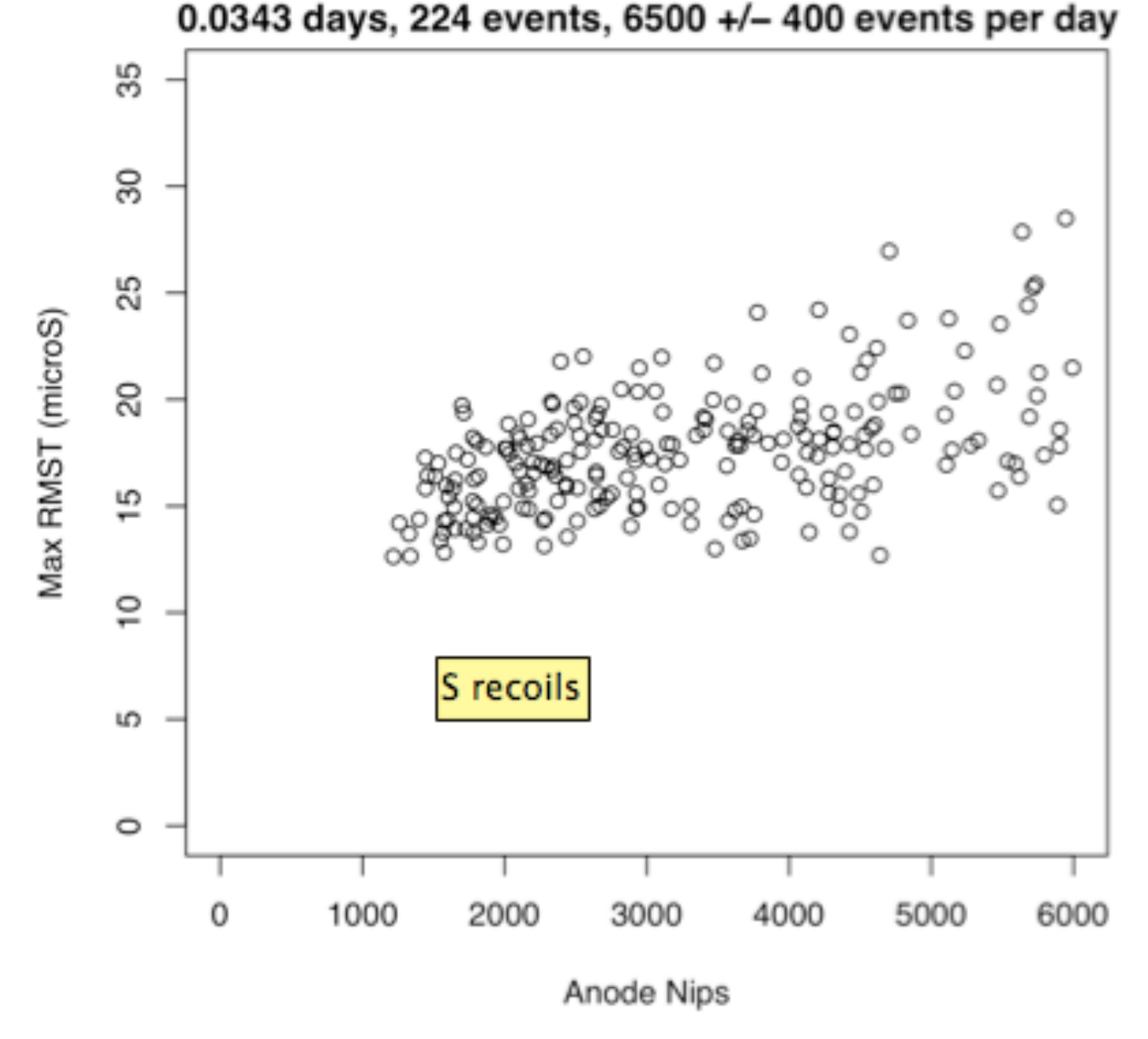}
\includegraphics[width=0.31\textwidth]{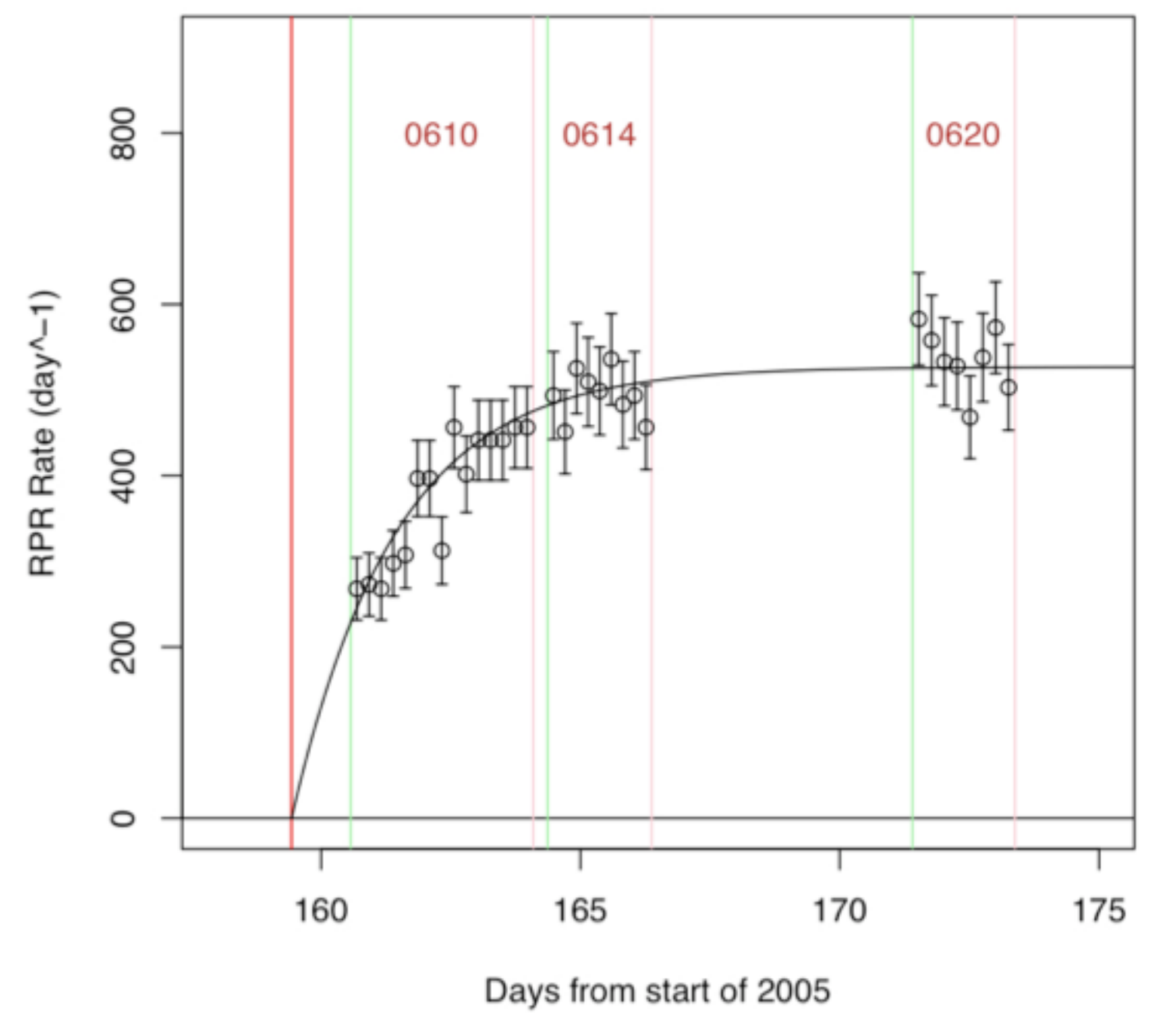}
\caption{(left) Identified RPR low background events in DRIFT IIb. (middle) Low energy ($>40$~keV) calibration sulphur recoils from a neutron exposure (gamma rejection is $>10^8$). (right) Rise in $^{210}$Po RPR events with time following new gas fill fitted to expectation of RPR hypothesis.  From Ref.~\protect\refcite{BurgosNIMA2008}.}
\label{fig:driftFig3}
\end{figure}

The strong gamma background rejection capability of the low-pressure
TPC technology with CS$_2$, typically $>10^8$, together with
maintenance of recoil detection efficiency $>50$\% (calibrated with
neutron sources) has been demonstrated by
DRIFT~II.\cite{BurgosAstroPart2007,CarsonNIMA2005} More significant is
the discovery, understanding and mitigation of specific radon related
background events that are an issue for the scale-up of any dark
matter directional TPC.\cite{BurgosNIMA2009} Called Radon Progeny
Recoils (RPRs), the origin of this important background in TPCs has
been identified by us as mainly $^{210}$Po recoils from the central
cathode (see Fig.~\ref{fig:driftFig3}).  We have developed new radon
reduction procedures via nitric acid
cleaning,\cite{driftNitricCleaning} new fiducialisation analysis
software, and a material selection process using a radon emanation
facility built for this purpose.  Our latest analysis demonstrates
that these techniques reduce the number of RPRs by a factor of 3000,
yielding a background rate in DRIFT IId of $\sim$1 event per week
above threshold (40~keV recoil) for a target mass of 167~g CS$_2$.
Further ideas for reducing the remaining RPRs to insignificant levels
are being investigated by DRIFT.  Elimination of RPRs would be a major
breakthrough for directional technology towards zero background.

\subsection{Directionality including recoil sense (head-tail)}
\begin{figure}
\centering
\includegraphics[width=0.9\textwidth]{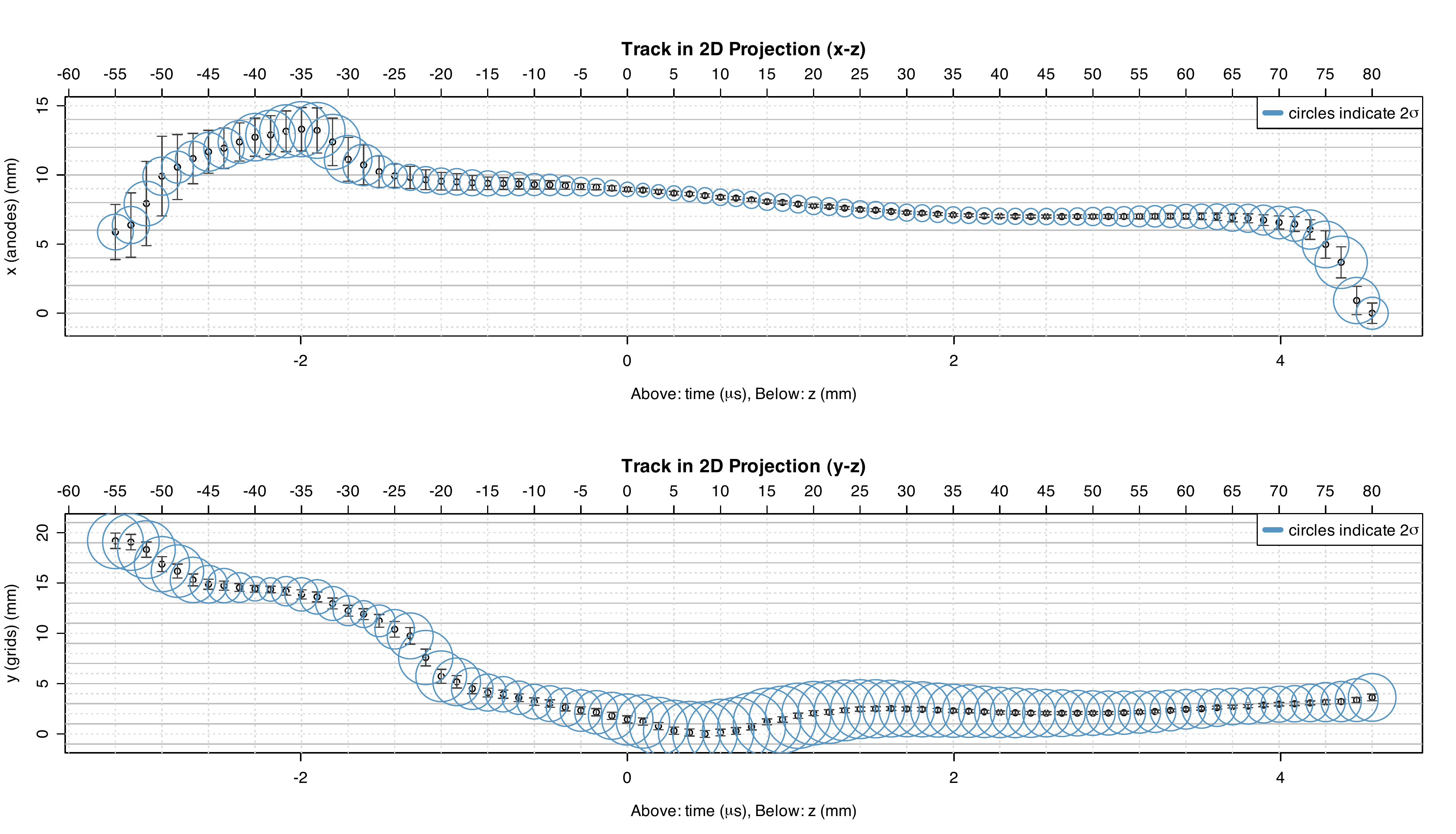}
\caption{Example 3D reconstruction (x-z and y-z projections) of a $\sim$100~keV S recoil in DRIFT IIb (circle sizes indicate the amount of deposited charge).}
\label{fig:driftFig4}
\end{figure}

\begin{figure}
\centering
\includegraphics[width=0.45\textwidth]{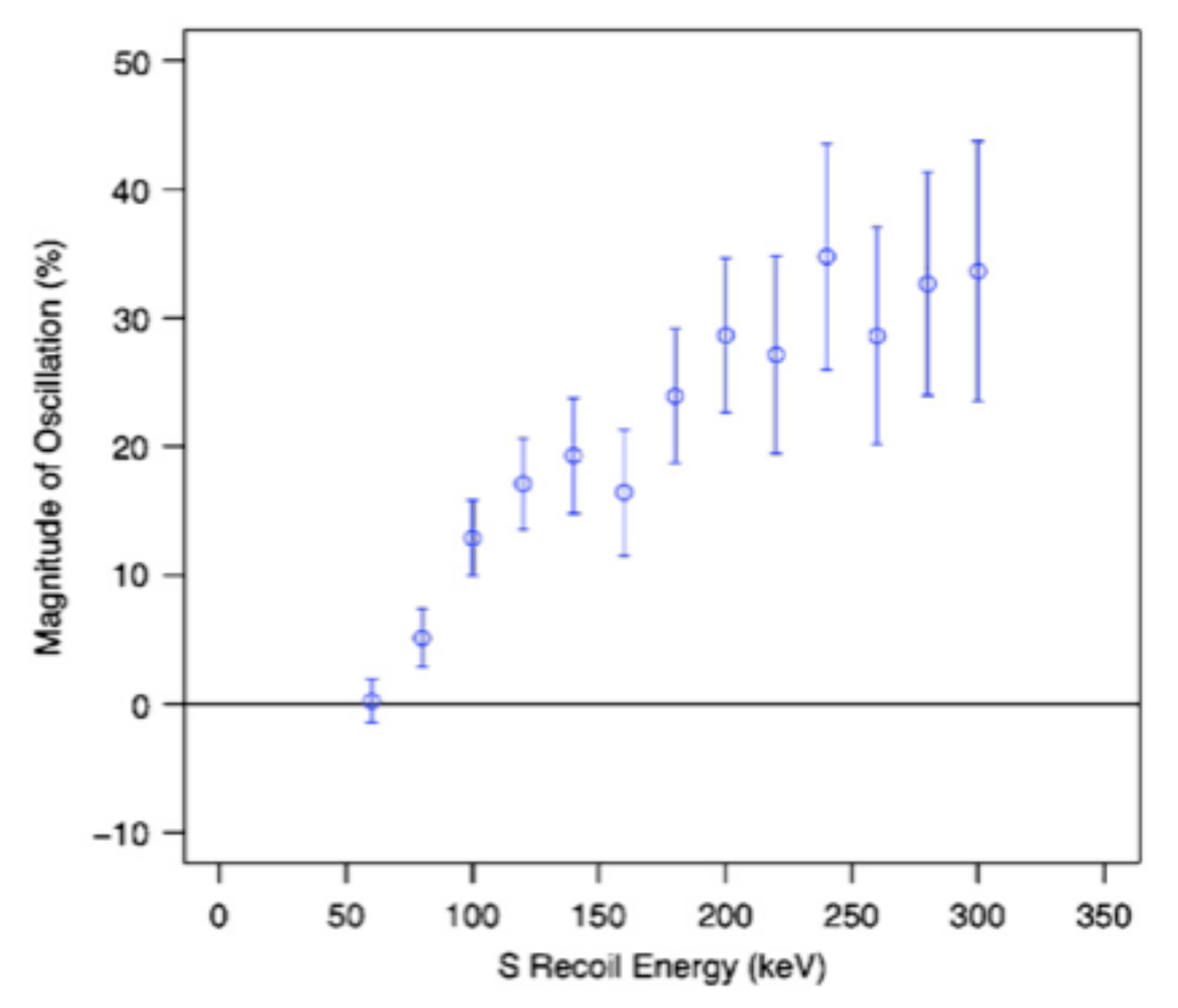}
\includegraphics[width=0.45\textwidth]{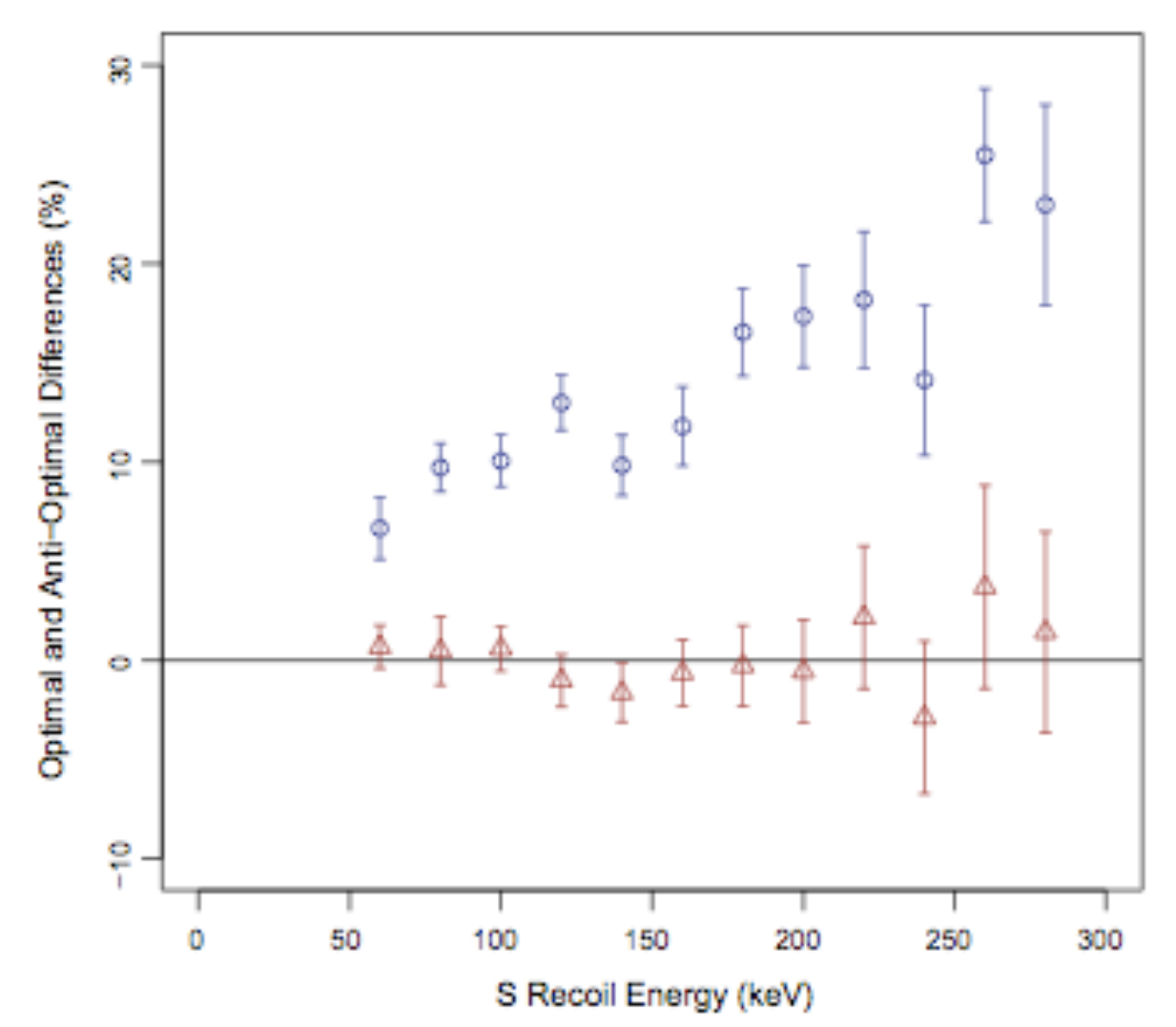}
\caption{Low energy (down to 40~keV S recoil) directional signals from DRIFT IIb: (left) ratio of $\Delta$z to $\Delta$x ranges vs energy and (right) head-tail asymmetry measured by the proportion of observed in first and second half of recoil tracks projected in the z direction.  From Ref.~\protect\refcite{BurgosNIMA2009}.}
\label{fig:driftFig5}
\end{figure}

\begin{figure}
\centering
\includegraphics[width=0.9\textwidth]{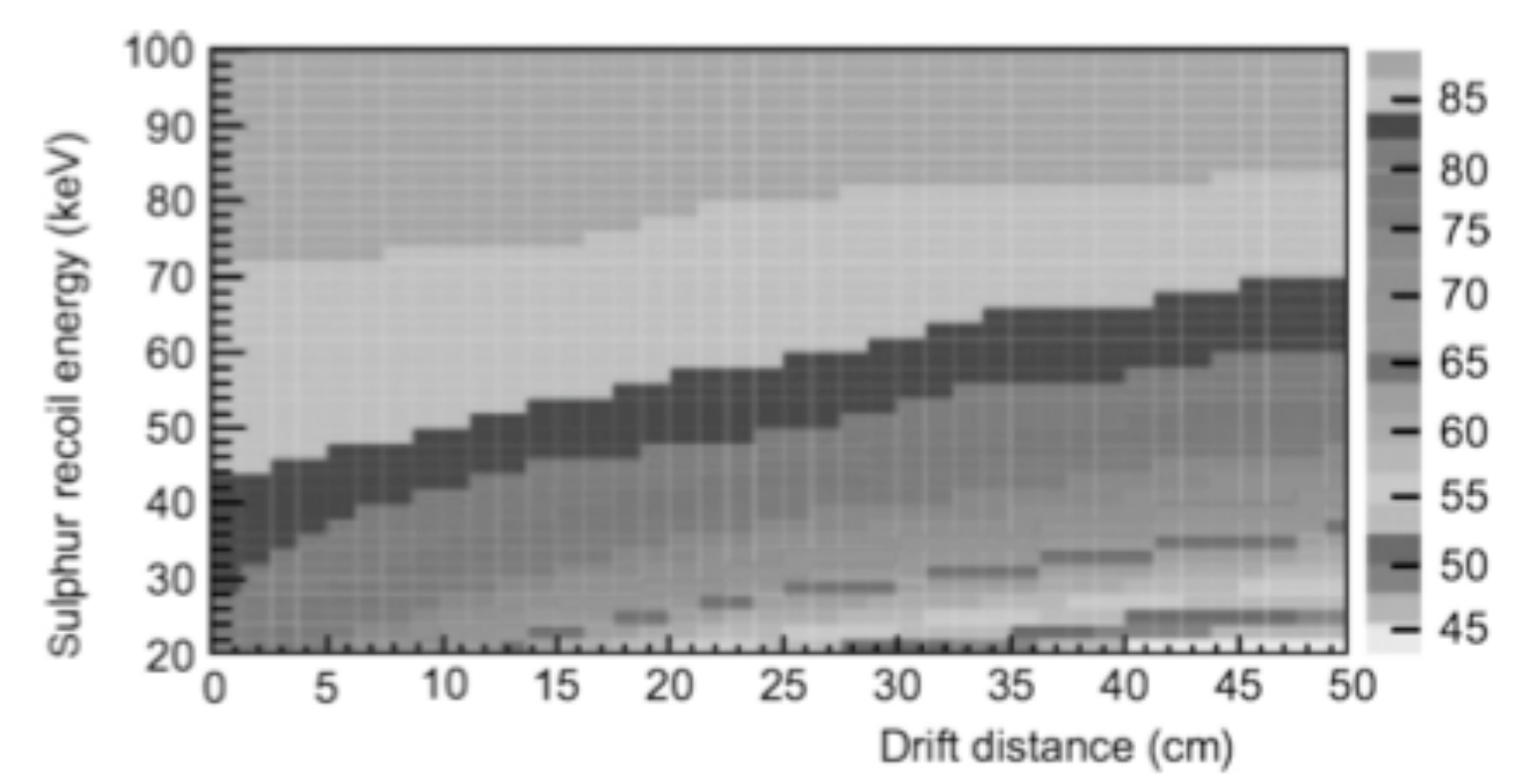}
\caption{Simulated angular reconstruction accuracy for DRIFT-II.
  Shading denotes the probability that the recoil direction is
  reconstructed within 30 degrees of the known initial direction.}
\label{fig:driftDirectionalResolution}
\end{figure}

3D track reconstruction of nuclear recoils, electrons and alphas has
now been demonstrated in DRIFT II (see
Fig.~\ref{fig:driftFig4}).\cite{BurgosNIMA2009,munaPhD2008,SpoonerPhysSocJap2007}
The latter has been shown also to provide a powerful diagnostic for
background identification via particle $dE/dx$ and range.  However, a
simpler 2D analysis also provides directional sensitivity (see
Fig.~\ref{fig:driftFig5} left).\cite{BurgosNIMA2009} There is a strong
dependence of the 3D directional resolution on the track orientation
relative to the x, y, z planes, on the absolute position in z, and on
the recoil energy.  Results of a simulation to illustrate this are
shown in Fig.~\ref{fig:driftDirectionalResolution}.  Here, shadings
represent the probability that the reconstruction is accurate to
within 30 degrees of the true recoil direction for events parallel to
the x-y plane.\cite{BurgosNIMA2009}

It has been a long-time goal of directional detectors to determine
(both theoretically and experimentally) whether the absolute direction
of low energy (10s of keV) recoils can be seen.  This has the
potential to improve directional sensitivity by a factor of 10,
through better correlation with galactic
motion.\cite{MorganPhysRevD2005} DRIFT has recently succeeded in this,
demonstrating for the first time at realistic recoil energies that a
clear head-tail asymmetry is observable (see Fig.~\ref{fig:driftFig5}
right).\cite{BurgosAstroPart2009} Our recent work with gas theorist
Akira Hitachi now also provides a rigorous theoretical basis for this,
in agreement with measurements.\cite{Majewski2009} 

\subsection{Low threshold operation (3 keV) and particle identification including axions}
\label{DRIFT:axions}
\begin{figure}
\centering
\includegraphics[width=0.45\textwidth]{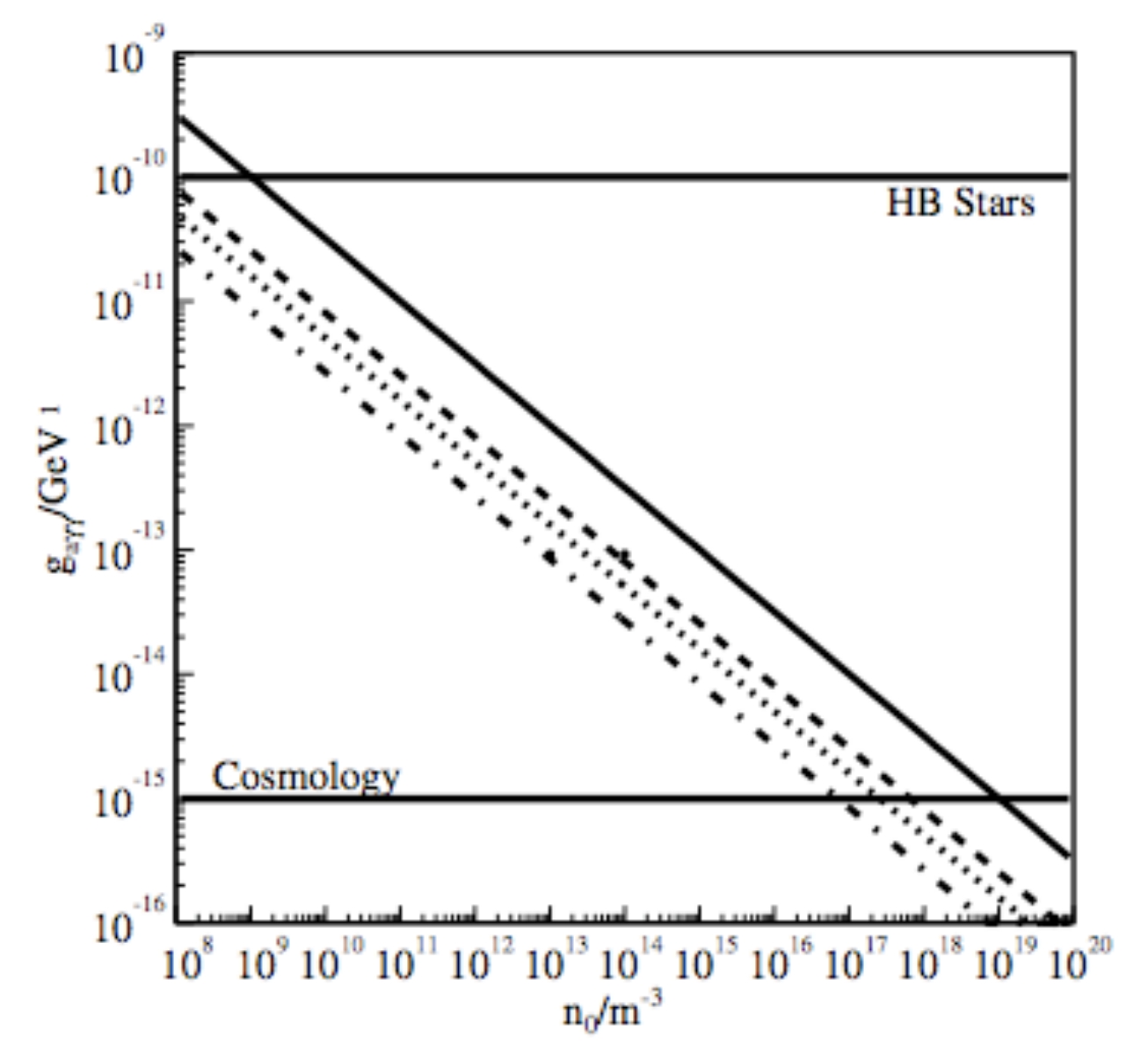}
\includegraphics[width=0.45\textwidth]{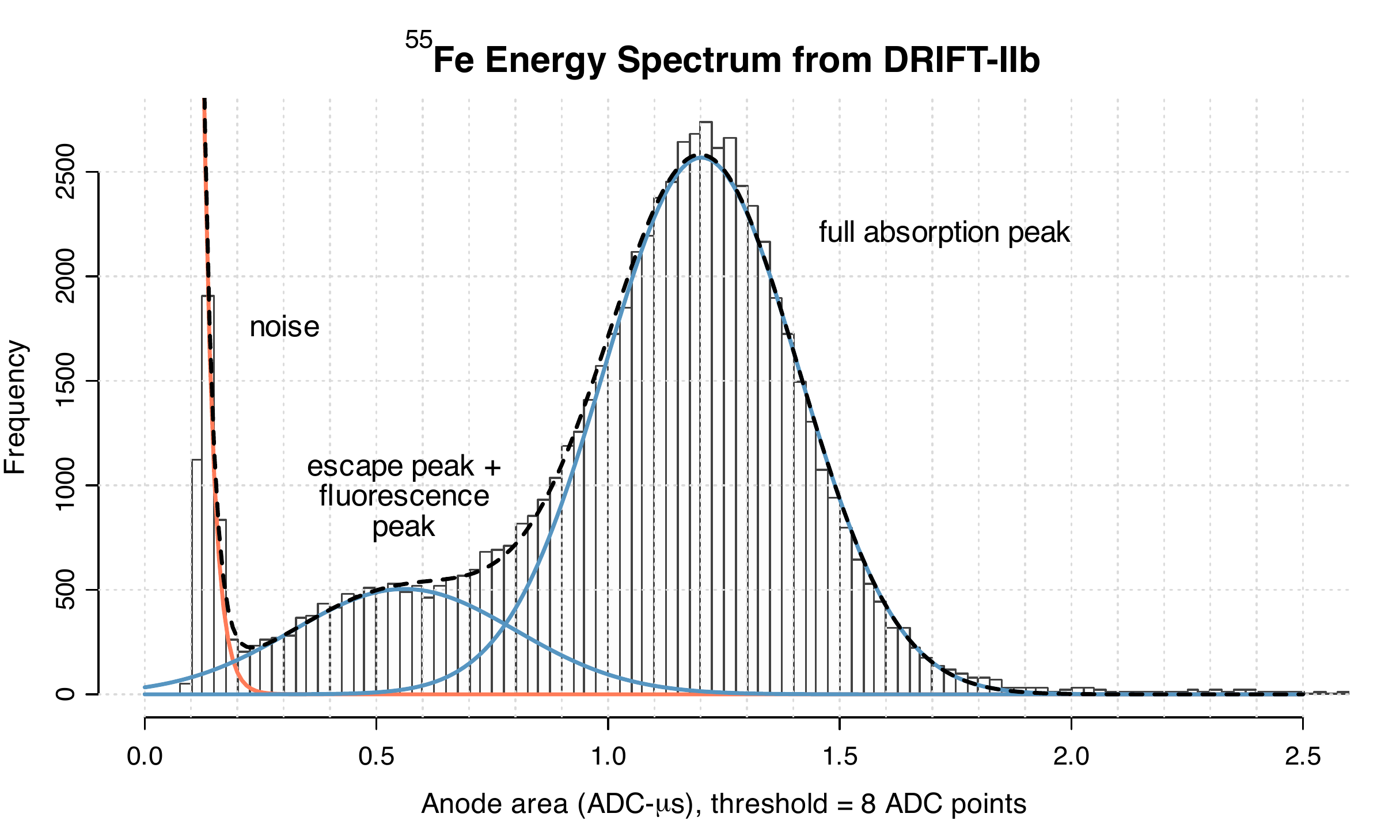}
\caption{(Left) predicted sensitivity of  1~m$^3$ CS$_2$ TPC at 0.2~bar to KK axions based on tracking of low energy electrons (see Ref. 8) and (right) measured energy threshold in DRIFT IIb using $^{55}$Fe.  From Ref.~\protect\refcite{BurgosJINST2009}.}
\label{fig:driftFig6}
\end{figure}

The potential advantages in WIMP detection of achieving a low
(sub-10~keV) recoil and electron energy threshold in a low pressure
TPC have been
recognized.\cite{AlnerNIMA2004,AlnerNIMA2005,MorganNIMA2003,SnowdenIfftNIMA2004}
However, simulation work through the Sheffield/Darmstadt collaboration
has also established the potential in this situation for axion
detection, specifically KK axions (see
Fig.~\ref{fig:driftFig6}).\cite{BattestiLectNotesPhys2008,MorganAstroPart2005}
More recently, we have shown experimentally that recoil thresholds
$<3$~keV (without directional information) are indeed feasible in the
1~m$^3$ DRIFT volume, even with the current commercial
electronics.\cite{BurgosJINST2009} Fig.~\ref{fig:driftFig6} shows a
sample $^{55}$Fe spectrum (6~keV photopeak) taken from a full-volume
exposure to the 1~m$^3$ DRIFT IIb CS$_2$ detector at Boulby.  Data
filtering algorithms were applied here, however new low-noise
electronics developed at Sheffield should eliminate the need for this,
opening prospects for S recoil trigger thresholds of 2--4~keV (lower
than achieved by the ionisation/phonon bolometric dark matter
experiments).\cite{BurgosJINST2009} The current results (see
Fig.~\ref{fig:driftFig6}) after pulse filtering indicate sensitivity
to electrons at 1.2~keV and hence S recoils at 3.5~keV, the former
with a resolution of 17.5\% at the 5.9~keV peak.

\subsection{Alternative target nuclei and spin-dependent sensitivity}
\begin{figure}
\centering
\includegraphics[width=0.45\textwidth]{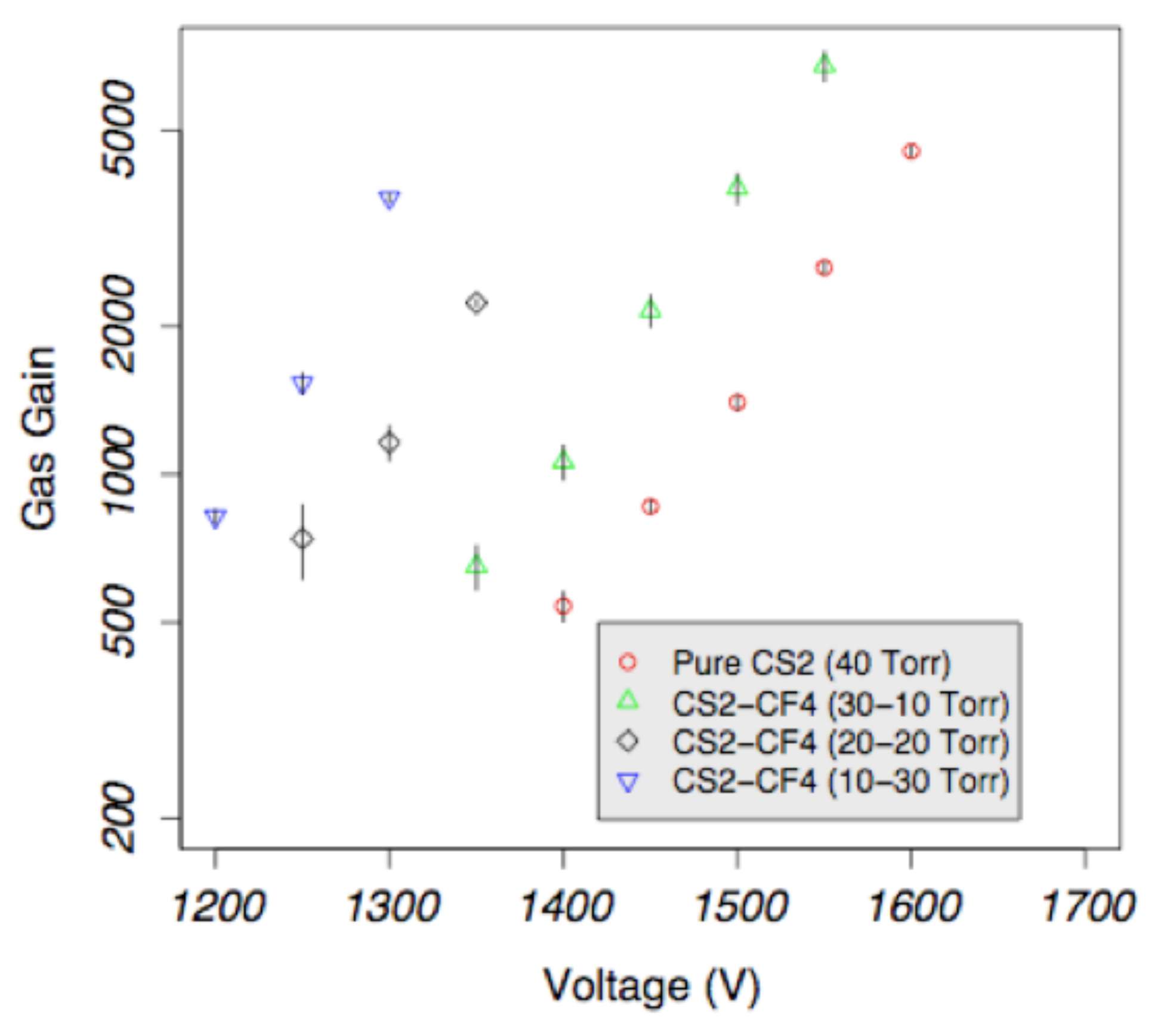}
\includegraphics[width=0.45\textwidth]{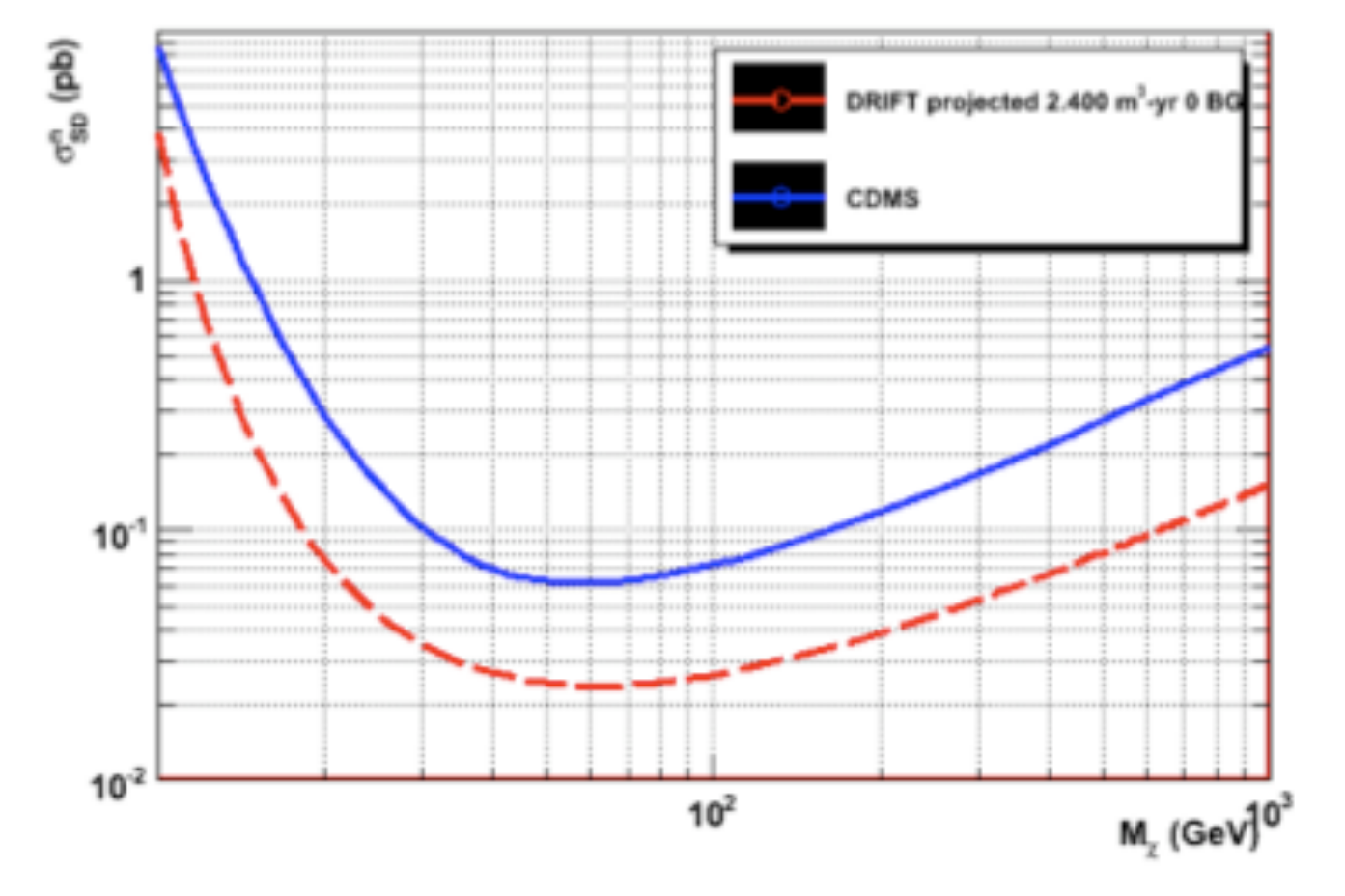}
\caption{(left) Gain curves measured for various \cstwo--\cffour~mixtures confirming operation in a small chamber; (right) predicted spin-dependent WIMP-neutron sensitivity with a 10~day exposure of 1~\mmm~with a 50\%--50\% mix of \cstwo~and \cffour.  The projected 90\% CL limits assume zero background and 100\% efficiency.}
\label{fig:driftFig7}
\end{figure}

It has long been postulated that CS$_2$ can be used as a ``carrier''
for other gases such that alternative target nuclei can be used while
maintaining the low diffusion advantages of the negative ion drift
CS$_2$.\cite{crane1961} Our studies with low pressure CS$_2$ mixtures
in small chambers have now shown excellent gain and stability behavior
with various gases including CF$_4$ and He (see
Fig.~\ref{fig:driftFig7}).\cite{MartoffNIMA2005,Pushkin2009} This
opens a clear avenue for operation with F and $^3$He targets for
spin-dependent sensitivity incorporating the reduced diffusion offered
by the CS$_2$.  In pure CS$_2$, the diffusion constant is reduced to
thermal levels.\cite{OhnukiNIMA2001} Recent measurements of $W$, the
average energy required to generate an eletron-ion pair, and mobility
with CS$_2$-CF$_4$ mixtures demonstrate operation in negative ion mode
with CF$_4$ and suggest near thermal diffusion, though more
measurement are needed to confirm this.\cite{Pushkin2009} Based on
these results, and those above, we show in Fig.~\ref{fig:driftFig7}
predicted limits for spin-dependent interactions for a 1~m$^3$
CS$_2$--CF$_4$ mixture.

\subsection{Towards scale-up with low background}
\begin{figure}
\centering
\includegraphics[width=0.4\textwidth]{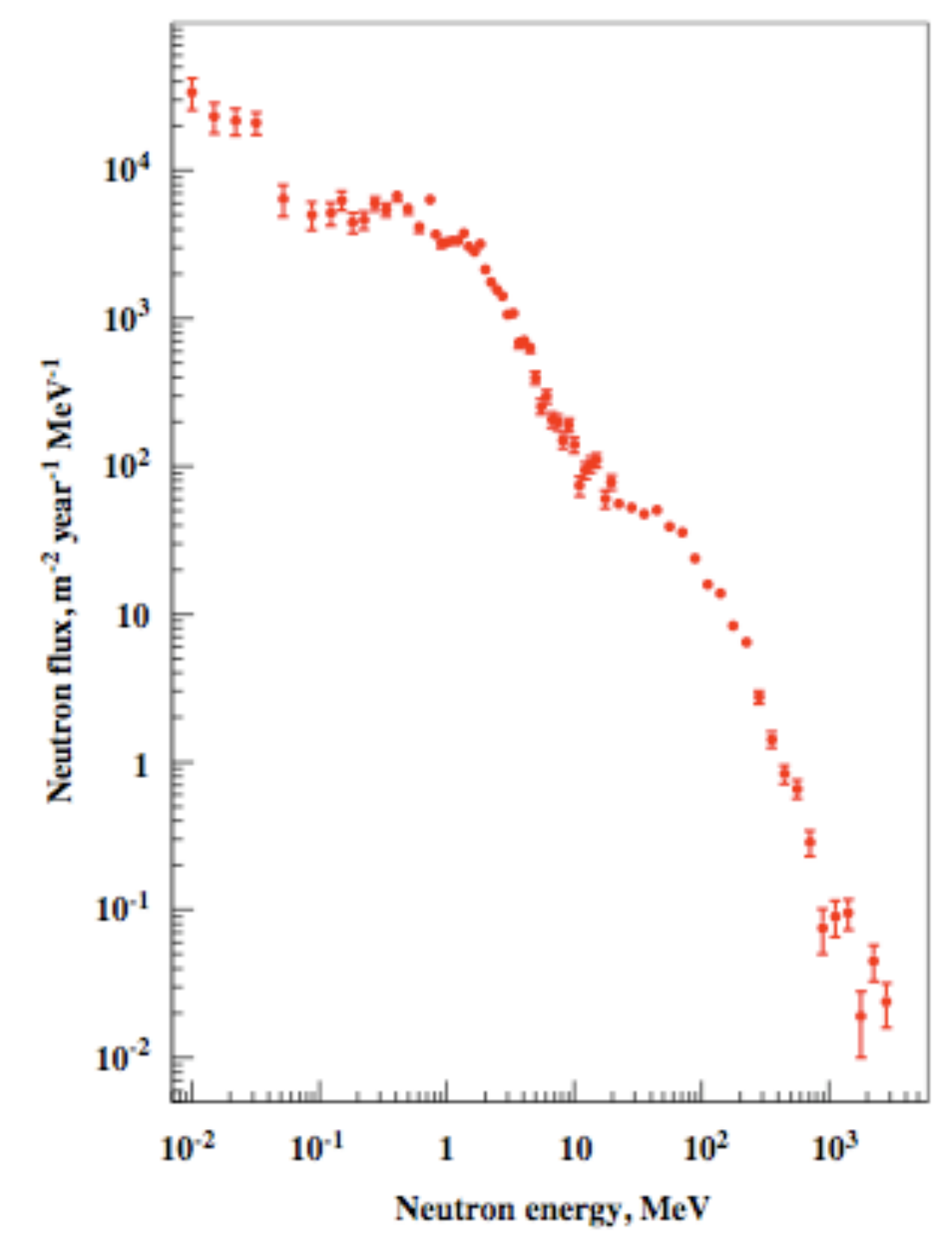}\hfill
\includegraphics[width=0.4\textwidth]{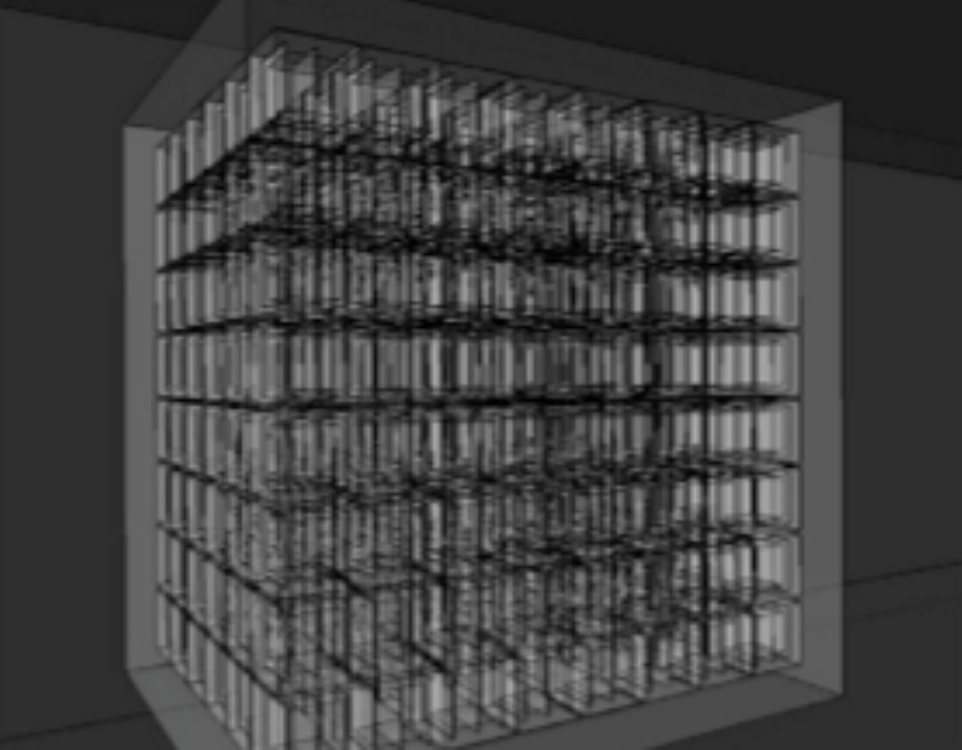}
\caption{(left) Simulated energy spectrum of muon-induced neutrons
  entering a (right) model multi-module low pressure TPC (vessel/gas
  boundary).  From Ref.~\protect\refcite{CarsonNIMA2005}.}
\label{fig:driftFig8}
\end{figure}

The DRIFT detectors at Boulby have also demonstrated that safe
operation with long term stability (operation for many months with
minimal gain shifts, trip-outs or alarms) is feasibile at the
\mmm~scale underground.  This is an essential step towards scale-up
designs.  However, of critical importance now is a deeper
understanding of backgrounds relevant to large volume, more sensitive
directional TPCs, particularly of muon, muon-induced neutron and
detector neutron backgrounds, for instance to determine the need for
an external bulk neutron veto.  Significant preliminary work has been
covered by our simulation activity (with GEANT4 and FLUKA) on
muon-induced
neutrons\cite{AraujoNIMA2005,KudryavtsevNIMA2003,LemraniNIMA2006}
backed by dedicated measurements at Boulby to understand the muon flux
at low and high energy,\cite{AraujoAstroPart2008,RobinsonNIMA2003} and
the relevant fast neutron flux.\cite{TziaferiAstroPart2007}
Fig.~\ref{fig:driftFig8} shows example simulation flux predictions for
a multi-module scaled-up TPC array.\cite{CarsonNIMA2005}

%% file: sectionDMTPC.tex
\section{The Dark Matter Time Projection Chamber (DMTPC) collaboration}
The Dark Matter Time Projection Chamber (DMTPC) collaboration has
developed and operated a 10-liter gas-based directional dark matter
detector.  The current instrument consists of a dual TPC, filled with
\cffour~gas at $\sim$75~Torr.  Proportional scintillation from the
avalanches is read out with two CCD cameras.  The charge on the TPC
anode is also measured.  With this instrument, DMTPC has demonstrated
head-tail sensitivity for neutron-induced recoils above 100~keV, and
an angular resolution for track reconstruction of 15$^\circ$ at
100~keV.  An 18-liter detector is currently under construction and
will be run underground at the Waste Isolation Pilot Plant (WIPP) at a
depth of 1600 meters water equivalent.

\subsection{Detector description and performance}
\begin{figure}
\centering 
\includegraphics[width=0.6\textwidth]{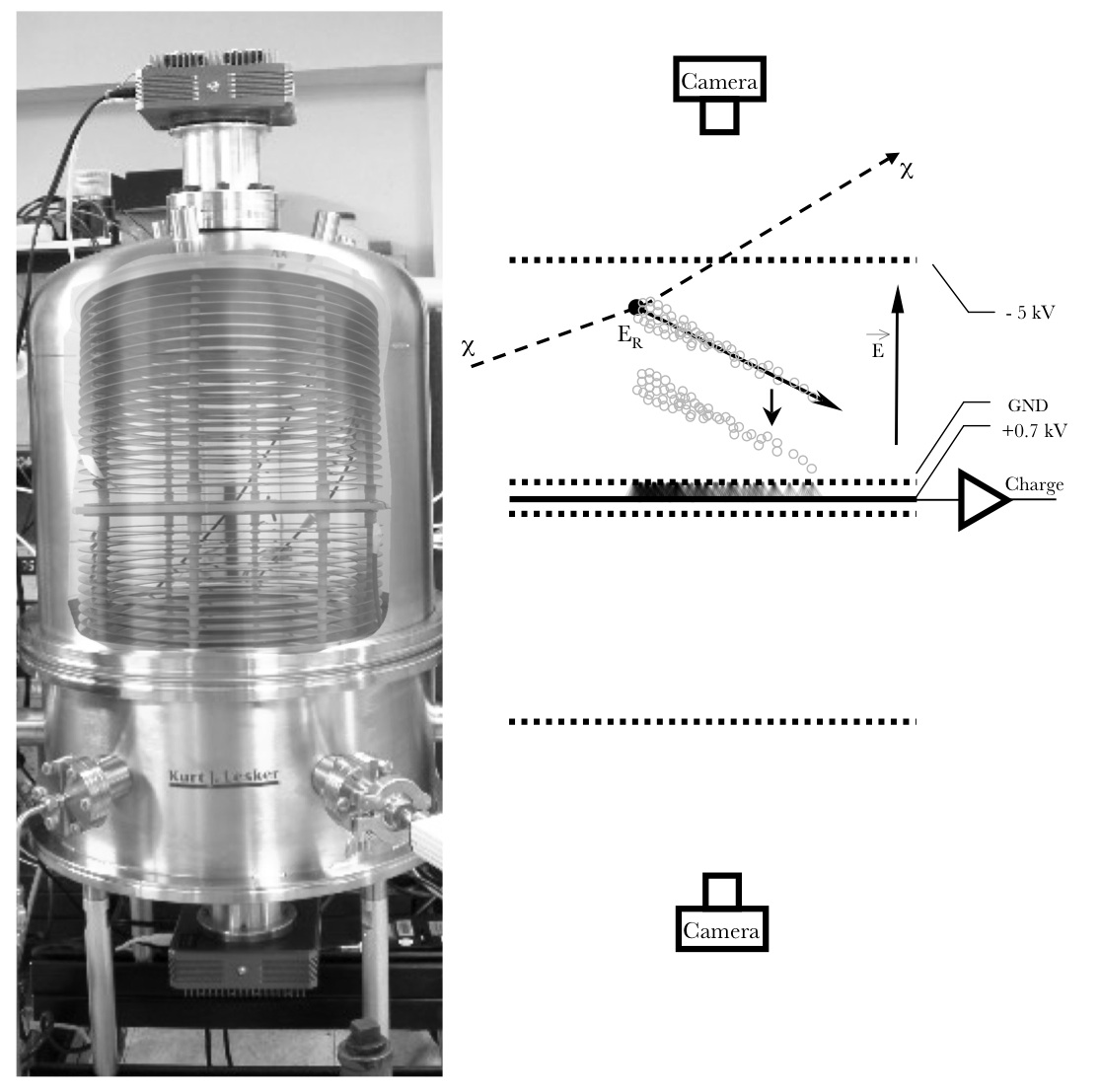}
\caption{(left) Photograph of the 10-liter DMTPC detector with an
  image of the dual TPC overlaid to provide an artificial glimpse
  inside the vacuum vessel.  The CCD cameras (top and bottom) each
  image an amplification region.  The stack of stainless steel field
  shaping rings condition the drift fields.  (right) A schematic
  representation of a WIMP-nucleus elastic scattering event in the
  detector.}
\label{fig:dmtpcDetector}
\end{figure}

The 10-liter DMTPC detector is shown in
Fig.~\ref{fig:dmtpcDetector}.  The dual-TPC is housed inside a
stainless steel vacuum vessel.  The drift region is defined by a
woven mesh cathode, typically at a potential of -5~kV, separated
from a wire mesh (28~$\mu$m wire, 256~$\mu$m pitch) ground grid 20~cm
away.  The vertical drift field is kept uniform to within 1\% by
stainless steel field-shaping rings spaced 1~cm apart.  An
amplification region is formed between the ground grid and a
copper-clad G10 anode plane (at 720~V) which are separated from each
other by 500~$\mu$m using resistive spacers (currently fishing line).
A charge amplifier connected to the anode measures the ionization
generated by a particle moving through the detector.  A CCD camera
images the proportional scintillation light generated in the
amplification region.  The CCD camera and readout electronics are
located outside of the vacuum vessel.  The mesh-based amplification
region allows for two-dimensional images of charged particle tracks.

\begin{figure}\centering
\includegraphics[width=0.45\textwidth]{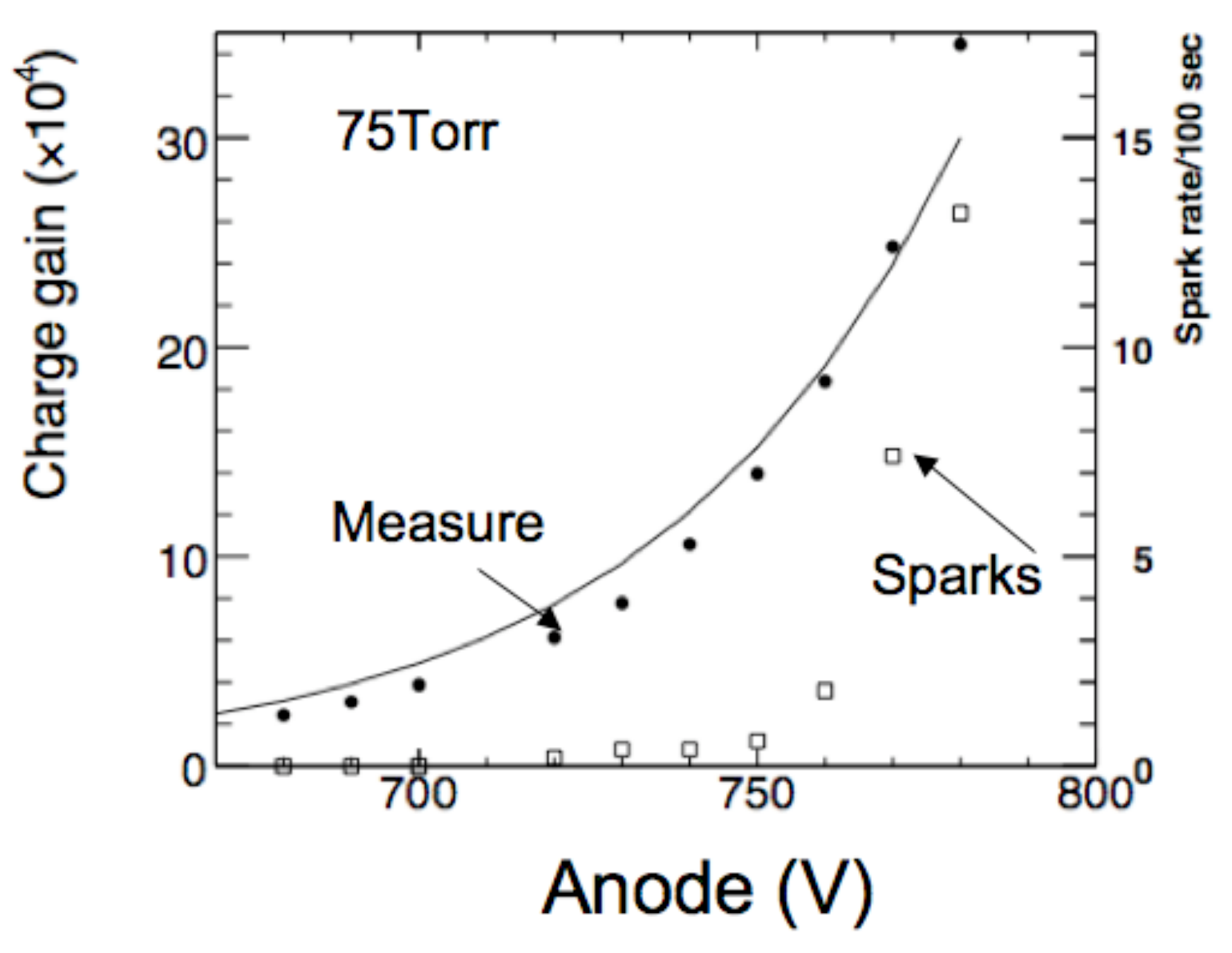}
\includegraphics[width=0.45\textwidth]{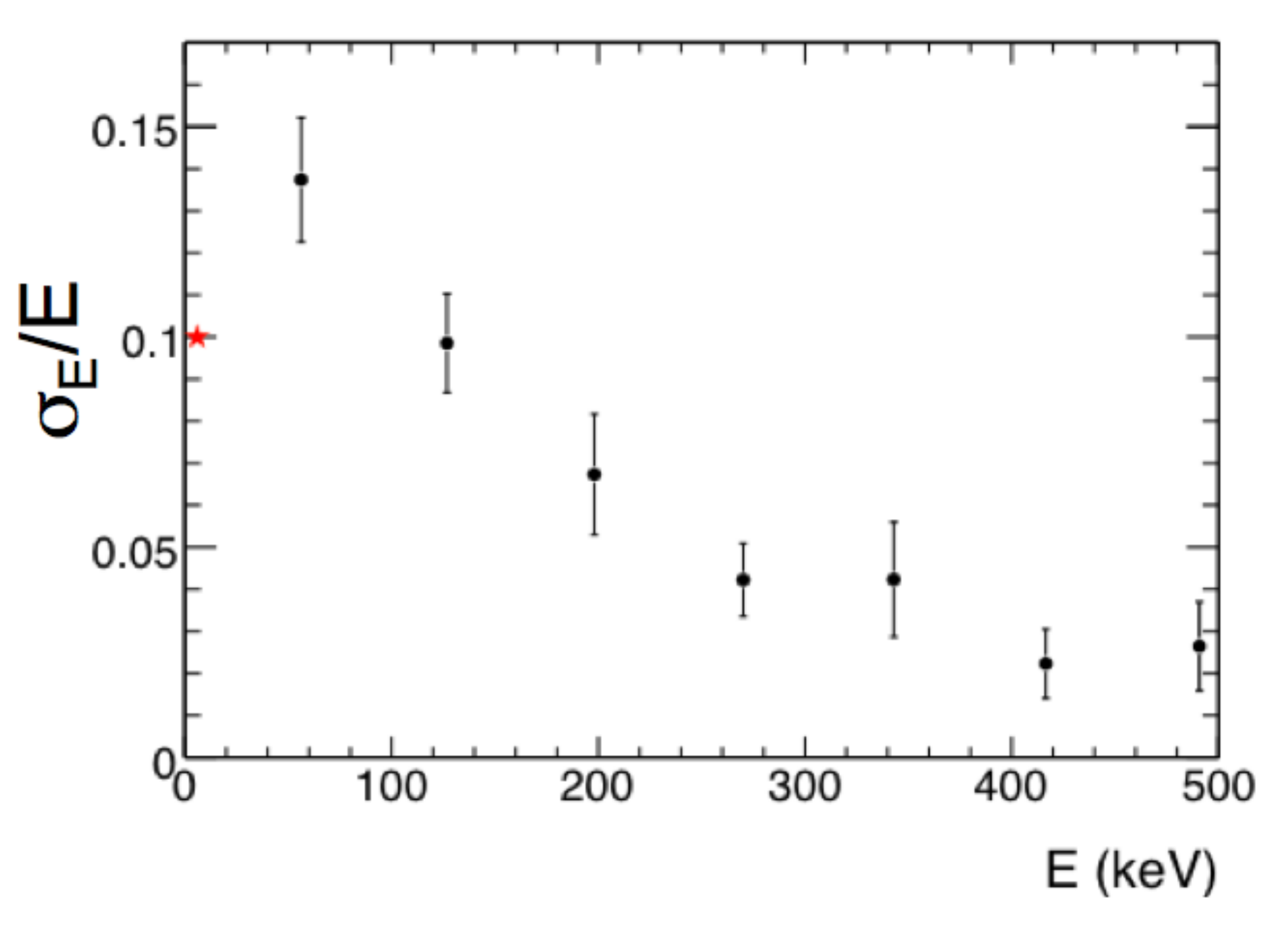}
\caption{(left) Gas gain and spark rate as a function of anode voltage for a mesh-based TPC filled with 75~Torr of \cffour.  The gas gain is measured from the charge collected from an $^{55}$Fe source.  (right) The energy resolution of the charge readout at 5.9~keV (star) and the CCD readout (circles) for energies above 50~keV.  Fig. from Ref.~\protect\refcite{caldwell2009}.}
\label{fig:dmtpcGainSpark}
\end{figure}

\begin{figure}\centering
\includegraphics[width=0.5\textwidth]{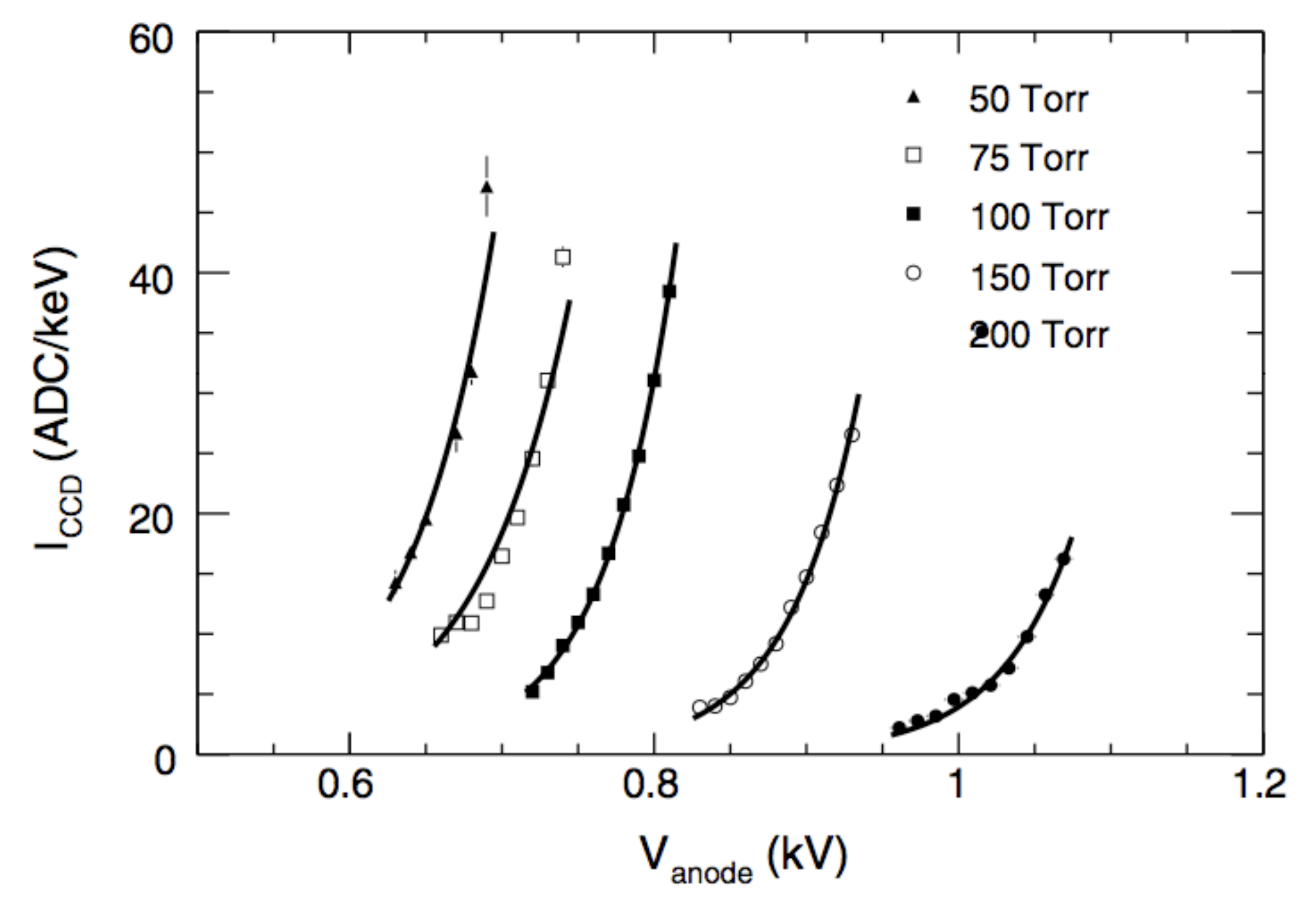}\\
\includegraphics[width=0.45\textwidth]{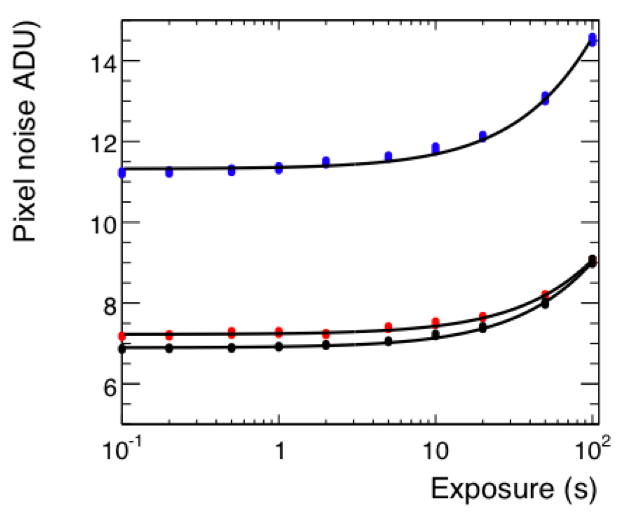}
\includegraphics[width=0.5\textwidth]{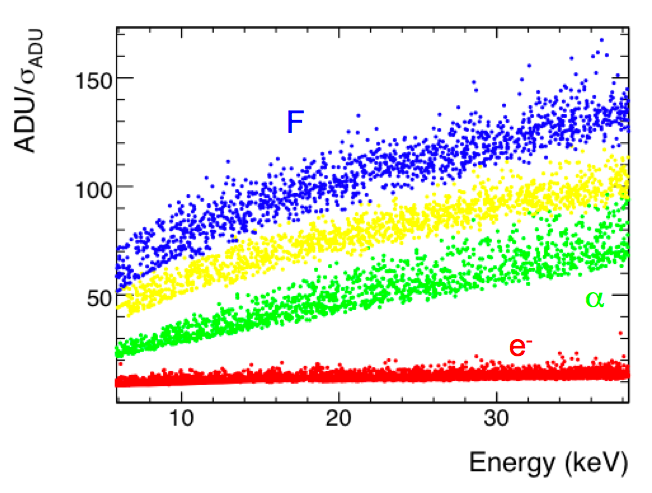}
\caption{(top) System gain for the TPC with CCD readout in units of
  ADU per keV as a function of amplification region voltage for
  \cffour~at various pressures.  (bottom left) CCD noise as a function
  of exposure time, obtained with the shutter closed.  The three
  curves correspond to different CCD cameras with different gain
  settings (photoelectrons per ADU).  In all cases, the CCD read noise
  (zero exposure) is approximately 12 electrons (rms).  For exposure
  times less than $\sim$10~seconds, the CCD dark noise is negligible.
  (bottom right) Monte Carlo predictions for the signal-to-noise ratio
  in a charged particle track.  Electrons, with smaller stopping,
  travel long distances and generate lower surface brightness tracks
  compared with nuclear recoils which lose their full energy over a
  shorter distance.}
\label{fig:dmtpcSNR}
\end{figure}

With a \cffour~pressure of 75~Torr, gas gains of approximately $10^5$
are routinely achieved with minimal sparking (see
Fig.~\ref{fig:dmtpcGainSpark}).  The energy resolution of the charge
readout is $10$\% at 5.9~keV (measured with an $^{55}$Fe source), and
is $15$\% at 50~keV for the CCD readout (measured with an alpha
source, see Fig.~\ref{fig:dmtpcGainSpark}).  Since the stopping
$dE/dx$ in the detector is much smaller for electrons than for nuclear
recoils, the surface brightness of an electron track is dimmer, and
electron tracks are easily distinguished from nuclear recoils.  This
is shown in Fig.~\ref{fig:dmtpcSNR}.  The gamma rejection of our
detector was measured to be $>10^6$ using an 8~$\mu$Ci $^{137}$Cs
source.\cite{dujmicTAUP2007}

\subsection{\cffour~gas properties}
\cffour~has many advantages as a target gas for a dark matter search
in a gaseous TPC.  First, it has good sensitivity to spin-dependent
interactions because of its unpaired proton.  In addition, it is a
good counting gas, allowing gas gains in excess of $10^5$.  The
scintillation spectrum of \cffour~has significant emission in a broad
($\sim$100~nm wide) peak centered at 625~nm.\cite{kabothCF4} This
spectrum is well-matched to the peak quantum efficiency of CCDs.  In
addition, the measured transverse diffusion of electrons in \cffour~is
less than 1~mm for a 20~cm drift length at $E/N = 12\times 10^{-17}$~V
cm$^2$ (e.g. 300~V/cm at 75~Torr in the detector).  Also, there is
negligible electron attachment over a 20~cm drift.\cite{caldwell2009}

\subsection{Head-tail measurements}
\begin{figure}\centering
\includegraphics[width=0.45\textwidth]{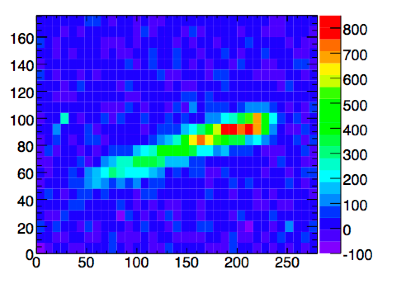}
\includegraphics[width=0.45\textwidth]{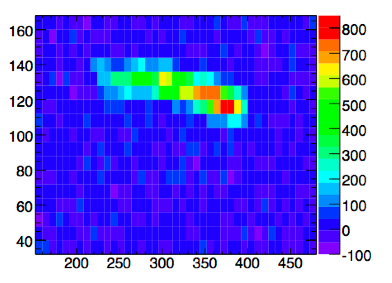}
\caption{Sample neutron-induced nuclear recoil candidates.  The neutrons were incident from the right.  The head-tail is evident from the light distribution along the track.  In these images, 100 pixels corresponds to 6~mm.  Figures taken from Ref.~\protect\refcite{dujmicAstroPart2008}.}
\label{fig:dmtpcRecoils}
\end{figure}

As described in Ref.~\refcite{dujmicNIMA2008}, the DMTPC collaboration has
demonstrated the ability to measure the head-tail effect (the vector
direction of a recoil) on an event-by-event basis for energies down to
100~keV.  In that work, a $^{252}$Cf neutron source irradiated a
mesh-based detector filled with \cffour~at 75~Torr.  The CCD camera
acquired 6,000 one-second-exposure images, and 19 of these images
contained a candidate nuclear recoil.  Two examples of these
neutron-induced nuclear recoils are shown in
Fig.~\ref{fig:dmtpcRecoils}.  In these images, the nuclear recoil
axis and direction (head-tail) is clearly visible for each event.

\begin{figure}\centering
\includegraphics[width=0.45\textwidth]{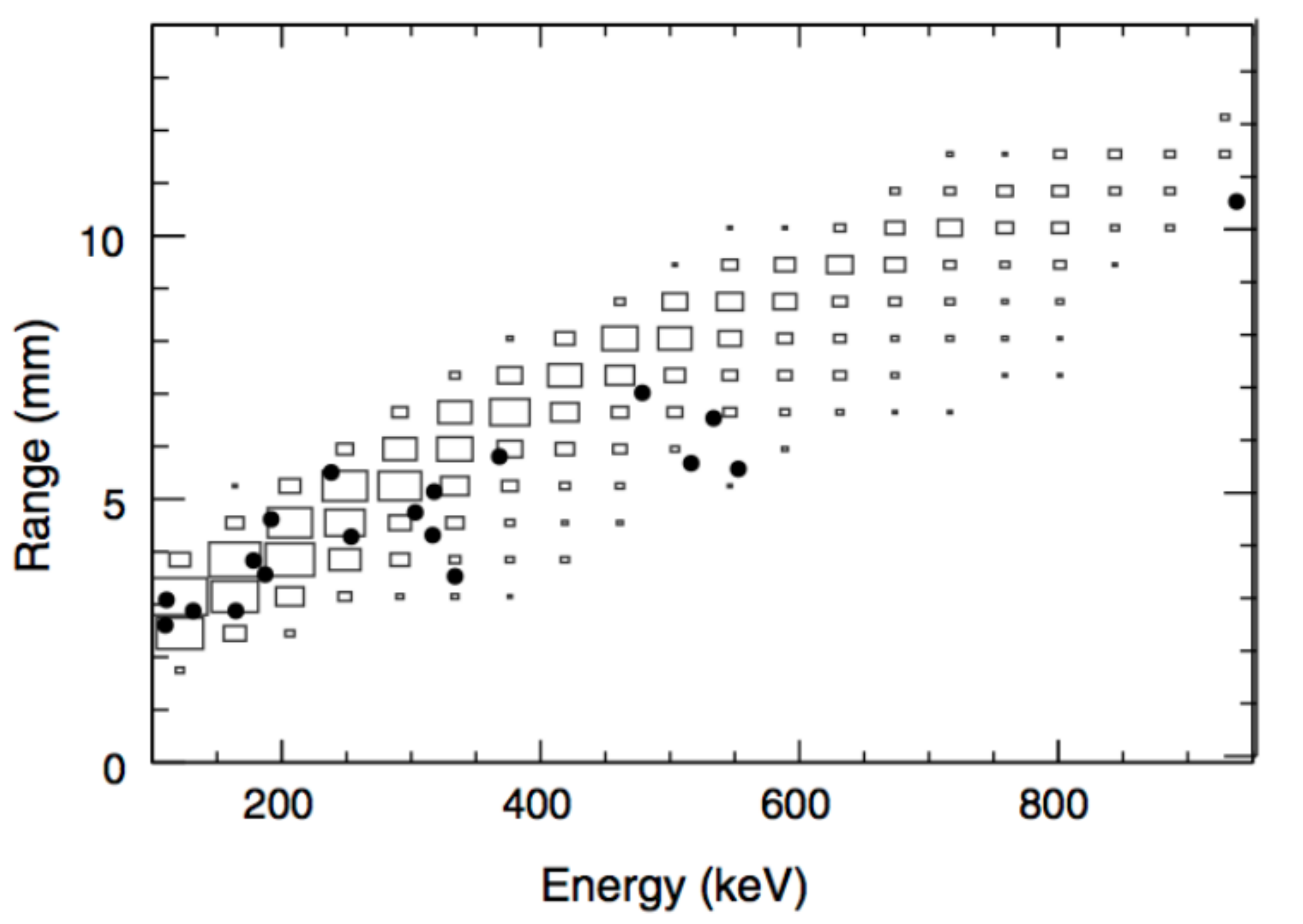}
\includegraphics[width=0.45\textwidth]{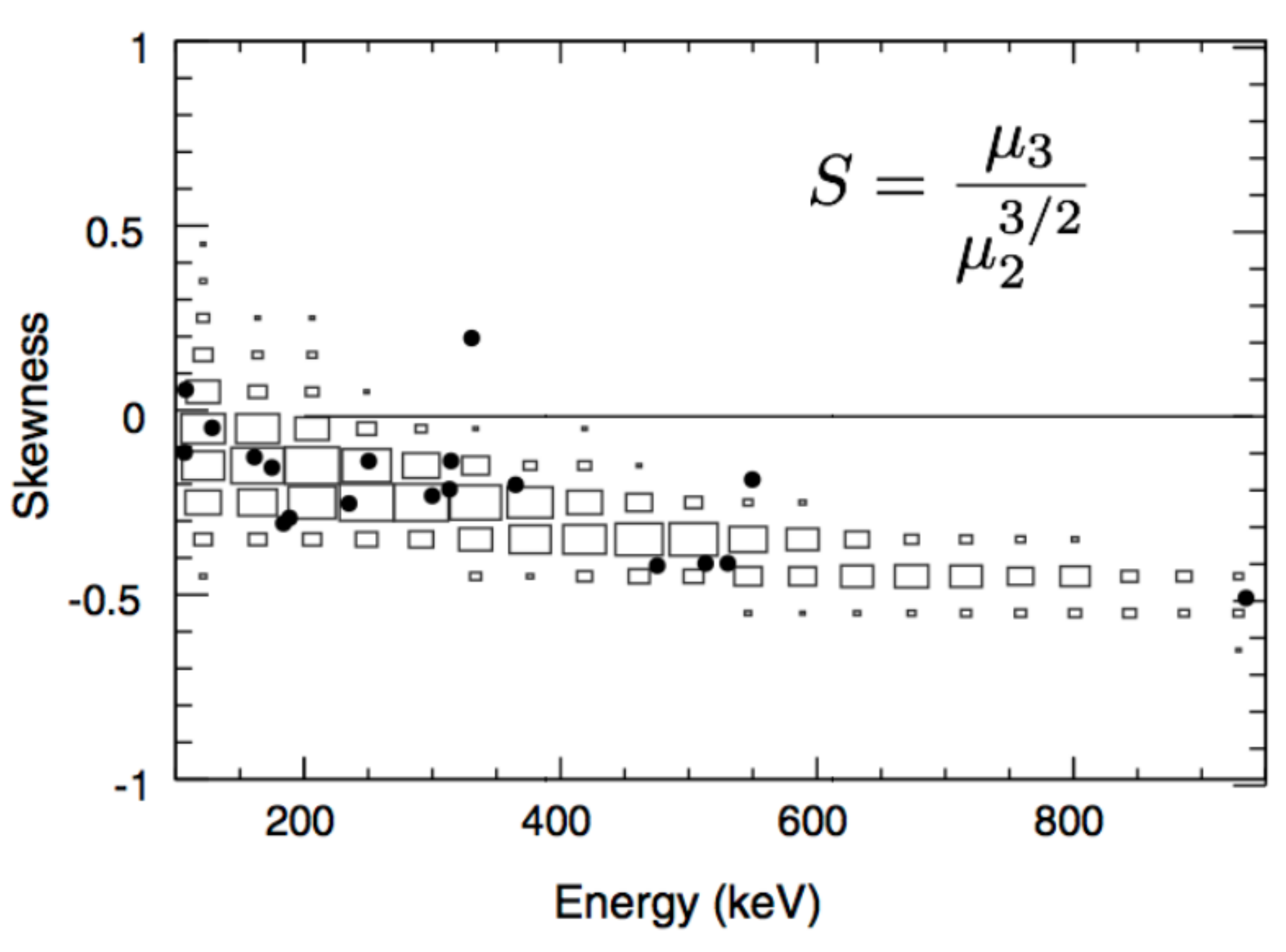}
\caption{(left) The observed (points) and predicted (box-histogram) range-energy relationship for candidate nuclear recoils induced by $^{252}$Cf neutrons in the detector; (right) the skewness parameter, a measure of the head-tail effect, for the candidate recoils.  A negative skewness parameter indicates that the head-tail was correctly reconstructed.  Figures taken from Ref.~\protect\refcite{dujmicAstroPart2008}.}
\label{fig:dmtpcRangeEnergySkewness}
\end{figure}

Fig.~\ref{fig:dmtpcRangeEnergySkewness} shows the measured and
predicted range vs. energy for these events.  The recoil direction can
be measured from the light profile along the recoil track.  For the
candidate nuclear recoils, a dimensionless skewness parameter
$S=\mu_3/\mu_2^{3/2}$ is constructed, where $\mu_2$ and $\mu_3$ are
the second and third moments of the light distribution.  In our data
set, kinematics constrain all nuclei to be forward scattered and
therefore have negative skewness.
Fig.~\ref{fig:dmtpcRangeEnergySkewness} shows that the skewness can
be correctly reconstructed down to 100~keV.

\begin{figure}\centering
\includegraphics[width=0.45\textwidth]{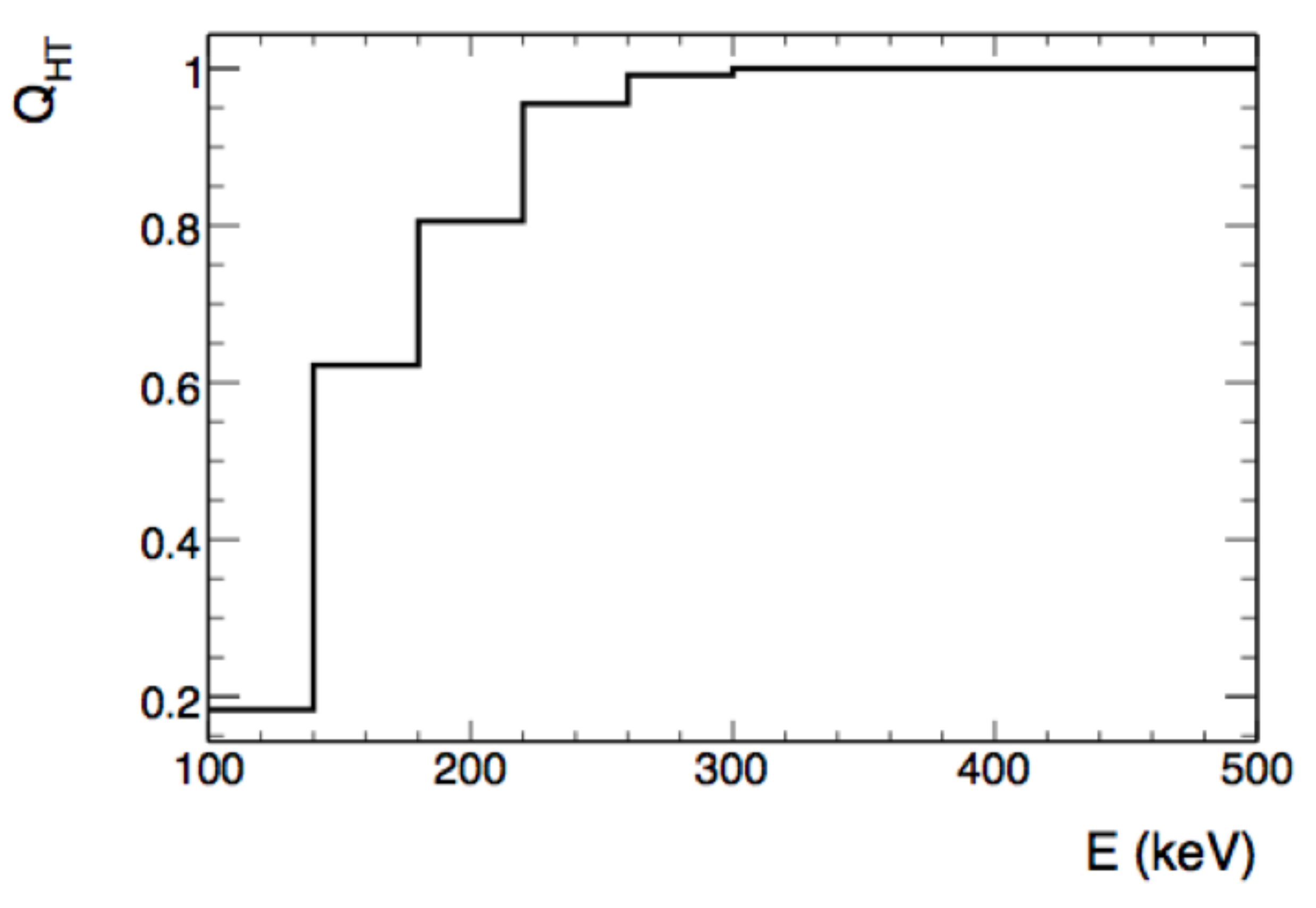}
\includegraphics[width=0.45\textwidth]{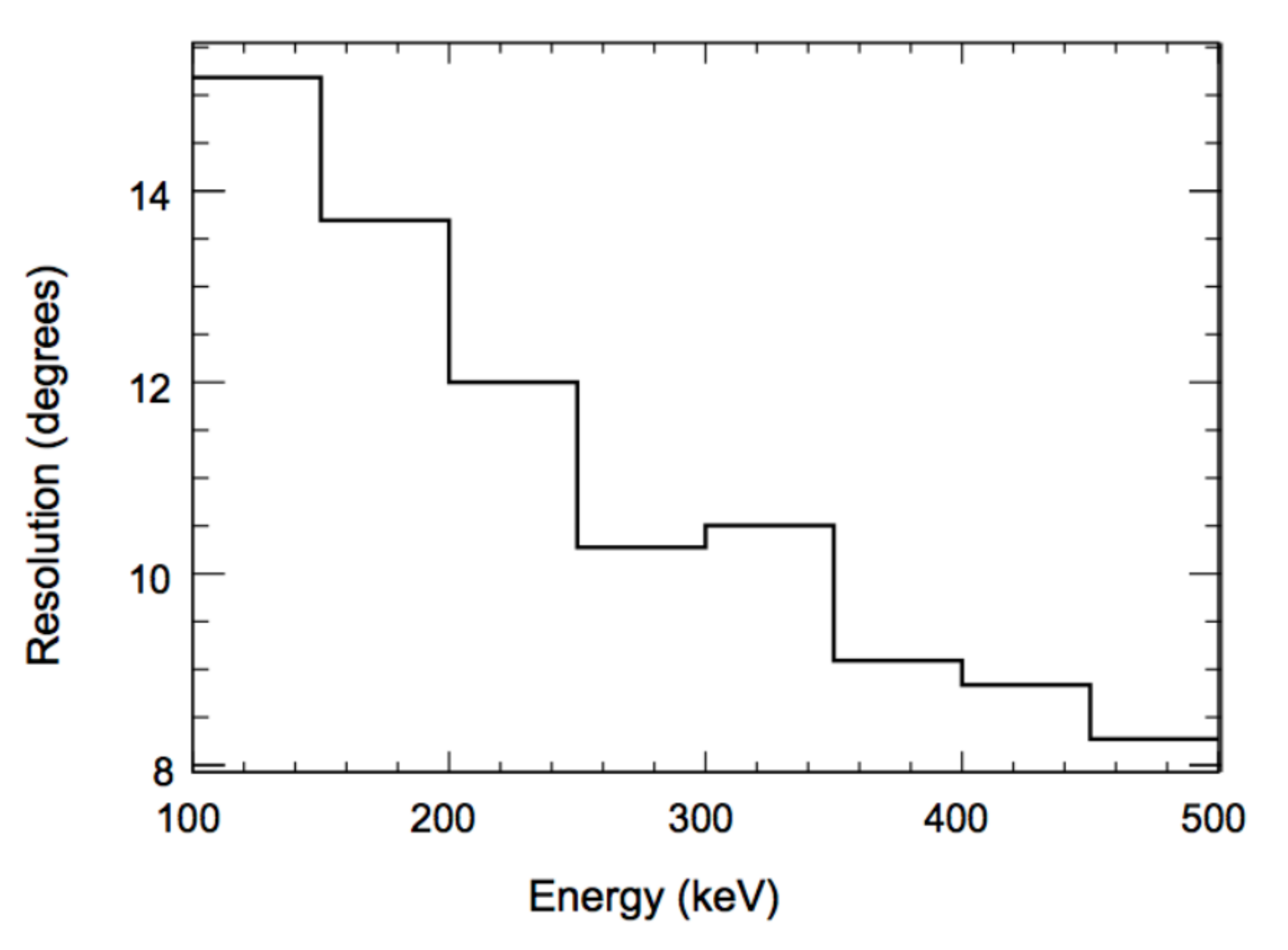}
\caption{(left) The head-tail quality factor $Q_{HT}$, as computed from Monte Carlo studies.  (right) The angular resolution for track reconstruction as a function of energy, also from Monte Carlo.  Figures taken from Ref.~\protect\refcite{dujmicAstroPart2008}.}
\label{fig:dmtpcAngularResQHT}
\end{figure}

For a set of nuclear recoils, the true forward-backward asymmetry is
$A=(F-B)/(F+B)$, where $F$ and $B$ are the number of forward and
backward recoils, respectively.  The measurement error on $A$ scales
like $\sigma_A\sim 1/\sqrt{N Q_{HT}}$, where $N$ is the total number
of measured recoils.  $Q_{HT}$ is a head-tail reconstruction quality
factor:
\begin{equation}
Q_{HT}(E_R)\equiv \epsilon(E_R) \left(\frac{N_{good}-N_{wrong}}{N_{good}+N_{wrong}}\right)^2
\end{equation}
where $E_R$ is the recoil energy, $\epsilon$ is the (recoil energy
dependent) head-tail reconstruction efficiency, and $N_{good}$ and
$N_{wrong}$ are the number of events with head-tail correctly and
incorrectly reconstructed.  Monte Carlo studies show that $Q_{HT}$
exceeds 50\% above 140~keV (see Fig.~\ref{fig:dmtpcAngularResQHT}).
The nuclear recoil direction can be reconstructed with an angular
resolution of 15$^\circ$ at 100~keV (see
Fig.~\ref{fig:dmtpcAngularResQHT}).

\subsection{From surface run to underground}
The 10-liter detector was operated for two long runs in a basement lab
at MIT for durations of six and three weeks.  To ensure gas purity
during the long runs, the vessel was evacuated and then re-filled with
75~Torr of \cffour~each day.  Over a 24~hour period, the system gain
was stable to 1\%.  An analysis of the data from these runs to study
the backgrounds in the detector will be the subject of an upcoming
publication.

The DMTPC collaboration is currently constructing a second detector,
similar in target mass to the existing detector, to deploy underground
to the Waste Isolation Pilot Plant (WIPP) in New Mexico, USA.  At WIPP
(1.6~km.w.e. depth), less than one neutron-induced event per year is
expected in the detector.  In addition, because the new detector will
be constructed from highly radiopure materials in a low-radon
environment, we expect a significant reduction of alpha backgrounds.

\subsection{DMTPC future goals}
In addition to the preparation for WIPP, DMTPC is working on several
aspects of detector R\&D.  A program is underway to achieve full
volume fiducialization by measuring the z-coordinate of an interaction
in the TPC.  This can be achieved through the detection of primary
scintillation light or from an analysis of the charge pulse profile on
the cathode.  Techniques to reconstruct the third dimension of tracks
($\Delta$z) from the charge or PMT signal at the amplification region
are also under development.

\begin{figure}\centering
\includegraphics[width=0.5\textwidth]{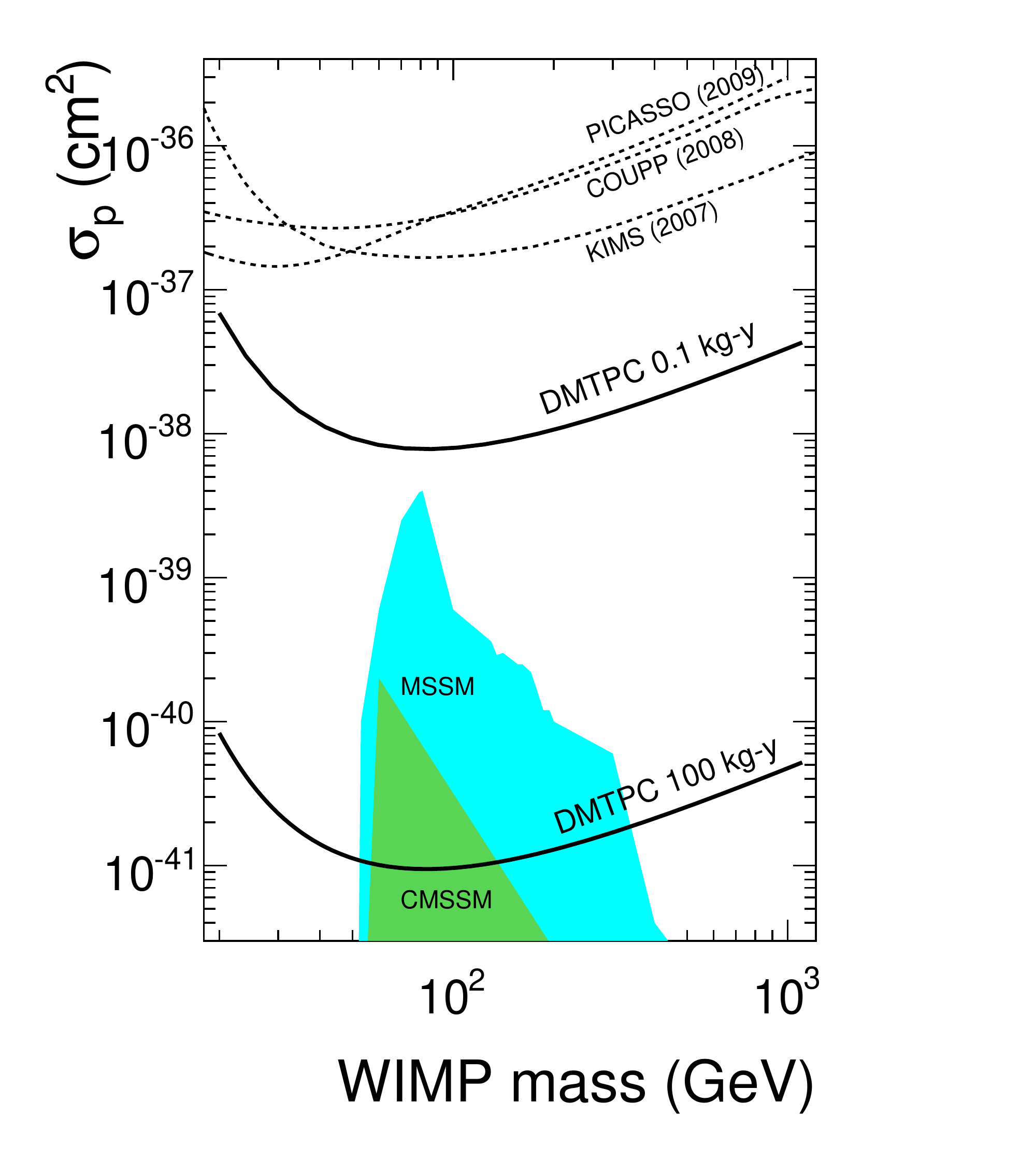}
\caption{Projected sensitivity for WIMP-proton spin-dependent interactions.  A cubic meter detector filled with \cffour~at 75~Torr with $\sim$3 months live time has an exposure of 0.1 kg-yr (upper solid line).  The lower solid line indicates the sensitivity for a 100 kg-yr exposure, which cuts deep into the preferred MSSM parameter space.  These projected sensitivities assume a threshold energy of 50~keV and less than one neutron-induced background event.  For reference, one picobarn equals $10^{-36}$~cm$^2$.}
\label{fig:dmtpcSensitivity}
\end{figure}

In addition, a cubic meter detector design is underway.  This design
consists of four TPC volumes and employs transparent mesh
anodes.\cite{dujmicAstroPart2008} The amplification region of each TPC
pair is read out with a single set of CCD cameras.  When filled with
75~Torr of \cffour, the cubic meter detector will contain 380~grams of
target material.  Given three months of live time (exposure
$\sim$0.1~kg-year), this detector is capable of achieving the most
stringent limit to date ($10^{-38}$~cm$^2$) on the spin-dependent
WIMP-proton interaction (see Fig.~\ref{fig:dmtpcSensitivity}).

%% file: sectionNEWAGE.tex
\section{NEWAGE -- Long-term observation tests 
with a practical-sized $\mu$-TPC in an underground laboratory}
\label{section:NEWAGE}
\begin{figure}
\includegraphics[width=1.\linewidth]{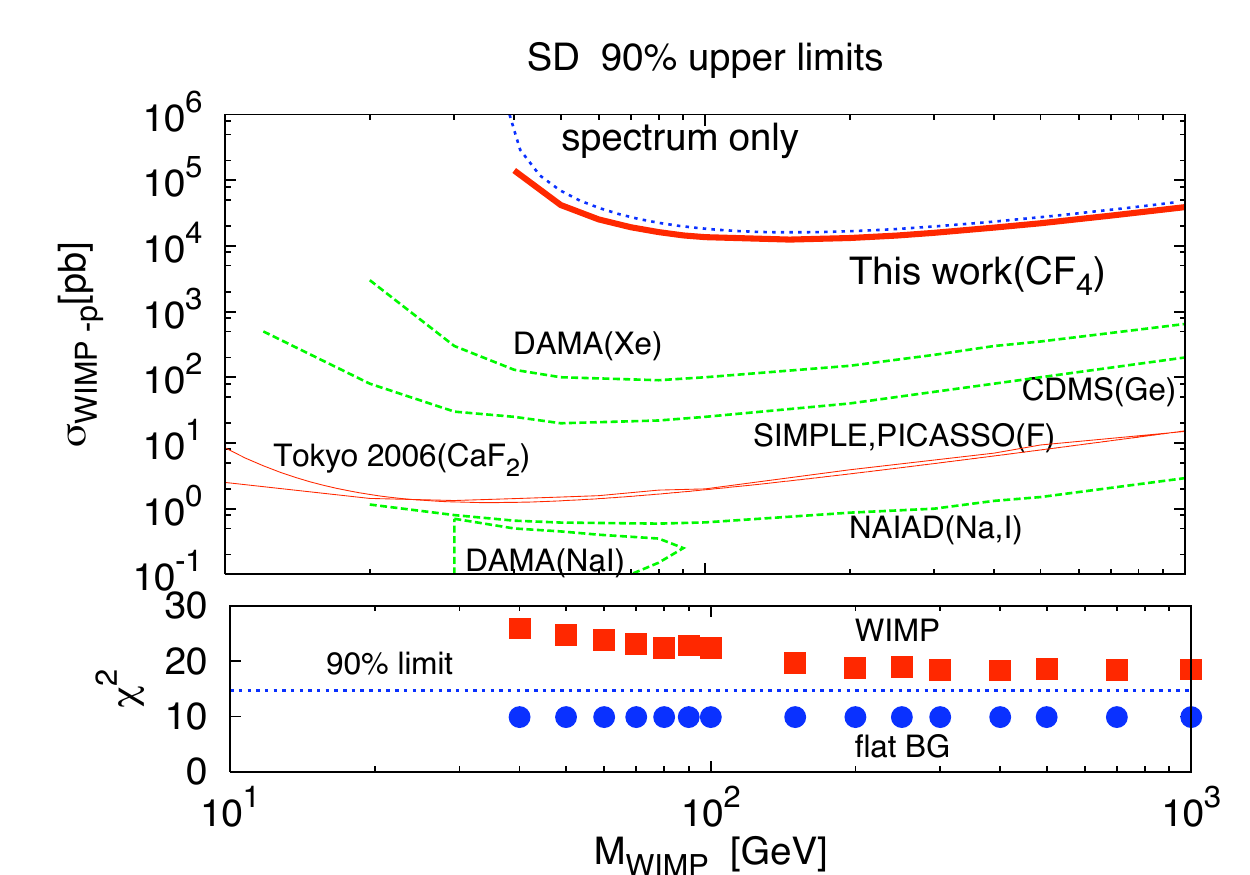}
\caption{\label{fig:NEWAGElimit2007} First direction-sensitive
  WIMP-proton cross section limits (spin-dependent) set by NEWAGE-0.3a
  detector.\protect\cite{MiuchiPLB2007} }\end{figure}

NEWAGE (NEw generation WIMP-search With an Advanced Gaseous tracking
device Experiment) is a direction-sensitive dark matter search
experiment with a gaseous micro-time-projection chamber ($\mu$-TPC)
that began detector R\&D in 2003,\cite{TanimoriPLB2004} and published
the first direction-sensitive dark matter limits in 2007 (see
Fig.~\ref{fig:NEWAGElimit2007}).\cite{MiuchiPLB2007} We have been
studying the detector background in the Kamioka Underground
Observatory since 2007.\cite{NishimuraAstroPart2009}

\subsection{Prototype detector and its performance}
\begin{figure}
\centering
\includegraphics[width=0.75\textwidth]{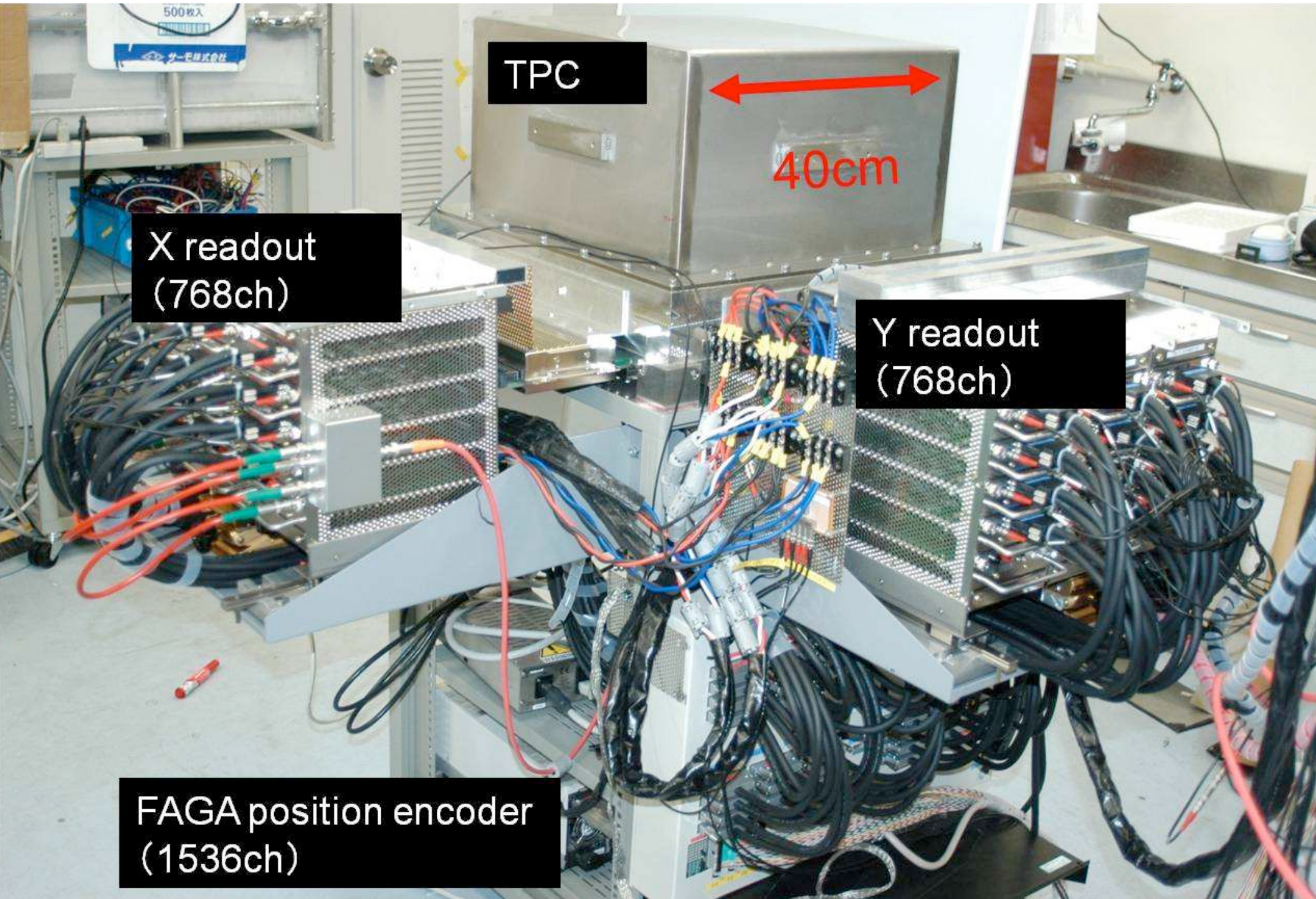}
\caption{\label{fig:TPCphoto}
Picture of the NEWAGE-0.3a detector.
}\end{figure}

\begin{figure}
\centering
\includegraphics[width=1.\linewidth]{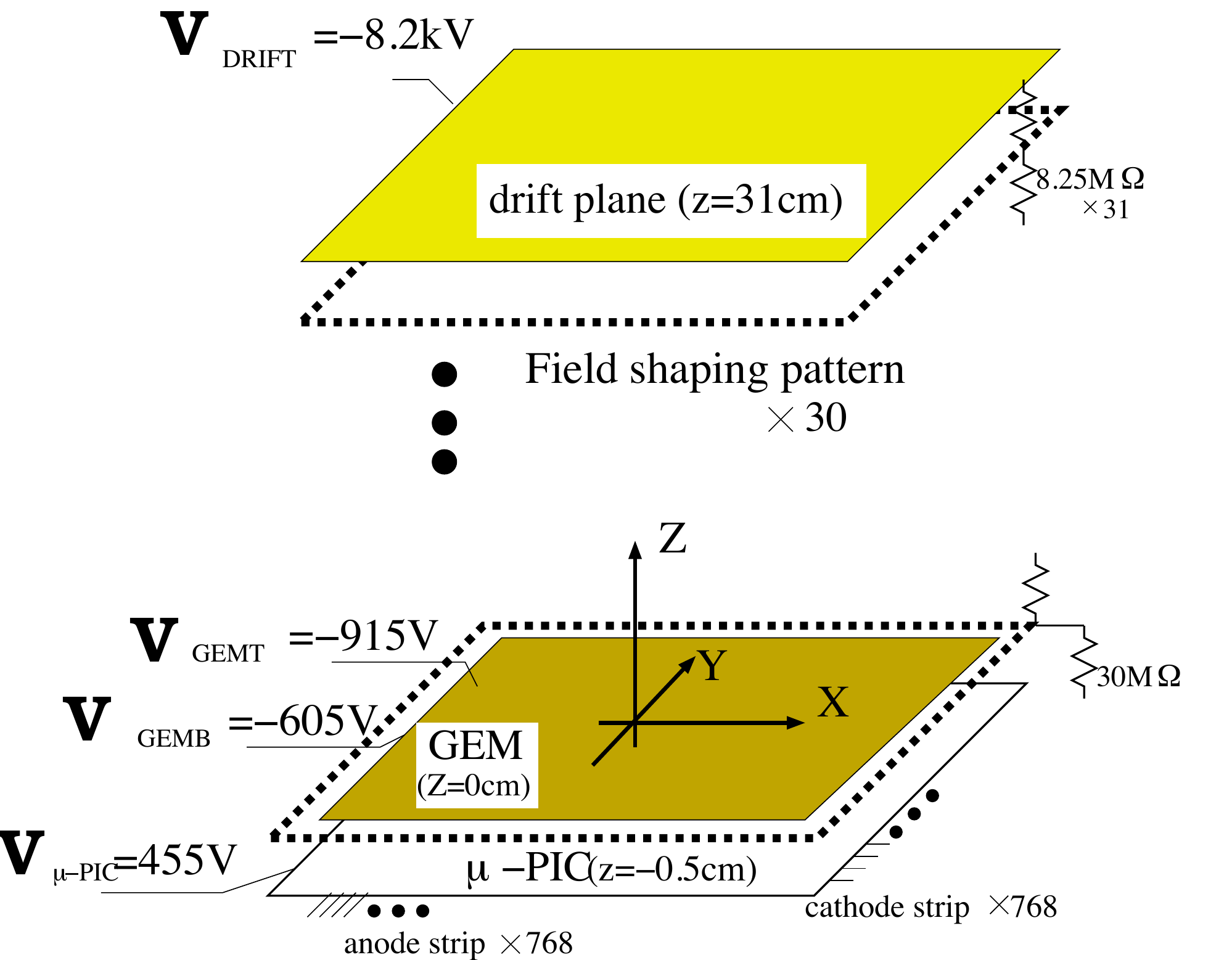}
\caption{\label{fig:TPC} Schematic view of the NEWAGE-0.3a detector.
  The volume between the drift plane and the GEM is the detection
  volume and is filled with CF$_4$ gas at 152~Torr.}
\end{figure}

The NEWAGE-0.3a detector, the first version of the $\rm(0.3m)^3$-class
prototypes, is a gaseous $\mu$-TPC filled with CF$_{4}$ gas at
152~Torr.  The effective volume and the target mass are 20 $\times$ 25
$\times$ 31 cm$^3$ and 0.0115~kg, respectively.  For details of the
detector system and performance studies, see
Ref.~\refcite{NishimuraAstroPart2009}.  A picture and schematic of the
NEWAGE-0.3a detector are shown in Fig.~\ref{fig:TPCphoto} and
Fig.~\ref{fig:TPC}, respectively.  The NEWAGE-0.3a detector is read
out by a 30.7~$\times$~30.7~cm$^2$ $\mu$-PIC. A $\mu$-PIC is one of
the several types of micro-patterned gaseous detectors.  By
orthogonally-formed readout strips with a pitch of 400~$\mu$m, the
$\mu$-PIC can generate two-dimensional images.\cite{TakadaNIMA2007}

\begin{figure}
\centering
\includegraphics[width=.5\linewidth]{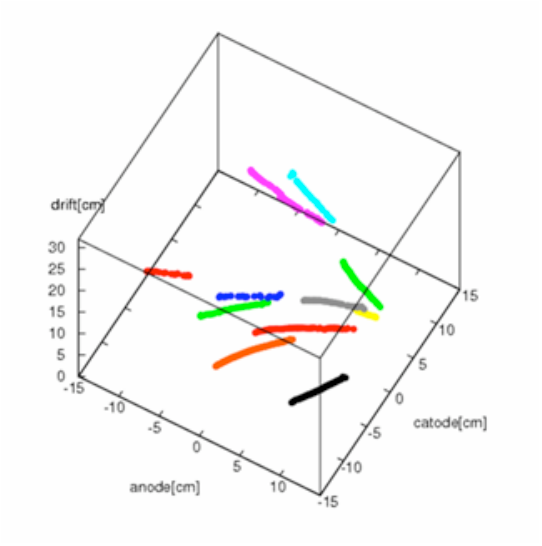}
\caption{\label{fig:ptracks}
Nuclear (proton) tracks detected with the NEWAGE-0.3a detector.
This measurement was performed with a CF$_4$-C$_4$H$_10$(9:1) gas mixture 
at 152~Torr.}
\end{figure}

\begin{figure}
\centering
\includegraphics[width=1.\linewidth]{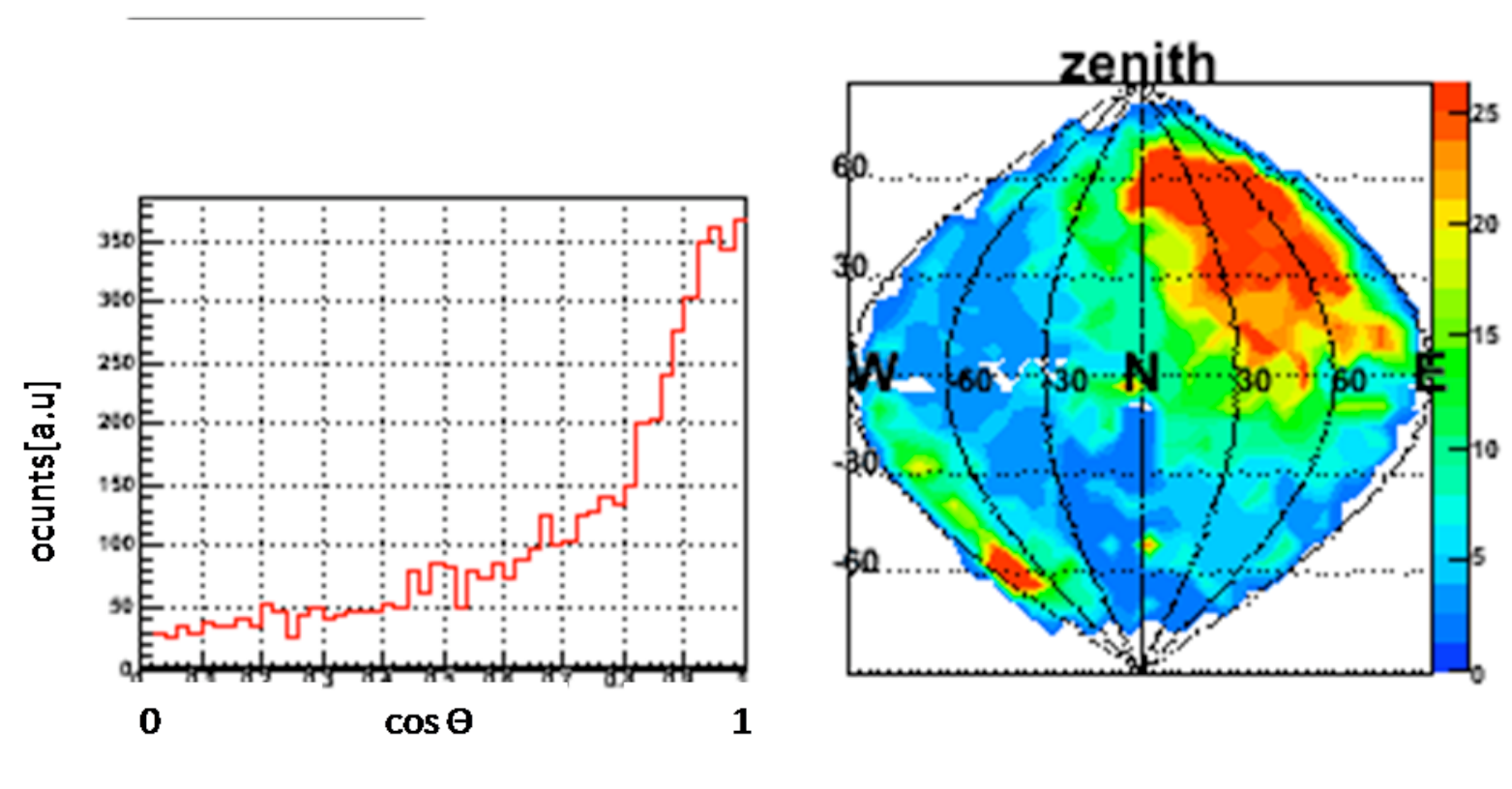}
\caption{\label{fig:NEWAGEskymap} Measured $\left|\cos\theta\right|$
  distribution (left) and the 2D map (right) of recoil tracks in the
  laboratory frame.  $\theta$ is the angle between the incoming
  neutron and the recoiling proton.  This reaction roughly emulates
  WIMP-fluorine elastic scatterings. The $^{252}$Cf neutron source was
  positioned at 45 degrees East of North at an elevation of 35
  degrees.  This measurement was performed with a
  CF$_4$-C$_4$H$_{10}$(9:1) gas mixture at 152~Torr.  }\end{figure}

The performance of the NEWAGE-0.3a detector measured in the
underground laboratory is listed in Table~\ref{tab:detector} together
with the projected goals.  We show typical results demonstrating the
capability of our method for direction-sensitive WIMP detection in
Fig.~\ref{fig:ptracks} and Fig.~\ref{fig:NEWAGEskymap}.  Three-dimensional
nuclear tracks detected with the NEWAGE-0.3a detector are shown in
Fig.~\ref{fig:ptracks}.  A dedicated data acquisition system with FPGA
chips realizes the detection of clear three-dimensional tracks as
successive digital hits.  The right panel of Fig.~\ref{fig:NEWAGEskymap}
shows a ``sky-map'' image drawn by the detected nuclear tracks.  We
simply traced back the detected nuclear tracks on the map and the
neutron source ($^{252}$Cf) was clearly reconstructed in the image.  The
recoil angle distribution is shown in the left panel of
Fig.~\ref{fig:NEWAGEskymap} and the distribution was peaked at cos$\theta=1$.
Detailed studies on fine tracks of fluorine nuclei are described in
Ref.~\refcite{NishimuraAstroPart2009}.

\begin{table}
\tbl{Current and goal performance of the NEWAGE-0.3a detector at the energy threshold.}
{\begin{tabular}{lrr} \toprule
Parameter                     & Value (present)  & Value (goal)\\ \colrule
Gas pressure [torr]           &   152            & 30\\
Energy threshold [keV]        &   100            & 35\\
Energy resolution [\%] (FWHM) &    70            & 30\\
$\gamma$-ray detection efficiency & $\rm 8.1 \times 10^{-6}$ & $\rm 1\times 10^{-8}$\\
Nuclear track detection efficiency [\%] & 80 & 80\\
Nuclear track angular resolution (RMS) & $\rm 55^\circ$ & $\rm 30^\circ$\\
\botrule
\end{tabular} \label{tab:detector}}
\end{table}

\subsection{Background studies in an underground laboratory}
\label{section:measurements}
We have been studying the detector performance and background in the
Kamioka Observatory located at 2700m water equivalent underground
since 2007.\cite{NishimuraAstroPart2009} The history of our
underground activities is listed in Table~\ref{tab:measurelog}.

\begin{table}
\tbl{Accumulated exposure of the surface-run and underground-runs with NEWAGE-0.3a as of June 2009.}
{\begin{tabular}{llll} \toprule
Run ID & Period & Exposure        & Comment\\ 
       &        & [kg$\cdot$days] & \\ \colrule
Surface run& 2006 Nov. 1st - 2006 Nov. 26th & 0.151 &first dark matter run\cite{MiuchiPLB2007}\\
\hline
Run1 & 2007 Mar.  6 - 2007 May. 15th & 0.08  & 1/4 volume\\
Run2 & 2007 May. 15 - 2007 Aug.  6th & 0.15  & 1/4 volume\\
Run3 & 2007 Dec.  7 - 2008 Jun.  9th & 1.744 & full operation\\
Run4 & 2008 Jun.  9 - 2008 Sep.  9th & 0.602 & BG study\\
Run5 & 2008 Sep. 11 - 2008 Dec.  4th & 0.524 & dark matter run\cite{NishimuraPhD2009}\\
Run6 & 2009 Mar.  2 - 2009 Jun. 24th & 1.039 & with gas circulation system (see Section~\ref{section:stability})\\
\hline
     & Underground total             & 4.139 & \\
\botrule
\end{tabular} \label{tab:measurelog}}
\end{table}

Based on the background studies performed in the underground
laboratory,\cite{NishimuraPhD2009} the most dominant source of
background is $\alpha$ particles that deposit a fraction of their
energy in the gap volume between the $\mu$-PIC and the GEM.  Although
the alpha particles have energies of several MeV, these events can
mimic low energy events because of the partial energy deposited in the
small gap region (5~mm) without the electron-multiplication by the
GEM.  We will suppress these background events by replacing the
detector components with radiopure materials.

\begin{table}
\tbl{Estimated background rates.  The rates at the energy threshold are shown in units of [counts/kg/days/keV].}
{\begin{tabular}{ll} \toprule
Source & Rate\\ \colrule
Ambient gammas&$\sim 10$\\
Ambient fast neutrons&$\sim10^{-1}$\\
Cosmic muon &$<10^{-1}$\\
\hline
Internal $\alpha$ (fiducial volume)&$<10^{-1}$\\
Internal $\alpha$ (gap volume)&$<40$\\
Internal $\beta$&$<5$\\
\hline
Measured (Run5) &50\\
\botrule
\end{tabular} \label{tab:BG}}
\end{table}

\subsection{Stability Improvement}
\label{section:stability}
\begin{figure}
\centering
\includegraphics[width=1.\linewidth]{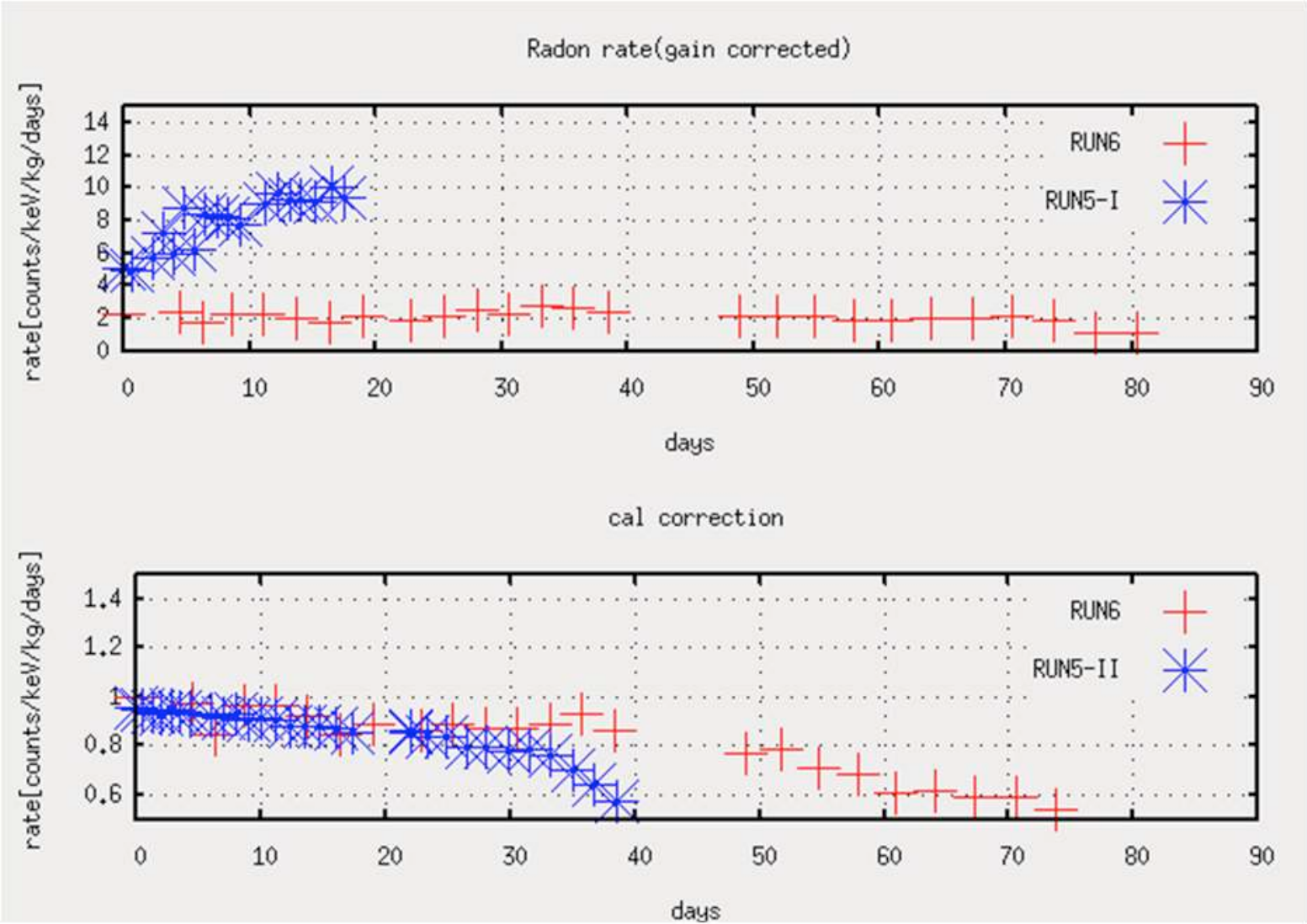}
\caption{\label{fig:radon}
Monitored radon progeny count rate and gas gains 
with (Run6) and without (Run5) the gas circulation system.}
\end{figure}

In the beginning of 2009, we installed a gas circulation system which
consists of a charcoal filter, getter pump, and Teflon bellows pump to
maintain the gas purity for more than one month.  We monitored the gas
gain and the radioactive radon ($^{220}$Rn and $^{222}$Rn)
contamination in the gas using the high energy ($\sim$6 MeV)
radon-progeny peaks.  Monitored radioactive radon rates and gas gains
are shown in the upper and lower panels of Fig.~\ref{fig:radon},
respectively.  Those monitored in the previous run (Run-5) are also
shown for comparison.  We found that the gas circulation system
reduced the radon rate by a factor of five in 20 days and maintained
the gas gain for twice as long.

\subsection{Sensitivity Improvement and Scaling-up}
\label{section:scalingup}

We are improving the tracking algorithm in the online data acquisition
system in order to improve the $\gamma$-ray rejection factor. Further
algorithm improvements are planned to reconstruct the head-tail of the
tracks.  Our next prototype, NEWAGE-0.3b, with a detection volume of
23~$\times$~28~$\times$~50~cm$^3$, has started performance
measurements in a surface laboratory.  Our next step will be to
assemble four $\mu$-PICs to form a 60$\times$60~cm$^2$ tiled read-out
area.  We started to develop a pixel-read out ASIC named ``QPIX,'' for
the ultimate readout system of the $\mu$-TPC.\cite{MatsuzawaTIPP09}

\subsection{NEWAGE Summary}
\label{section:summary}
Notable R\&D results contributing
to this international study so far are:
\begin{itemize}
\item We performed a direction-sensitive dark matter search
  measurement, analyzed the data, and set the first
  direction-sensitive limits.  This means ``the whole-chain'' was
  performed not by a simulation but by a real experiment and supports
  our feasibility study in the final section.
\item We employed a micro-patterned gas detector (MPGD), which is one
  of the strong candidates for the final huge detector, of a practical
  size (30.7$\times$30.7~cm$^2$) for the first time.  We also have
  shown three-dimensional tracks can be detected with a dedicated
  electronics system.  We are studing the scaling-up and background
  reduction issues of these MPGDs in a few years considering the
  cost-versus-sensitivity issues discussed in the final section.
\item We have performed underground measurements for two years.  R\&D
  results, especially gas-circulation studies for long-term
  measurement, should be indispensable for a huge detector
  development.
\end{itemize}
 
We plan to develop a detector with a $\sim 1$~m$^3$ fiducial volume in
a few years, and to determine the final design for a very large
detector (multi-modules of the 1~m$^3$ detector) taking account of the
sensitivity and its cost.

%% file: sectionMIMAC.tex
\section{A directional detector for non-baryonic dark matter search:  MIMAC:  (MIcro-tpc MAtrix of Chambers)}
\label{section:MIMAC}

The MIMAC project is a multi-chamber detector for dark matter search,
which aims to measure both tracks and ionization with a matrix of
micromegas $\mu$TPC filled with $^3$He, $\rm CF_4$, $\rm CH_4$ or/and
$\rm C_4H_{10}$.  A 10 kg $^3$He dark matter detector, or the
equivalent mass of $\rm CF_4$, with a 1~keV threshold (MIMAC) would
be sensitive to SUSY models allowed by present cosmological and
accelerator constraints. This study highlights the complementarity of
this experiment with most current spin-dependent experiments: proton
based detectors and neutrino telescopes.

Using both $^3$He and $\rm CF_4$ in a patchy matrix of $\mu$TPCs opens
the possibility to compare rates for two atomic masses, and to study
separately the neutralino interaction with neutrons and protons, as
the main contribution to the spin content of these nuclei is dominated
by one nucleon. With low mass targets, the challenge is also to
measure low energy recoils, below 6~keV for Helium, by means of
ionization measurements.  Low pressure operation of the MIMAC detector
will enable the discrimination of the neutralino signal from 
backgrounds on the basis of track features and directionality.

The electron/nuclear recoil discrimination is based on track length,
which is expected to be about ten times longer for an electron than
for a nuclear recoil of the same energy.  Identification of neutrons
will be done by event time correlations between chambers, assuming
that a WIMP will not interact twice in the whole MIMAC detector.
Ultimately, the last background rejection tool is the reconstruction
of the incoming direction of the particle.  There are two steps when
preparing a dark matter detector aiming at directional search:
\begin{itemize}
\item first, the energy of the recoil must be measured with
accuracy, which implies a  precise knowledge of the quenching factor, 
\item second, the possibility to reconstruct a 3D track must be shown.  This is a key point as the required exposure  is decreased 
by an order of magnitude between  2D read-out and 3D read-out.\cite{GreenAstroPart2007}
\end{itemize}
Nonetheless, the energy threshold must be as low as possible, in the keV range or even sub-keV, owing to the exponential  feature of the
recoil spectrum.

We developed a $\mu$TPC prototype with a 16.5~cm drift length, read out
by a bulk micromegas.  At first, we used a standard 128~$\mu$m bulk
micromegas (non-pixellized anode plane) in order to measure the energy
resolution of our detector.  To correctly assess the real recoil
energy on the nuclei, we performed a complete measurement of the
ionization quenching factor in the energy range of dark matter search,
i.e. below 10~keV.\cite{GuillaudinJPhys2009,MayetJPhys2009} In the
following, we describe the Ionization Quenching Factor (IQF)
measurements\cite{Santos2008} and 3D track measurements realized by
the MIMAC collaboration.

\subsection{Ionization quenching measurements}

\begin{figure}
\centering 
\includegraphics[height=0.3\textheight]{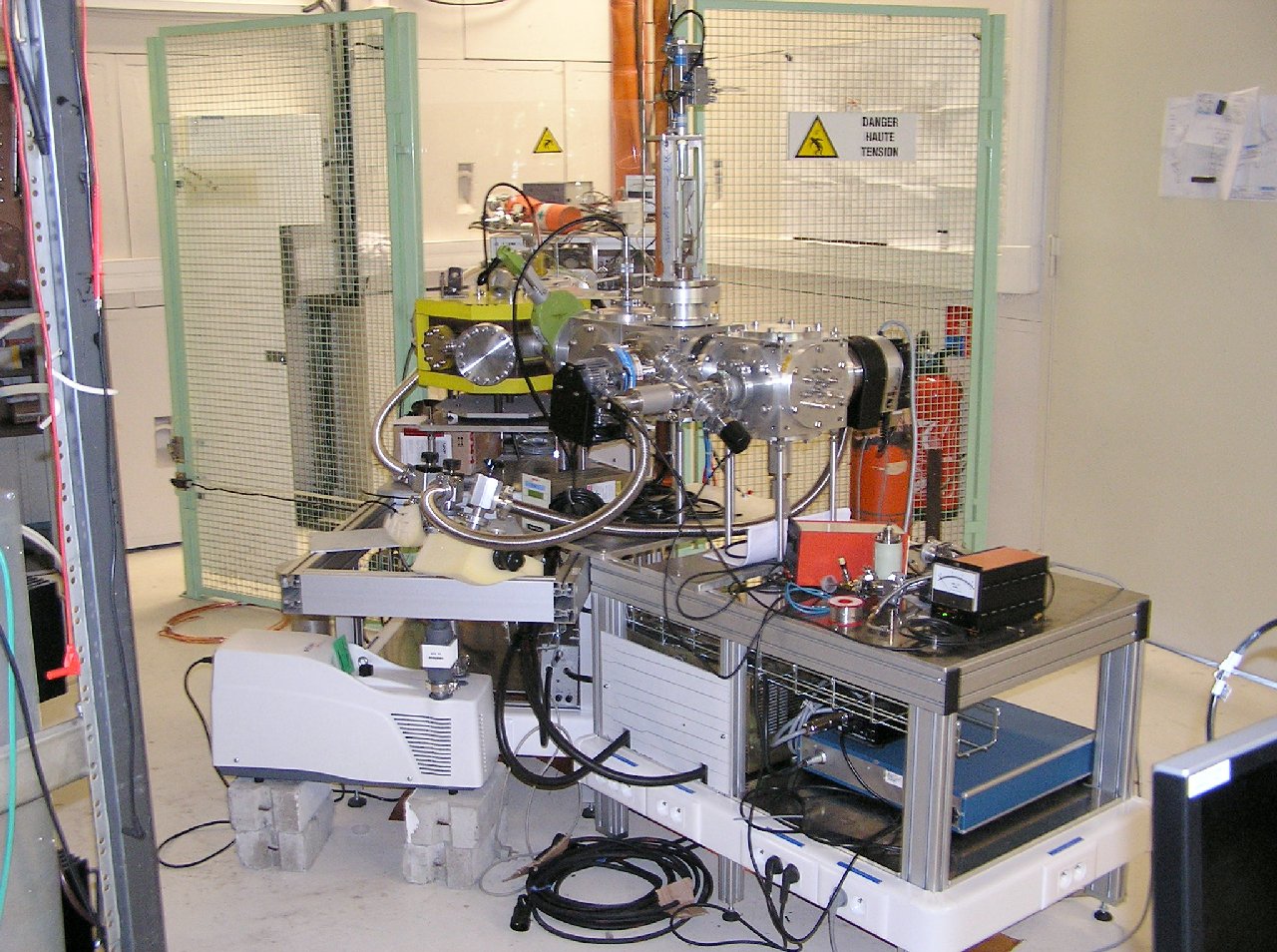}
\caption{Ionization Quenching Factor Facility at the LPSC Grenoble. }  
\label{fig:mimac2}
\end{figure}

\begin{figure}
\centering
\includegraphics[angle=0,width=0.5\textwidth]{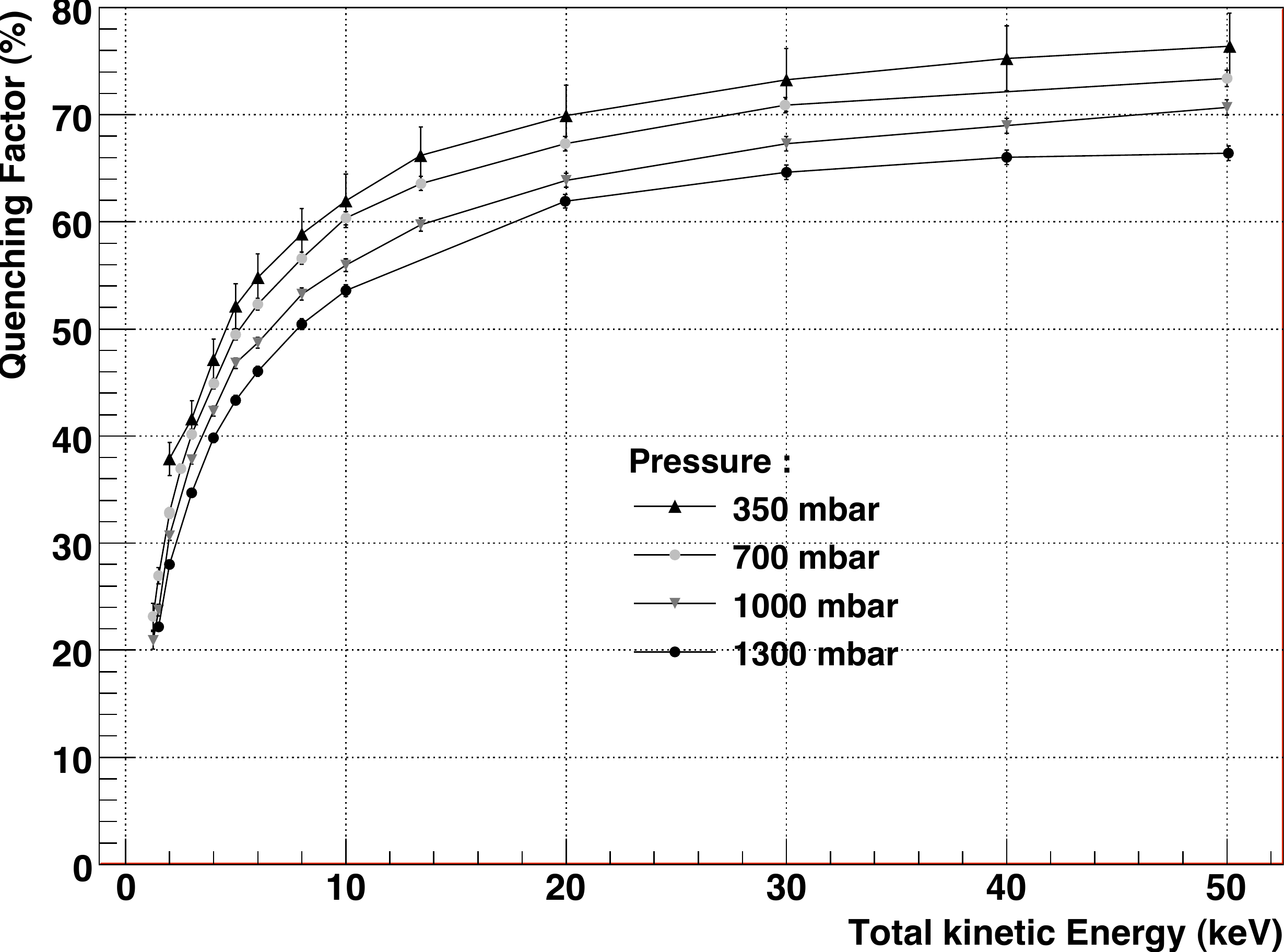}
\caption{ \he quenching factor as a function of \he total kinetic
  energy (keV) for 350, 700, 1000 and 1300 mbar, in the case of \he in
  \he +5\% \iso mixture. Measured data are presented with error bars
  included mainly dominated by systematic errors. Straight line
  segments between experimental points help to separate the different
  series of measurements. From Ref~\protect\refcite{Santos2008}.}
\label{fig:mimac3}
\end{figure}

The energy released by a nuclear recoil in a medium produces in an
interrelated way three different processes: i) ionization, producing a
number of electron-ion pairs, ii) scintillation, producing a number of
photons coming from the de-excitation of quasi-molecular states and
iii) heat produced essentially by the motion of nuclei and electrons.
The way in which the total kinetic energy released is shared between
the electrons and nuclei by interactions with the particle has been
estimated theoretically four decades ago,\cite{Lindhard1963} for very
specific cases (those in which the particle and the target are the
same). Since then, phenomenological studies have been proposed for
many (particle, target) systems.\cite{Hitachi2008,srim} At low
energies, in the range of a few keV, the ionization produced in a
medium is strongly energy dependent, and systematic measurements are
needed.  In order to measure the ionization quenching factor for \hee,
\hett, \hid and \fl, the LPSC has developed a dedicated facility
producing light ions at energies of a few keV. This facility, called
MIMAC's source (shown in Fig.~\ref{fig:mimac2}), uses an Electron
Cyclotron Resonance Ion Source (ECRIS)\cite{Lamy2009,Geller1996} and a
Wien filter to select the desired charge-to-mass ratio with a high
voltage extraction going up to 50~kV. This facility has enabled the
first measurements shown in Fig.~\ref{fig:mimac3} confirming that
there is sufficient ionization at low energies with \he and
\hett.\cite{Santos2008}

In Fig.~\ref{fig:mimac3}, the measurements at 350, 700, 1000 and 1300
mbar are shown.\cite{Santos2008} We observe a clear, roughly linear,
dependence of the IQF on the pressure of the gas that will be reported
in a future study down to less than 100 mbar.

In summary, we have measured for the first time the ionization
quenching factor of \he down to very low energies showing the amount
of ionization available from recoils of \hee. The IQF dependence on
the pressure of the gas has been described for four different
pressures.  An estimation of the scintillation produced in the gas
mixture as a function of the energy of the particles has been
done. The IQF variation as a function of the quencher has been
presented. These measurements are particularly important for searching
WIMPs using \het and in general to better understand the ionization
response of helium gas detectors. For more details see
Ref.~\refcite{Santos2008}.

\subsection{Low recoil energy threshold}

\begin{figure}
\centering
\includegraphics[angle=0,width=0.6\textwidth]{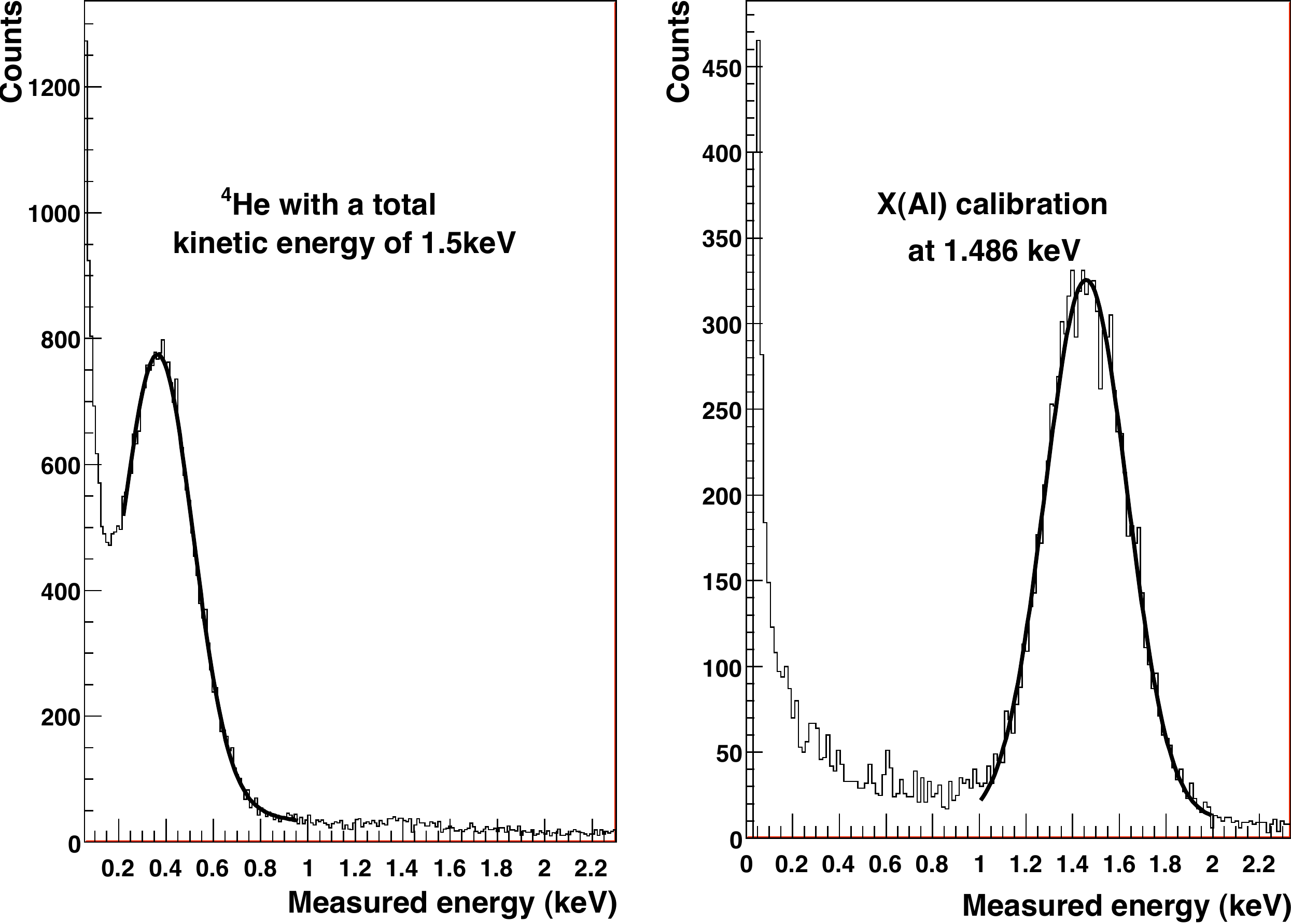}
\caption{Spectra of 1.5~keV total kinetic energy \he (left) and 1.486~keV X-ray of \alu (right) in \he +5\% \iso  
mixture at 700 mbar. The comparison of the energy measured between these two spectra gives  the quenching factor of 
Helium in \he +5\% \iso mixture at 700 mbar. From Ref.~\protect\refcite{Santos2008}.}
\label{fig:mimac4}
\end{figure}

Because the number of expected WIMP events increases exponentially at
low energies,\cite{LewinAndSmith1996} one of the most important
detector parameters is the recoil energy threshold. We have measured,
with a micromegas detector\cite{GiomatarisNIMA2006,GiomatarisNIMA1996}
taking into account the IQF previously measured, \he recoils down to
energies of 1~keV as shown in Fig.~\ref{fig:mimac4}.

The electron energy resolution ranges from $\sim 12\%~(\sigma)$ with
an \alu source, down to $\sim 5\%$ for 5.9~keV electrons.  Although a
precise measurement of the energy of the recoil is the starting point
of any background discrimination, the 3D reconstruction of the track
is necessary to do dark matter directional detection.

\subsection{Three dimensional tracks with micromegas}

\begin{figure} 
\centering
\includegraphics[width=0.9\textwidth]{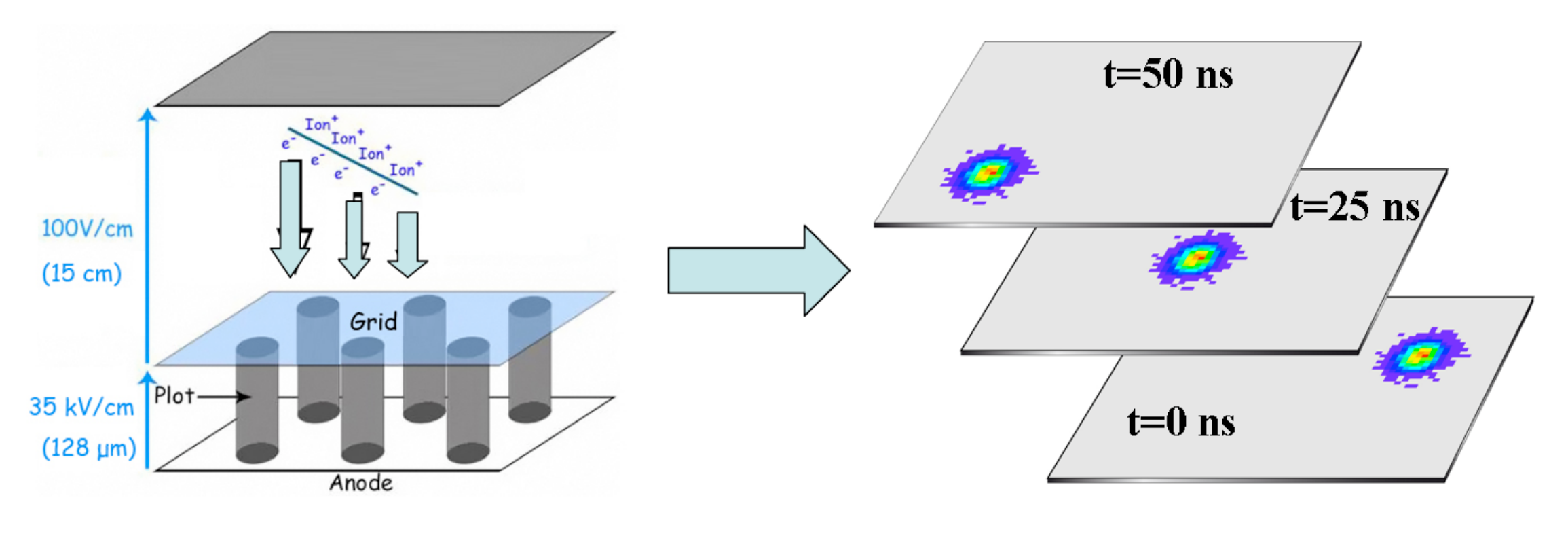}
\caption{Track reconstruction in MIMAC. The anode is scanned every 25~ns and the
3D track is recontructed, knowing the drift velocity,  from the series of images
of the anode.}
\label{fig:mimac5}
\end{figure}

The 3D reconstruction strategy chosen for the MIMAC project is the
following: i) the electrons from the track are projected on the anode
thus allowing to access information on x and y coordinates, ii) the
anode is read every 25 ns, iii) knowing the drift velocity, the series
of images of the anode allows to reconstruct the 3D track.

In order to reconstruct a few mm track in three dimensions, we use a
bulk micromegas\cite{GiomatarisNIMA2006,GiomatarisNIMA1996} with a
3$\times$3~cm$^2$ active area, segmented in 300~$\mu$m pixels.  We use
a 2D readout with 424~$\mu$m pitch, in order to read both dimensions
(x and y).  This bulk is provided with a 325 LPI (Line Per Inch) woven
micro-mesh made from 25~$\mu$m stainless steel wire.  The angular
resolution of the recoil track reconstruction is $\sim$15~degrees
(with energy dependence).

3D track reconstruction requires dedicated, self-triggering
electronics, able to sample at a frequency of 40 MHz.\cite{Richer2009}
To do so, the LPSC electronic team designed an ASIC in a 0.35~$\mu$m
BiCMOS-SiGe technology.  With an area approximatively equal to
1.5~mm$^2$, this ASIC contains 16 channels, each having its own
charge-sensitive preamplifier, current comparator and 5 bit coded
tunable threshold.  The 16 channels are sent to a mixer and a shaper
to measure the energy in the ASIC.  Each of the 12 ASICs is connected
to FPGAs programmed to process, merge and time sort data.  Finally,
the electronics board is connected to an ethernet microcontroller
which forwards the data via a TCP socket server to the acquisition
station.  This first version of the MIMAC ASIC has been running in the
MIMAC prototype since May 2008.  The next version, with 64 channels,
is currently under development at the LPSC and will be ready before
the end of 2009.

\begin{figure*}
\centering
\begin{tabular}{cc}
\includegraphics[height=0.3\textheight]{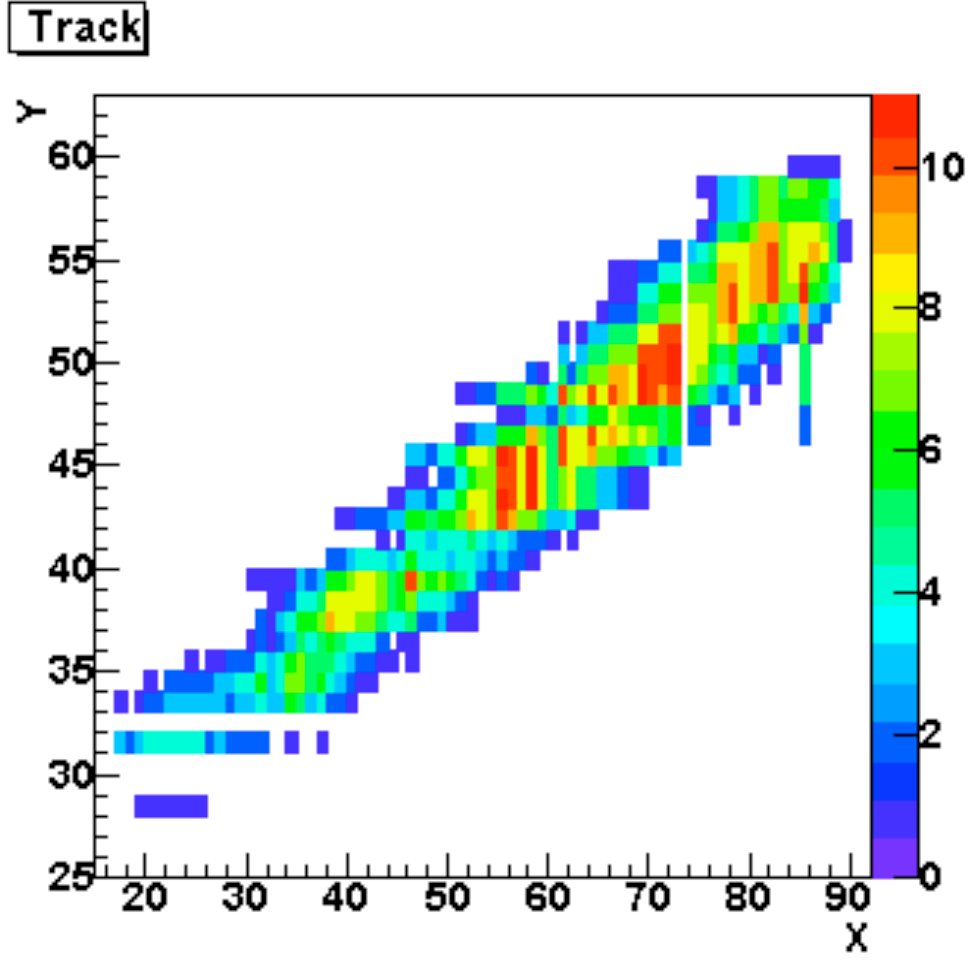}&
\includegraphics[width=0.3\textheight]{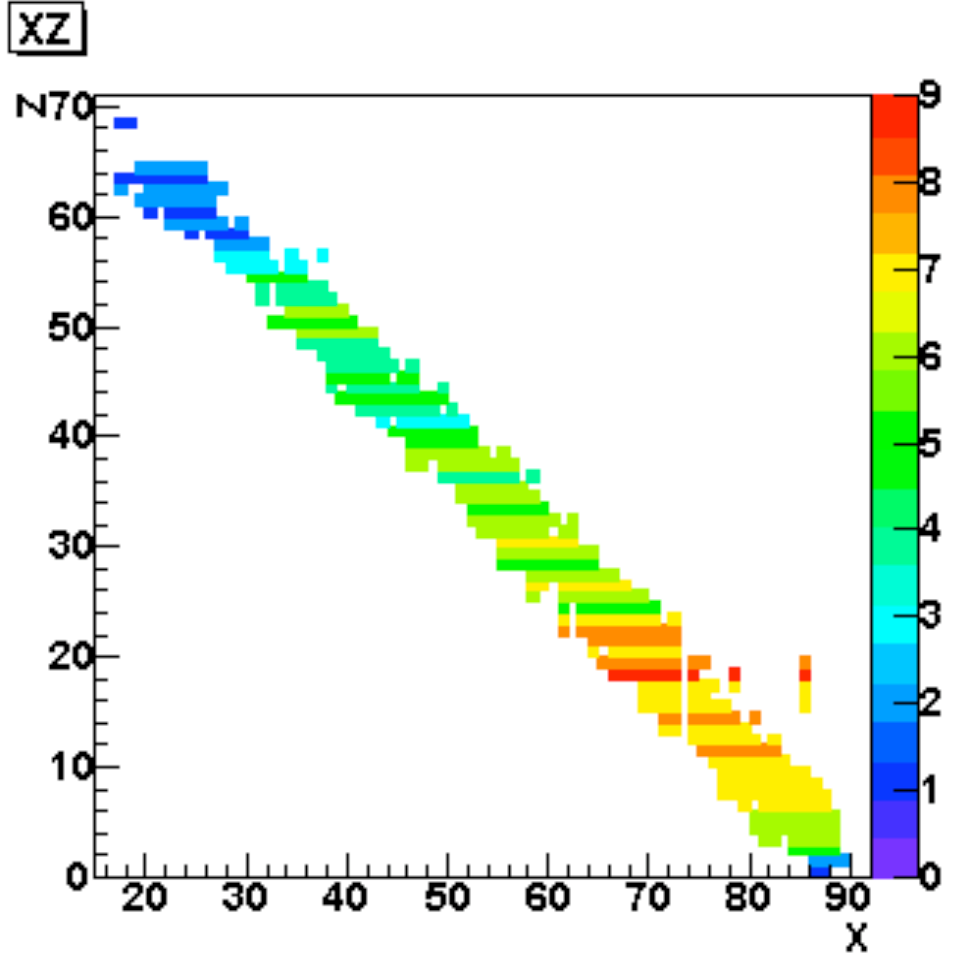} \\
\includegraphics[height=0.3\textheight]{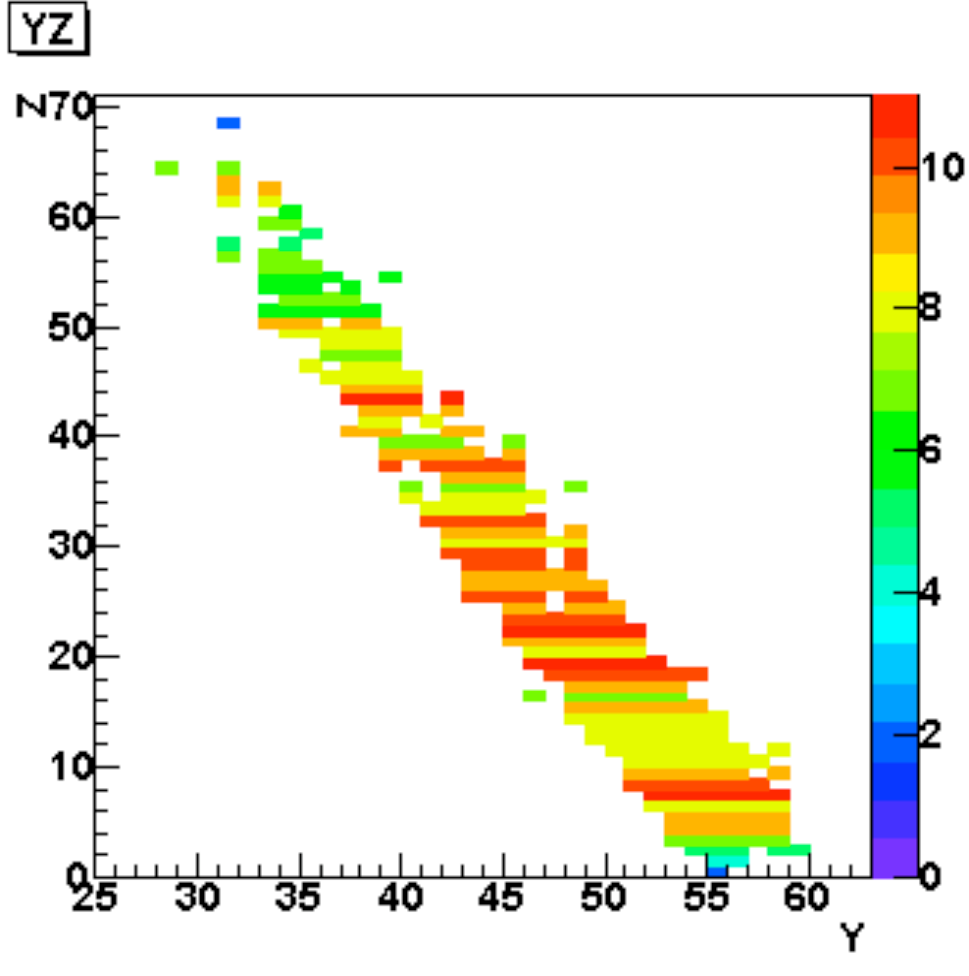}&
\includegraphics[width=0.2\textheight]{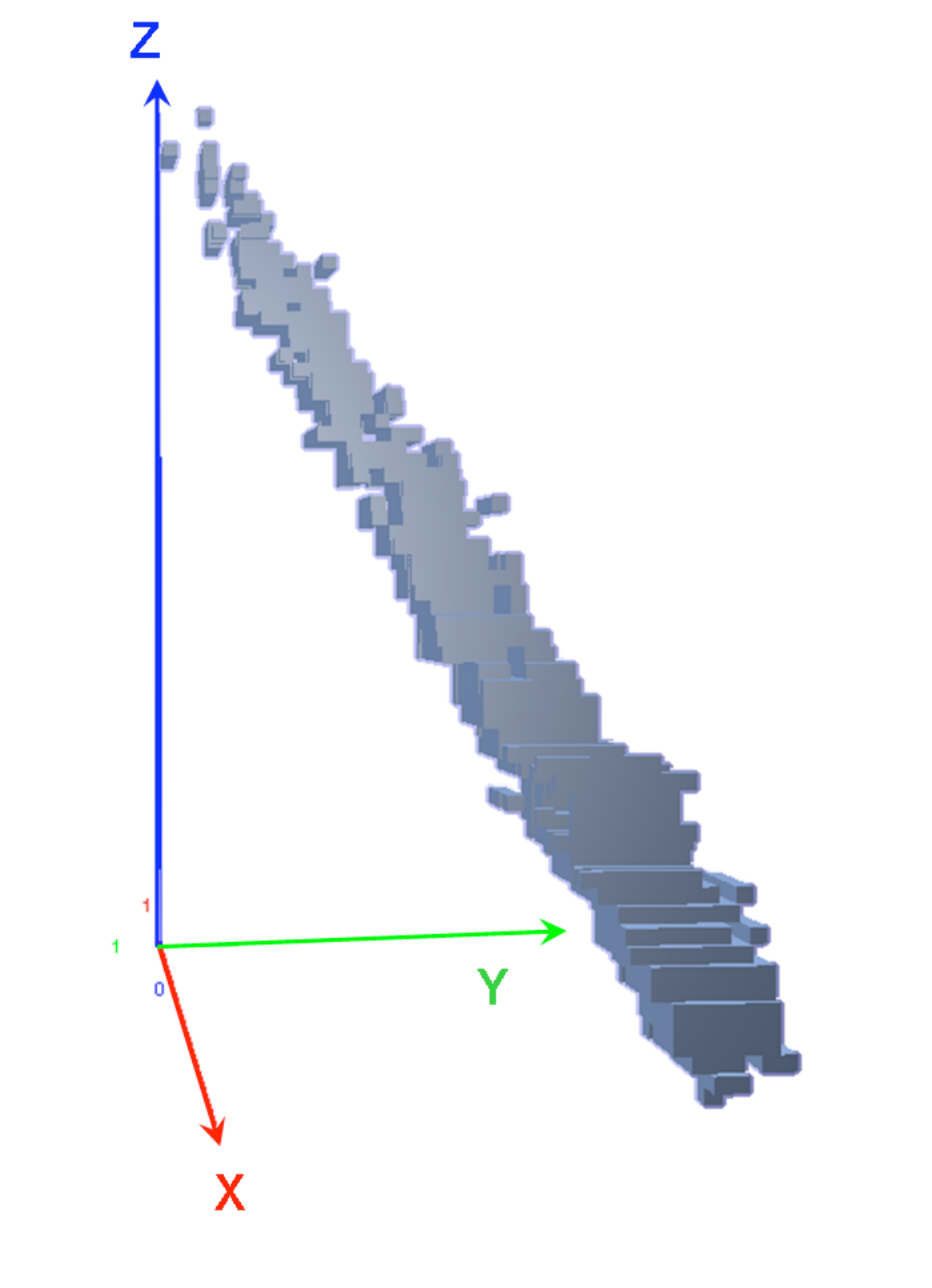}
\end{tabular}
\caption{The x-y, y-z and x-z projections of a 5.5~MeV alpha track
  (from $\rm ^{222}Rn$) as seen by the MIMAC electronic and
  acquisition system. The z-coordinate is reconstructed from 25~ns
  scans of the x-y anode.  The lower right panel presents a 3D view of
  the track. This high energy event will be used to evaluate the drift
  velocity in the gas mixture.\protect\cite{Mayet2009}}
\label{fig:mimac6}
\end{figure*}

\begin{figure*}
\centering
\begin{tabular}{cc}
\includegraphics[height=0.3\textheight]{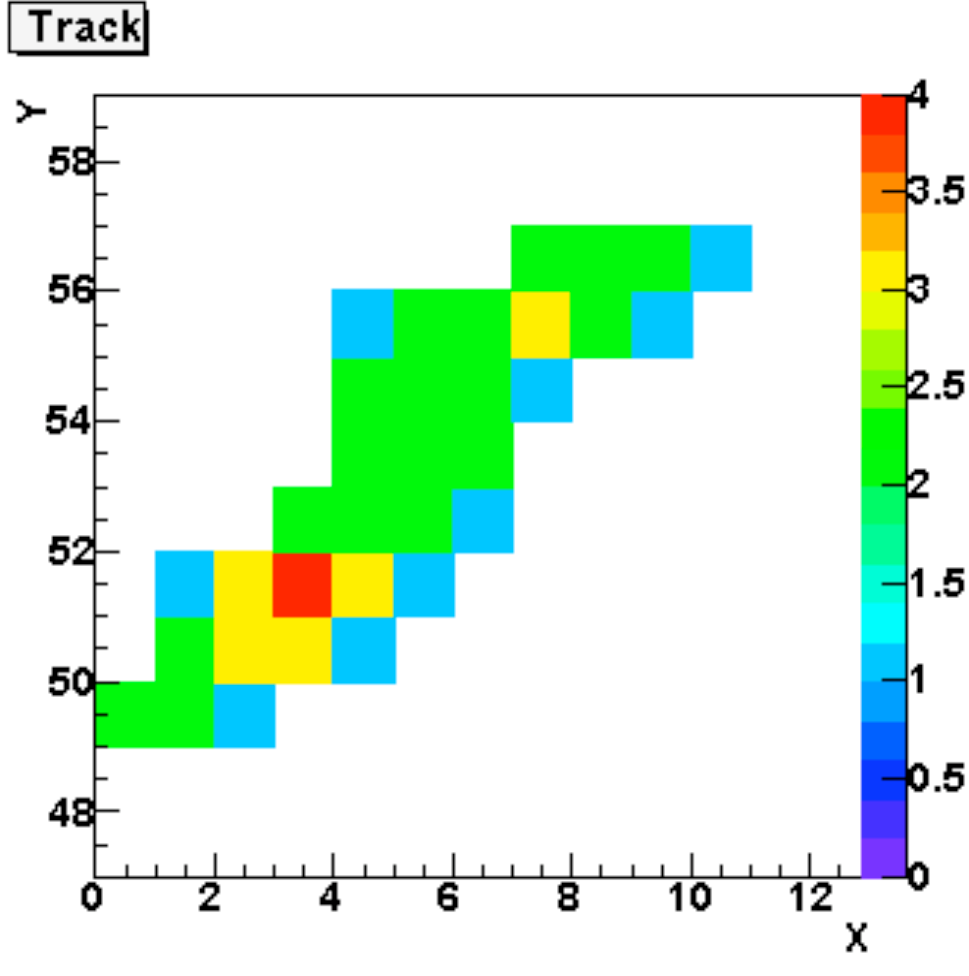}&
\includegraphics[width=0.3\textheight]{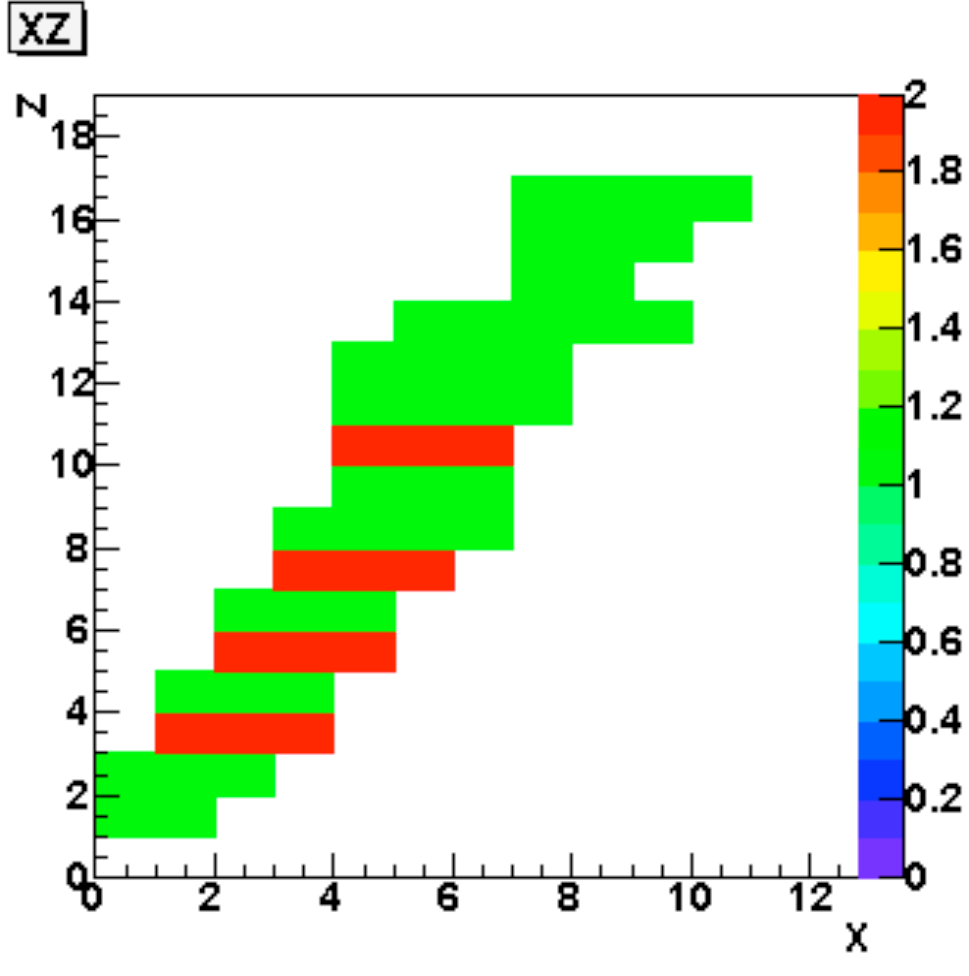} \\
\includegraphics[height=0.3\textheight]{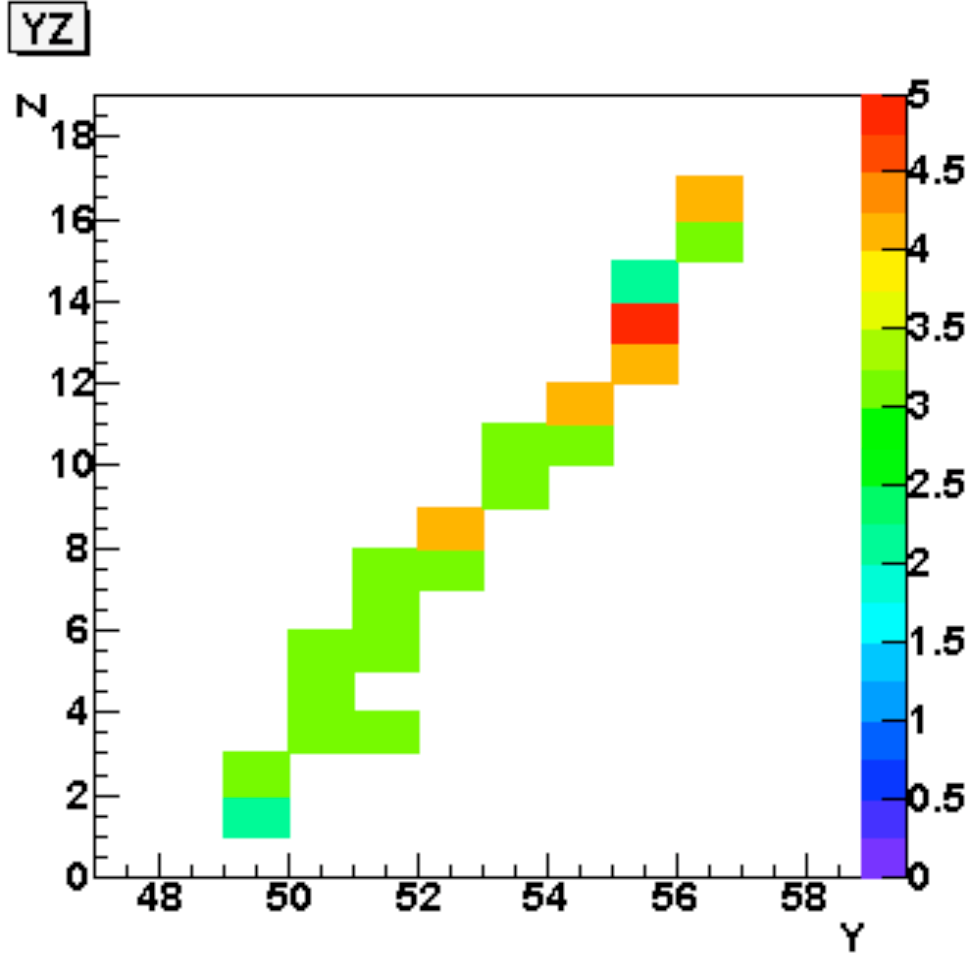}&
\includegraphics[width=0.2\textheight]{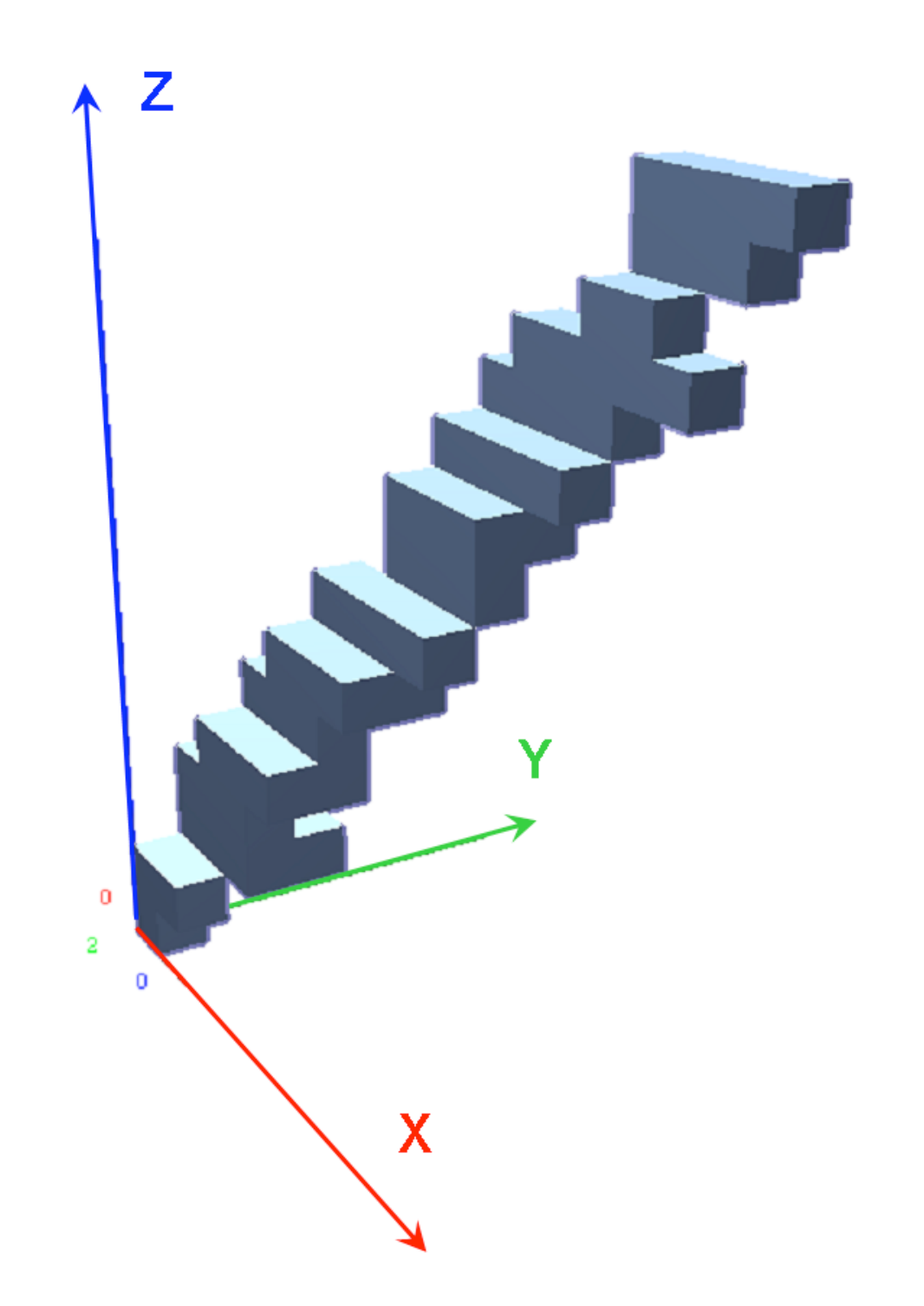}
\end{tabular}
\caption{The x-y, y-z and x-z projections of a 5.9~keV electron track
  as seen by the MIMAC electronic and acquisition system. The
  z-coordinate is reconstructed from 25~ns scans of the x-y anode.  The
  lower left panel presents a 3D view of the track. The volume of a
  pixel is $424 \times 424 \times 400$~$\mu$m$^3$, the size in z
  being driven by the scanning time ($\rm$25~ ns) and the electron
  drift velocity ($\rm 16 \ \mu m /ns$ in this case).}
\label{fig:mimac7}
\end{figure*}

Fig.~\ref{fig:mimac6} presents a 5.5~MeV alpha track (from $\rm
^{222}Rn$), obtained in \he +5\% \iso mixture at 350~mbar.  Although
at higher energy than needed for dark matter detection, this event
shows the quality of the 3D reconstruction.  The upper left panel
presents the projection on the anode (x-y plane) and the lower right
panel presents the 3D track, the z-coordinate being reconstructed with
the strategy described above.  5.5~MeV alpha tracks from $\rm
^{222}Rn$ will be used to evaluate the drift velocity in various
mixtures of interest for directional dark matter detection.

Fig.~\ref{fig:mimac7} presents a reconstructed 5.9~keV electron track,
obtained in \he +5\% \iso mixture at 350~mbar. This is the first
low-energy track in MIMAC. Both the length (8~mm), and the angle
($\phi,\theta$) are reconstructed. The lower right panel presents a 3D
view of the track. The volume of a pixel is $\rm 424 \times 424 \times
400 \ \mu m^3$, the size in z being driven by the scanning time ($\rm
25 \ ns$) and the electron drift velocity ($\rm 16 \ \mu m /ns$ in
this case). This event is of particular interest as it is a typical
background event for dark matter.  The correlation of the measured
energy and the 3D reconstructed track length allows us to discriminate
nuclear recoils from background.

\subsection{MIMAC summary}
MIMAC has successfully measured recoils through ionization at the low
energies of interest for WIMP detection.  We have developed a specific
chip giving access to a 3-D track reconstruction with a 300 microns
spatial resolution.  We are ready to scale-up the prototype to a
1~m$^3$ in the framework of an international collaboration.

%% file: sectionEmulsions.tex
\section{Nuclear Emulsions}
\label{sec:emulsions}

Nuclear emulsions allow for both tracking resolution and large target
mass, which has great potential for directional dark matter detection.
Emulsions are photographic films composed of AgBr and gelatin that can
be used as 3D tracking detectors with $\sim 1$~$\mu$m resolution. This
resolution is essential for the detection of short life-time
particles, for example the tau neutrino,\cite{kodamaPRD2008} double
hyper nucleus\cite{aokiProgTheorPhys1991} and so on.  Recent analyses
of nuclear emulsions are all automated,\cite{aokiNIMB1990} which
enables large-scale experiments.  For example,
OPERA,\cite{AcquafreddaJINST2009} a neutrino oscillation experiment at
Gran Sasso, Italy, uses 30,000~kg of emulsions.

\begin{figure}
\centering
\includegraphics[width=0.9\textwidth]{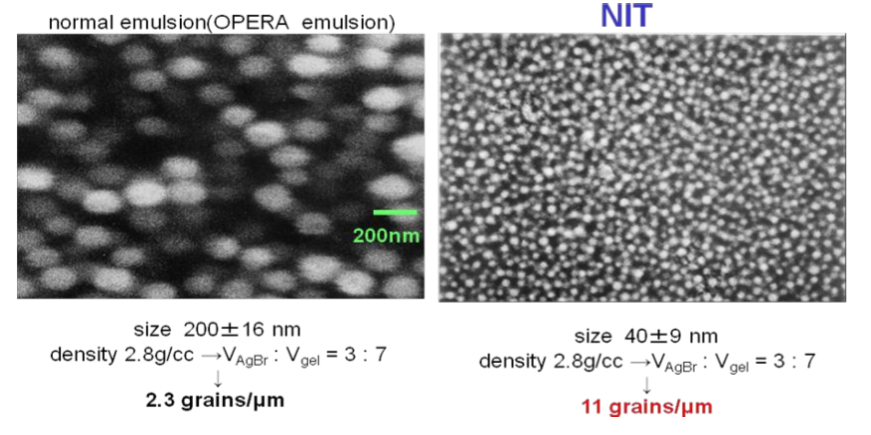}
\caption{Electron microscope image of standard emulsion (left) and NIT emulsion (right)}
\label{fig:emulsionsFig1}
\end{figure}

Nuclear emulsions may be useful in a directional dark matter search if
they can detect the nuclear recoil tracks from WIMP interactions with
sufficient accuracy.  The high density ($\sim$3~g/cm$^3$) of
emulsions, and extremely high resolution are the strongest points.
From SRIM, the expected range of a WIMP-induced nuclear recoil track
is on the order of 100~nm.  However, it is difficult to detect nuclear
recoil tracks in standard emulsions because the maximum resolution is
about 1~$\mu$m.  Therefore, we developed a new high-resolution nuclear
emulsion, called the ``Nano Imaging Tracker''
(NIT).\cite{NatsumeNIMA2007} In the NIT, the AgBr crystal size is
40$\pm$9~nm and the density is 2.8~g/cm$^3$.  The density of AgBr that
an incoming particle can penetrate is 11~AgBr/$\mu$m
(Fig.~\ref{fig:emulsionsFig1}).

\subsection{Test of nuclear recoil track direction}
\begin{figure}
\centering
\includegraphics[width=0.9\textwidth]{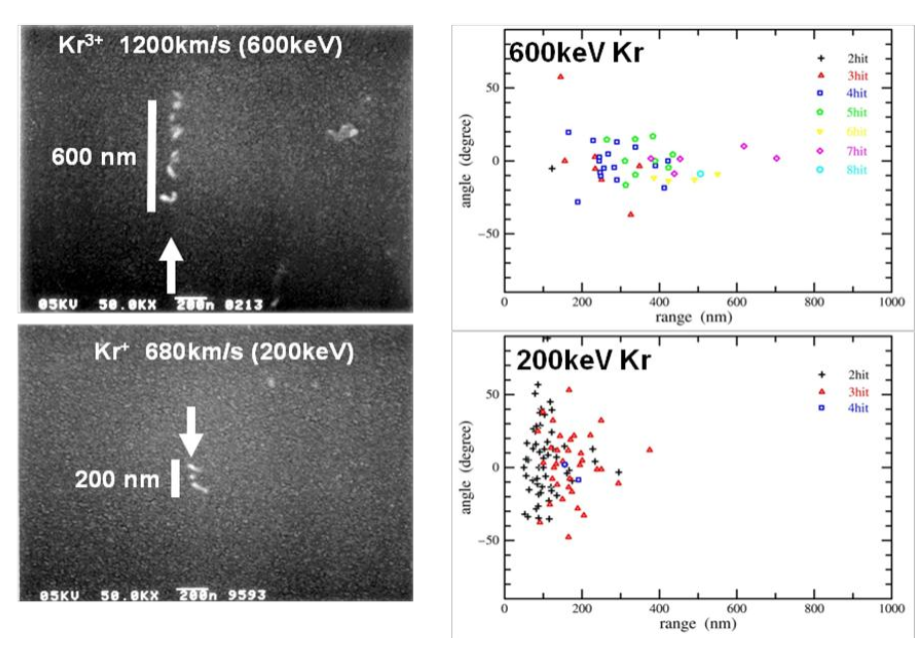}
\caption{(left) Electron microscope image of Kr ion track (600~keV and 200~keV) and (right) track data.}
\label{fig:emulsionsFig2}
\end{figure}

By using low-velocity Kr ions instead of nuclear recoils, the ability
to reconstruct the nuclear recoil detection was determined.  With an
electron microscope, the track, which extended over several emulsion
grains, could be observed (Fig.~\ref{fig:emulsionsFig2}), and the
detection efficiency was measured to be more than 90\%.  The measured
range was consistent with SRIM simulation.\cite{NatsumeNIMA2007}

\subsection{Optical detection of nuclear recoil track}
\begin{figure}
\centering
\includegraphics[width=0.9\textwidth]{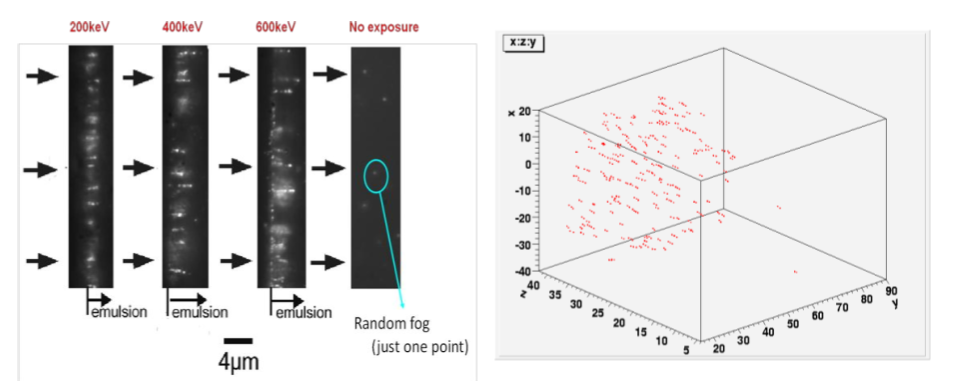}
\caption{(left) Optical microscope image of expanded Kr ion track (200, 400, 600~keV and reference).  (right) 3D tracking data for expanded Kr ion track (600~keV).}
\label{fig:emulsionsFig3}
\end{figure}

When we consider a dark matter search, it is not realistic to use an
electron microscope to scan a large volume of emulsions.  Furthermore,
nuclear recoil tracks which are less than 1~$\mu$m long (smaller than
the optical resolution) cannot be identified as tracks by an optical
microscope.  To resolve this problem, a method was developed to expand
the tracks.  Nuclear recoil tracks consist of grains spanning roughly
100~nm. If the emulsion is then expanded, the inter-grain spacing
grows and the track length is expanded to several $\mu$m.  With this
technique, nuclear recoil tracks may be identified by an optical
microscope.  Here, we used a pH-controlled chemical treatment to
expand the emulsion.  As a result, tracks from Kr ions with
$E>200$~keV attained lengths of several $\mu$m, and could be
recognized as tracks by an optical microscope.\cite{NakaNIMA2007} Such
tracks could be distinguished from random noise (fog) because the fog
consists of single grain events (see
Fig.~\ref{fig:emulsionsFig3}). The angular resolution of the original
state for NIT is about 12~degrees or better. However, the angular
resolution is expected to be about 45~degrees with the expansion
technique. In practice, if the expansion technique is used, two or
more NIT emulsion detectors should be mounted in the directions
horizontal and vertical to Cygnus on an equatorial telescope.

\subsection{Background rejection}
We can discriminate electrons and $\gamma$ from nuclear recoils by
their different $dE/dx$ values.  Because the sensitivity of nuclear
emulsions usually depends on the $dE/dx$ of the incoming particle, by
controlling the sensitivity of the emulsion itself and the power of
the development treatment, electrons with low $dE/dx$ will not appear
in the emulsion.  In the present state, the estimated electron
rejection power is 10$^{5}$ or better using new sensitized chemical
treatment (Halogen Acceptor sensitization).\cite{KugeJIST2009}
Moreover, we expect to improve this sensitivity control of the
development treatment.  For alphas, serious background sources are U,
Th chain and Rn. Since these energies are on the order of MeV, nuclear
emulsions can identify the alpha track with 3D range
discrimination. This means that alphas are not backgrounds for
emulsions by at least fiducial cut.  Neutrons are expected to be the
main background. With sensitivity control of the development
treatment, neutron recoil tracks which have low $dE/dx$ may be
rejected, but, finally they are rejected by directionality.

\subsection{Sensitivity for WIMP searches}
The sensitivity of nuclear emulsions to WIMPs is estimated with a
Monte Carlo simulation. For zero background events, a 1000~kg-year
exposure, and 100~nm range threshold ($\sim$100~keV energy threshold
for Ag, Br recoils), the sensitivity is expected to be better, by
about one order of magnitude, than the current XENON10
limit.\cite{AnglePRL2008} In addition, by using higher resolution NIT
(for example, 20~nm AgBr size), the range threshold would be 50~nm and
the sensitivity would improve further by one order of magnitude.

\subsection{Future Planning}
Currently, we are developing the new readout system and emulsion
production facility. In addition, in recent R\&D efforts, we are
studying large nuclear stopping power using a dedicated development
treatment. This will enable more background rejection and head-tail
discrimination.  We aim to put these developments into practice and
start running a 1~kg prototype within two years.

%% file: sectionVahsen.tex
\section{TPC Readout with Gas Electron Multipliers and Silicon Pixels}
\label{sec:pixelChips}
Directional dark matter detection may benefit from recent advances in
detector technologies at particle colliders.  One option is to read
out Time Projection Chambers with silicon strip or silicon pixel
detectors. This approach is used by NEWAGE
(Section~\ref{section:NEWAGE}), MIMAC (Section~\ref{section:MIMAC}),
and Lawrence Berkeley National Laboratory (LBNL) (this section). While
a silicon based readout is costly per unit readout area, it can offer
full 3-D reconstruction of nuclear recoils, ionization measurement in
each detector element, and low threshold operation. As a result this
approach to TPC readout should improve both background rejection and
WIMP signal sensitivity, and may prove competitive in terms of total
detector cost for comparable sensitivity.

\subsection{LBNL TPC Prototype Device}

The LBNL group has developed a small TPC prototype device where the
ionization charge is amplified with Gas Electron Multipliers (GEMs)
and read out with pixel Integrated Circuit (IC)
chips.\cite{KimNIMA2008} This allows 3-D reconstruction of tracks in
the TPC drift volume.  The GEMs used were purchased from CERN, while
the pixel chip used was the FE-I3 pixel chip from the ATLAS experiment
at the LHC.  This pixel chip was designed for the high event rates and
high radiation dose at the LHC, but its pixel size ($50\times400$~$\mu$m)
and operating frequency (40 MHz) happen to match the spatial and
timing resolution requirements set by directional dark matter
detection with low pressure gas TPCs.  By measuring ionization in each
pixel, the pixel chip also provides a measurement of $dE/dX$, which is
needed for background suppression when detecting WIMPs,

\subsection{Measured Performance with Cosmics Rays}

\begin{figure}
\centering 
\includegraphics[height=0.3\textheight]{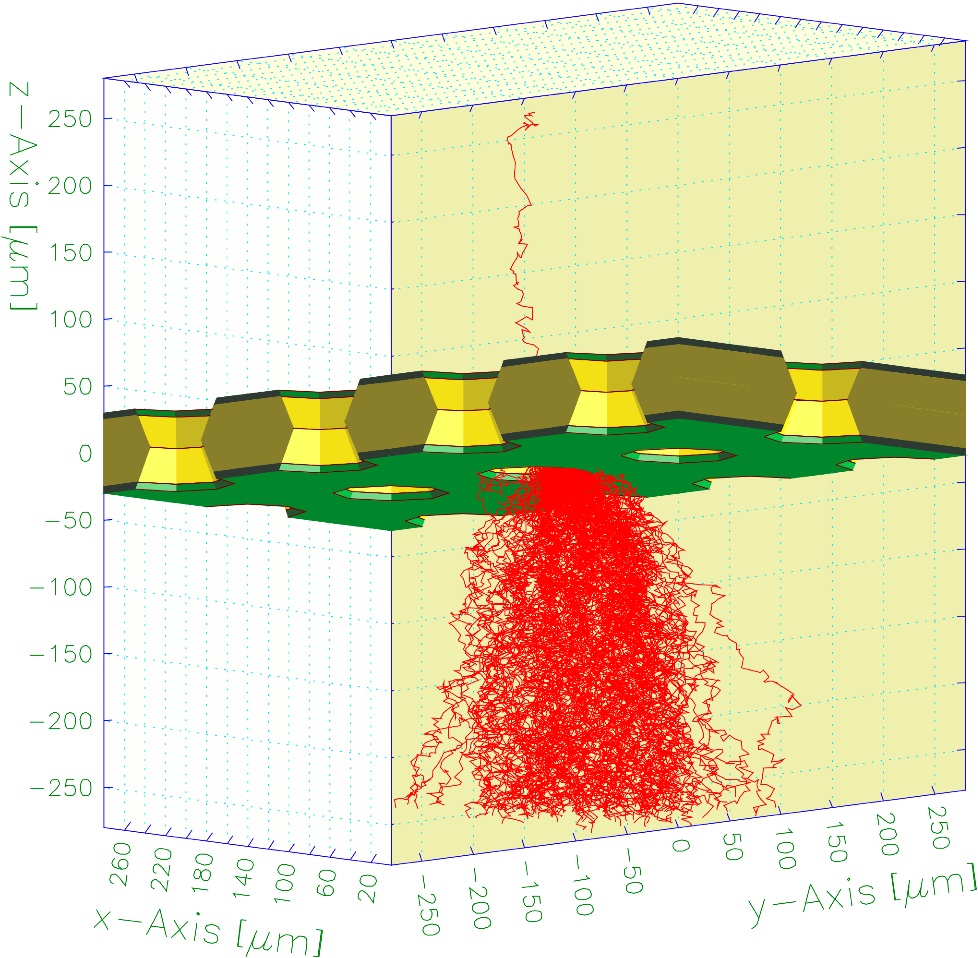}
\caption{A simulated avalanche produced by a single electron passing
  through a single GEM. With 500 V across the GEM, this event has a
  multiplication factor of 258. Only the electrons contributing to the
  effective avalanche gain are shown.}
\label{fig:vahsen1}
\end{figure}

\begin{figure}
\centering 
\includegraphics[height=0.3\textheight]{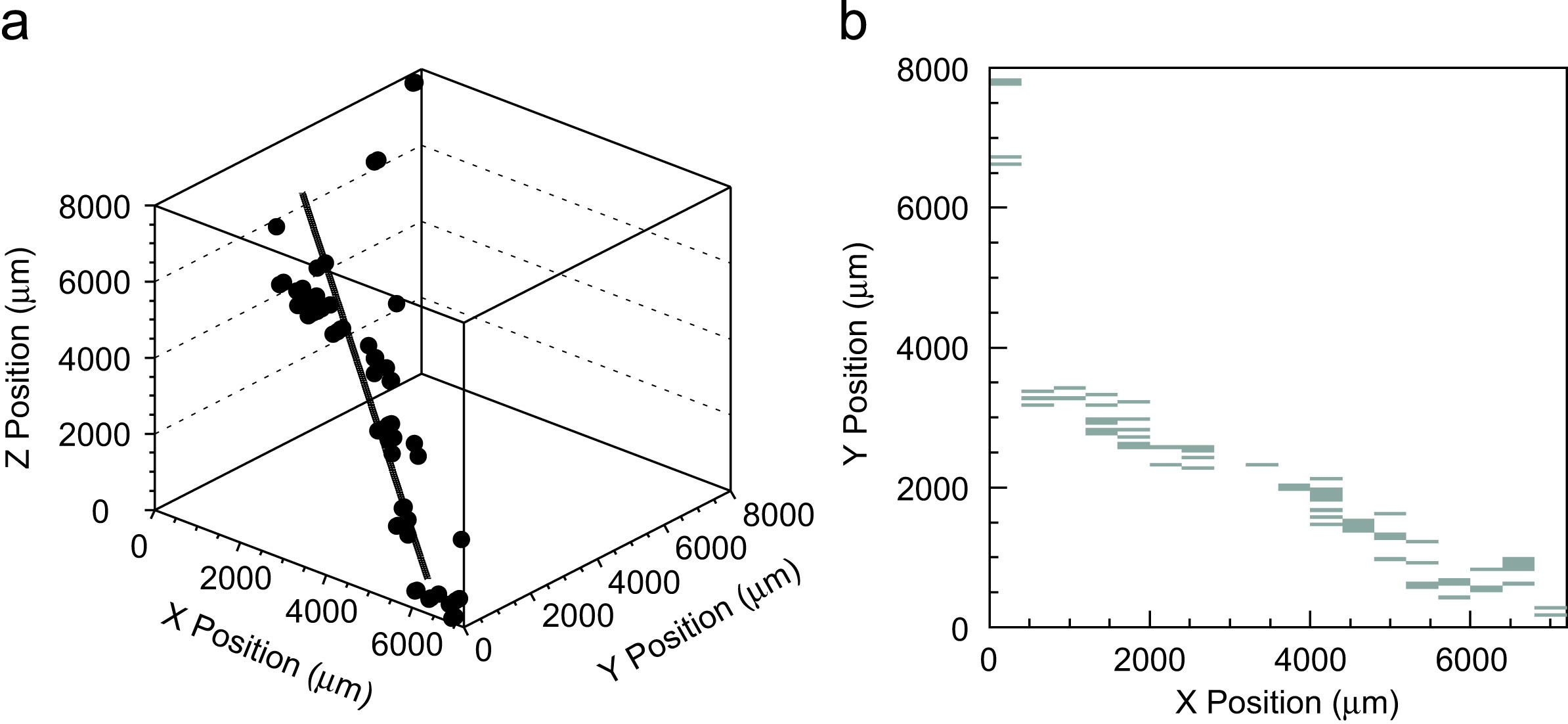}
\caption{A typical cosmic ray event. (left) The event reconstructed in
  3-D space. Each point represents a hit in the pixel chip, and the
  line is a result of a fit to these points. The x-y position of each
  point is given by the center of mass of the corresponding hit pixel,
  and the z position given by the arrival time of the hit. (right) Raw
  readout of the pixel chip for the same event. Each filled rectangle
  represents a $50\times 400$~$\mu$m silicon pixel where a hit was
  recorded.}
\label{fig:vahsen2}
\end{figure}

The performance of the prototype device was measured with cosmic rays,
and seems very promising. By using two GEM layers, it was possible to
achieve both large gains (up to 40,000 with Ar/C0$_2$) and stable
operation.  Fig.~\ref{fig:vahsen1} shows a simulated avalanche in
one of the GEM layers.  The pixel chip has very low noise ($\sim$120
electrons), compared with a typical operating threshold of 2-5000
electrons. Reconstructed charged tracks yield an x/y/z point
resolution of 130/70/150~$\mu$m when operating with Ar/C0$_2$ gas at
atmospheric pressure (these are the readout resolutions, after
subtracting the estimated diffusion in quadrature).
Fig.~\ref{fig:vahsen2} shows a typical example track found in a
cosmic ray event.

\subsection{Applicability to WIMP Searches}

Of particular relevance to dark matter detection is that the prototype
device appears capable of reading out all the primary ionization
charge, even single electrons, with efficiency close to unity.  That
means one should be able to reach a very low energy threshold when
reconstructing nuclear recoils from WIMP/nucleon collisions. The FE-I3
pixel chip buffers hits until a trigger occurs, and pixels without
hits do not generate data. Thus demands on a downstream data
acquisition system should be relatively light in the case of WIMP
detection.

Several critical issues need more study and work before this
technology can be applied to directional WIMP dark matter detection.
For instance, the ATLAS pixel chip is currently limited to reading out
16 consecutive time intervals of 25~ns, which limits the size of drift
volume (along the drift direction) that can be read out.  This part of
the chip design may need to be modified for directional dark matter
detection.  One also needs to demonstrate that the GEMs can operate
without sparking or serious aging problems when using gases and
pressures more suitable for directional dark matter detection.

%% file: sectionZaragoza.tex
\section{Development of low-background micromegas readout planes for directionality experiments}
\label{sec:zaragoza}
In micropattern readouts, metallic strips or pads, precisely printed
on plastic supports using photolithography techniques (much like
printed circuit boards), are used in place of wires to receive
drifting charge produced in the gas.  Micropattern readouts are more
simple, robust and mechanically precise than conventional wire planes.

Detectors based on this concept, first introduced by Oed in
1988,\cite{OedNIMA1988} are called Micro-Pattern Gas Detectors (MPGD).
Thus far, several designs employing different multiplication
structures have been proposed, with varying levels of success
(Microstrip, Microwire, Microgap, Microdots (micropin), Microwell, Gas
Electron Multiplier (GEM), etc.).  Here, we focus on one of the most
promising MPGDs: the Micromesh Gas Structure or
micromegas.\cite{GiomatarisNIMA1996}

The micromegas concept, created about 14 years ago and actively
developed since then by the CEA/Saclay group led by I. Giomataris,
consists of a micromesh held $\sim$50-100~microns away from the strip
plane by insulating spacers.  This geometry defines a high electric
field gap in which an electron avalanche is produced, as in parallel
plate chambers, inducing signals in both the mesh and the strips.  The
resulting avalanche topology can be imaged with unprecedented spatial
resolution: a 2D readout pitch of 300~$\mu$m is used in the CERN Axion
Solar Telescope (CAST), and small prototypes with 50~$\mu$m pixels
exist.  The micromegas concept is already employed in many particle
physics experiments, and unique temporal, spatial and energy
resolutions have been demonstrated (temporal resolutions below 1~ns,
and spatial resolutions of at least 17~$\mu$m have been
achieved,\cite{Derre2002} and an energy resolution of less than 11\%
FWHM for the 5.9~keV $^{55}$Fe peak is routinely achieved with
state-of-the-art micromegas planes).

In the field of rare events, the use of micromegas readouts have been
pioneered by the Saclay and Zaragoza groups for CAST.  Since 2002,
CAST has employed a low-background micromegas detector.  In 2007, two
additional ones replaced the multiwire TPC.  CAST has been a testing
ground for micromegas technology.  Ongoing efforts have produced three
generations of detectors (regarding the fabrication method, detector
materials, shielding and electronics and data reduction treatment),
each with improved background levels.

Current micromegas readouts profit from the \emph{bulk} method of
fabrication,\cite{GiomatarisNIMA2006} in which the micromesh is
imprinted into the readout itself using photolithography techniques.
This improves the device's durability.  In the latest generation of
CAST detectors, the so-called \emph{microbulk} micromegas developed by
CERN/Saclay, the readout plane is fabricated from double-clad kapton
foils, thereby increasing the gap precision and the gain stability and
energy resolution of the detector.  These improvements come at the
cost of losing a bit of robustness and potential maximum area of
fabrication with respect to the bulk fabrication technique.

An R\&D program dedicated to developing high radiopurity micromegas is
currently being pursued by the Zaragoza group.  In the first phase,
measurements of micromegas planes are being performed in the ultra-low
background facilities of the Canfranc Underground Laboratory (LSC).
Promising preliminary results indicate that \emph{microbulk} readouts
fabricated from radiopure raw materials exhibit lower levels of
radiopurity per unit surface area (at the level of 0.1~mBq/cm$^2$ for
U/Th) than competing readouts used in very low background experiments
(e.g. low-background photomultiplier tubes or avalanche photodiodes),
even in the absence of specific radiopurity development.  In
subsequent phases of this project, the aim is to identify traces of
radioactivity and try to further reduce them by better choosing raw
materials or by reducing contamination introduced by the manufacturing
processes.

These tasks are part of a larger R\&D effort, led by the Zaragoza and
Saclay groups, which studies several aspects of the new generation of
detectors for their application in other rare event experiments.
Among other things, exhaustive studies of the energy resolution
capabilities of microbulk micromegas are being
done.\cite{DafniNIMA2009} In addition, the development of multiplexed
patterns is under exploration.  This would reduce the number of
channels required to read out large detector areas with high spatial
resolution.

This work is part of the demonstration of scaling-up strategies which
are essential for large directional WIMP detectors, and which includes
the possibility of building single readouts of medium-large size,
without losing the performance shown in the small prototypes.  Along
this line, \emph{bulk} micromegas have been already built at the m$^2$
scale in the context of other projects,\cite{AnvarNIMA2009} though
operational \emph{microbulk} micromegas have been built only at the
few tens of cm$^2$ scale.\cite{AbbonNJP2007,AuneJPCS2009} Currently, a
$30 \times 30$~cm$^2$ \emph{microbulk} is being built to test and push
the capability of current manufacturing techniques towards larger
areas.  Alternatively, \emph{mosaic} micromegas schemes are being
devised to instrument large readout areas with minimal dead zones.
Finally, the previously mentioned multiplexing schemes are to be
implemented in these large area readouts.

To conclude, micromegas detectors are under active development.  An
increasingly important portion of this R\&D effort focuses on low
background systems that would be a promising readout option for
large-scale gas TPCs for directional WIMP detection.

%% file: sectionOutlook.tex
\section{Directional detector scale-up feasibility study}

We now provide a qualitative discussion of the feasibility of a large
scale directional TPC in relation to the state of the art.  We
summarize the main factors concerning a design, and highlight some
critical-path issues needing further R\&D or design work.
\subsection{Cost Arguments}
\label{outlook:costArguments}
First, it is important to note that the total cost per unit
sensitivity must be a primary factor in a design strategy.  This point
is particularly important for directional dark matter detectors
because, when comparing with non-directional large-scale experiments,
the cost drivers are not necessarily obvious.  For instance, the need
to use low pressure gas (leaving aside for now any prospect of solid
state devices, such as emulsions) implies a detector of relatively
large volume (possibly up to 10$^3$-10$^4$ m$^3$ for spin-independent
sensitivity) and hence an underground cavern larger than is within the
experience of the conventional non-directional dark matter community.
However, it is likely incorrect to assume that size itself is a major
issue of concern, let alone a showstopper.  Neither cavern excavation
nor mechanical construction are likely to be major cost drivers for a
large-scale directional detector.

Quantifying the sensitivity goal for a directional concept requires a
different philosophy since, unlike non-directional detectors, the
objective is not just to achieve a hint of a signal (or a limit) but
to show clearly that events are of galactic origin.  This would be
unambiguous evidence for the existence of WIMP dark matter.  A useful
sensitivity parameter for comparison purposes is to determine, for a
given WIMP type, interaction cross section, and distribution in the
galactic halo, the minimum number of WIMP events N$_W$ required to
demonstrate to 95\%~C.L. that the distribution of recoil directions is
not isotropic in the galactic frame, and hence cannot be of
terrestrial origin.  For instance, for the standard halo model,
assuming 3D reconstruction with sense determination in an idealised
detector with 20~keV recoil threshold, and accounting for basic gas
physics such as straggling, it has been shown that N$_W \sim
10$.\cite{MorganPhysRevD2005} For a practical detector it is likely
that $N_W >> 10$ (if, for instance, there is only 2D reconstruction or
no sense determination, depending on the detector design).

There is a cost tension between $N_W$ and target mass, two parameters
that likely scale very differently.  For instance, investing in a new
technology that reduces $N_W$ by a factor of 2, such as by improving
the position resolution with better readout technology, is clearly
only reasonable if the cost is less than that of doubling the target
mass with the same readout technology.  The latter might be a better
option, depending on the absolute sensitivity required.  Either way,
the scale required implies that more work is needed to produce lower
cost readout, electronics and mechanical infrastructure, though less
so if the DAMA signal proves correct, in which case a much smaller
detector would be sufficient.  In fact, in this situation (e.g. if the
DAMA signal can be explained by inelastic scattering), then the
low-pressure gas TPC technology has significant advantages thanks to
its particle tracking and identification ability for both electrons
and nuclear recoils.

\subsection{Directional capability cost benefit}
Recent advances, many described in this work, have shown that there is
some head-tail discrimination in low pressure gases.  Thus, any of the
current readout technologies, be it MWPCs, micropixel or CCD optics
etc, could in principle cover a full spectrum of directionality
capability as follows: (i) simple 1D readout with sense (head-tail
discrimination), (ii) 2D readout with or without sense, (iii) 3D
readout, with or without sense, (iv) all of the previous with or
without absolute 3D event location (fiducialisation), i.e. $\Delta x$,
$\Delta y$, $\Delta z$ and absolute $x$, $y$, $z$.  The first option
may appear a poor bet.  Indeed, 1D sensitivity (say $z$-projection in
a gas TPC via timing information only) without sense, is probably
ruled out completely. However, the forward-backward asymmetry signal
for dark matter is so powerful that, depending on the extent to which
head-tail discrimination can be realised, it is possible to envisage a
scenario in which the funds that would have been used to build
expensive 2D ($x$-$y$) readout planes are better used simply to build
more target mass with 1D sensitivity, overcoming the implied loss in
$N_W$.  The main issue with this option is then how to maintain the
necessary high background rejection (see Section
\ref{outlook:bkgRejection}).  Pure 2D readout alone, as provided by
current generation CCD technology, may also not be sufficient without
the introduction of head-tail discrimination.  Full 3D reconstruction
of tracks provides a potential option even without head-tail.
However, while the MWPC technology has approached 3D reconstruction,
at least one of the dimensions is restricted in resolution by the wire
spacing, currently $2$~mm but potentially $1$~mm.  The micropixel
readout has demonstrated full 3D reconstruction with position
resolution $< 1$~mm in all directions, but only at small scale ($\sim
10$~cm) and relatively larger cost, without head-tail discrimination.

\subsection{Background rejection}
\label{outlook:bkgRejection}
While it may be possible to reduce the scale-up cost for a given
sensitivity by saving on the readout tracking capability, compensating
this by increased target mass, in practice will depend also on the
background rejection capability, particularly of gammas, low energy
alphas and, most notably, radon progeny recoils (RPRs).  The extreme
case is 1D readout (case (i) above).  For instance, the use of simple
non-pixellated anode planes severely reduces the $dE/dx$
discrimination capability and introduces a degeneracy for certain
orientations of particle tracks.  The situation is less severe with
simple 2D projection readout, as in the basic CCD option.  However,
one remaining issue is the degeneracy between short high $dE/dX$
tracks and long low $dE/dX$ tracks that happen to be oriented nearly
perpendicular to the readout plane.  Full 3D reconstruction, as
provided by (e.g.) measuring the temporal profile of the induced
charge on the 2D readout plane can break this degeneracy and improve
background rejection.  However, there still remains the issue of
identifying low energy events from the detector walls.  Radon progeny
recoils, the low energy recoils from decay of radon daughters like
$^{218}$Po deposited on surfaces by radon, have been identified as a
severe issue here, since they can mimic WIMP
events.\cite{BurgosNIMA2008} Low energy alphas, from U, Th
contamination, arising when most of the alpha energy has been lost in
the walls prior to emerging into the gas, are also a challenge.

Elimination of such events requires measurement of the absolute 3D
position of events within the target volume, i.e. $x$, $y$ and $z$
information - i.e. 3D fiducialisation.  Implementation of such
fiducialisation is well established in 2~dimensions ($x$-$y$), because
the track position is measured relative to the $x$-$y$ edges for any
pixilated 2D $x$-$y$ readout.  However, without a measurement of the
absolute $z$ of an event, RPRs from the central high voltage cathode
are the most insidious background events in directional TPCs.

Significant $z$-fiducialisation can be obtained from the pulse shape
information encoded in the induced charge signals on the TPC
electrodes.  Ionization products from RPR events at the cathode must
travel the entire drift length and will suffer greater diffusion, and
therefore produce longer pulse shapes.  This method of RPR rejection,
however, is not perfect since there is a degeneracy with track
orientation.  New techniques are under investigation by several groups
to achieve the necessary full $z$-fiducialisation.  In summary, taking
account of the concepts in Sections
\ref{outlook:costArguments}-\ref{outlook:bkgRejection}, it is likely
that a large-scale directional detector will need 3D fiducialization,
at least 2D directional reconstruction, and head-tail discrimination.

\subsection{Intrinsic background, shielding, and site choice}
The efficiency required to achieve the background discrimination and
$z$-fiducialisation described in Section \ref{outlook:bkgRejection}
depends on the intrinsic level of background achievable.  This has two
aspects: (i) passive and active shielding to reduce external
backgrounds, including the rock overburden in an underground
laboratory, and (ii) material purification to reduce internal
backgrounds, particularly from radon and neutrons.  The requirements
for the former have been well investigated.\cite{CarsonNIMA2005} The
intrinsic insensitivity of the low pressure TPC technique to gammas
means requirements for external high-Z passive shielding (Pb, Cu) are
much reduced compared to conventional detectors.  It is for this
reason that first and second generation DRIFT detectors were
constructed with no gamma shielding.  However, the more sensitive
detectors of the future will likely require some gamma shielding,
depending on the capability to measure $dE/dx$.  Hence, given the low
mass-to-volume ratio of a low pressure TPC and the absolute size
needed, and unlike the case for conventional detectors, it is worth
specifically seeking an underground site that has intrinsically the
lowest possible gamma background, to save on gamma shielding costs.
This means a site in salt is rather well suited to a large TPC.  Salt
sites have the lowest U, Th content but also, unlike the requirements
for other proposed very large detectors such as for proton decay, a
large directional TPC can be designed with an elongated shape, the
type preferred in salt for geotechnical reasons.  So a deep salt site
could be rather well suited to a TPC.  On the other hand, conventional
sites, such as the proposed DUSEL, likely have other benefits that can
impact positively on cost, for instance lower maintenance
requirements.  These issues need careful study.

As with all dark matter detectors, neutron backgrounds are likely the
greatest challenge.  However, the low mass-to-volume ratio for a TPC
relative to conventional detectors makes neutrons a particular issue
because the possibility of rejection by detection of multiple scatters
is reduced, and backscattering within the volume is
increased.\cite{KudryavtsevNIMA2003} Concerning muon-induced neutrons,
this implies that the underground site should likely be deeper than
required for conventional detectors, or alternatively that an external
neutron veto be used.  This requirement may be partially mitigated if
sensitivity to EM radiation (low $dE/dx$) is maintained as a means of
internally vetoing muon induced showers.  Reduction of the neutron
flux from radioactivity in the surrounding rock is better understood
and can be achieved at modest cost using either polyethylene pellets,
high-H walling material, or perhaps water shielding.  However,
neutrons from contamination of detector construction materials and
components are more problematic, as in conventional direct detection
experiments.  The vessel structure is likely an important factor,
however other components will also need careful selection, for
instance the insulator material used in micromegas, and the optics in
the case of CCD readout.  Concerning radon and RPRs, the active
fiducialisation being developed to mitigate this will not have 100\%
efficiency.  Hence selection of detector materials with very low radon
emanation rates as well as radon getters in the gas system will be
necessary.

\subsection{Engineering and infrastructure limits}
Excavation of large caverns underground is well understood even at the
scale required for a megaton proton decay experiment.  A directional
TPC would likely not require such a large cavern.  Taking the example
of SuperK, such a cavern could accommodate a directional target mass
of order 10~tons, even at 40~Torr.  At the known typical excavation
cost of \$20-50 per m$^3$, this implies a cost of $\sim\$250$k per ton
of directional target.  Excavation is thus unlikely to be a cost
driver.  Taking as the limit to feasibility an excavation comparable
with a future proton decay experiment, we can obtain an upper limit
target mass of order 400~tons, allowing for some increase in pressure.
This could be capable of directional detection at 10$^{-11}$~pb dark
matter interaction cross section without sense determination.  More
realistic in the near-term are the proposed excavation modules at
DUSEL, which can provide sufficient space for a competitive first
stage target mass.

Regarding engineering structures, though large by dark matter
standards, there are no cryogenics necessary and few physics
constraints on geometry.  Most important is the need for radiopurity
and capability to limit outgassing of impurities, including radon
daughters.

\subsection{Data Acquisition and readout}
This area is likely to be a (or the) major cost driver.  For example,
a 1~ton target at 40~Torr and 2~mm resolution would require $\sim 10^7$
readout channels for a charge readout concept, each requiring fast ADC
output. Allowing for increased pressure and better position resolution
could see this rise to $10^9$ channels per ton.  However, use can be
made of the expected short recoil track lengths to allow grouping of
channels, yielding a reduction to say $\sim 10^6$ channels. In this
scenario, track reconstruction is maintained and an internal fiducial
volume defined to exclude edge events, but there is degeneracy in the
absolute position of accepted events.  Grouping of this form is
performed in DRIFT and is under development for micropixel readout.
Development of new chips could be important here to suppress costs per
channel, but careful design with current technology may provide
significant cost savings.  The CCD option would require of order
10,000 CCD cameras per ton plus associated optics, unless some similar
form of multiplexing or grouping can be found. Otherwise, the concept
would maintain absolute position information in 2D.  A significant
issue for charge readout is also the design of the cabling and
associated feeds to pre-amplifiers outside the vessels.

\subsection{Target gas systems, remote operation and search strategy}
The use of CF$_4$, CS$_2$ negative ion gases, plus mixtures and
additives of other target gases including He, is quite well understood
at small scale now.  It has been demonstrated that CF$_4$ works with
CS$_2$, such that the low diffusion capability of the negative ion gas
technology works with F as a target.  Remote and safe operation has
been shown with CS$_2$ and CF$_4$ in 1~m$^3$, demonstrating that
requirements for operations staff can be minimal.  So far gases have
generally been flowed through vessels and vented to atmosphere through
filters.  However, for large-scale detectors, in order to reduce costs
and to reduce the need for disposal systems, recirculation, with radon
and impurity removal, and/or long-term operation with vessels sealed
is probably essential.  Work is underway on this by multiple groups.

One advantage of using a gas TPC is the possibility of easily changing
the target nuclei and target density (pressure) while using the same
detector apparatus.  This opens the possibility of various search
strategies, including both spin-dependent and spin-independent in the
same detector.  One possibility also is to start at high pressure
first, say 400~Torr, to increase the target mass and hence sensitivity
to events, maintaining recoil discrimination but without directional
information.  Work is needed here to understand the discrimination
capability vs. pressure and gas mixture.  Use of multiple targets and
multiple sites simultaneously may have advantages.

\subsection{Health and safety}
Health and safety is a major consideration for any experiment, however
no showstoppers have so far been identified for a large directional
TPC.  We can make use of extensive experience from previous and
planned large underground experiments.  The main novel issue is gas
handling and high voltages, with their (relatively low) potential for
asphyxiation or accidental creation of poisonous or explosive
mixtures.  All these issues have been addressed by existing
experiments.  Larger volumes will require larger ventilation systems,
but these are well within current mine technology.  A road tunnel site
might require more stringent investigation because of the proximity of
the general public.  The use of large quantities of CF$_4$ needs
special care given its nature as a greenhouse gas.

\subsection{Costs, timescales and main areas of future work}
It is possible to estimate the cost and timescale for construction of
a large-scale experiment by extrapolating from current technology.
For instance, construction of a single 1~m$^3$ DRIFT module of the
current design, including commercial DAQ and shielding is now
$\sim\$50$k.  This is dominated by the electronics and vessel
construction.  Extrapolating from this, making reasonable assumptions
on cost savings through use of mass produced electronics, larger unit
vessels, gas recirculation and other scaling factors, yields
$\sim$~$\$150$M per ton of target.  Additional shielding, such as a
neutron veto or passive gamma shield, could significantly increase
this if a suitable deep and low background site is not available.
Development of the field indicates that such a detector will be
required within $6-8$ years, either to prove that a signal seen in a
non-directional detector is not simply an unidentified terrestrial
background, or because directionality will be needed to reject
possible neutrino-induced backgrounds.

To meet this challenge we can identify several key issues from above
that require further research before a large scale design can be
finalized:
\begin{itemize}
\item demonstration of a technique to completely fiducialize the detector volume
\item demonstration that intrinsic radon related background can be sufficiently reduced
\item demonstration that low background outer vessels are possible
\item demonstration of bulk gas recirculation and cleaning
\item optimization of readout vs. target mass for best cost/sensitivity benefit
\end{itemize}

%% file: sectionAcknowledgements.tex
\section{Acknowledgements}
The production of this document was inspired by the Cygnus 2009
workshop on directional dark matter detection at the Massachusetts
Institute of Technology (MIT) in Cambridge, Massachusetts, USA.  We
would like to thank the National Science Foundation, the MIT Kavli
Institute for Astrophysics and Space Research and the MIT Laboratory
for Nuclear Science for their generous financial support of the
workshop.

The DMTPC collaboration acknowledges the support of the Advanced
Detector Research Program of the U.S. Department of Energy, the
National Science Foundation, the Reed Award Program, the Ferry Fund,
the Pappalardo Fellowship program, the MIT Kavli Institute for
Astrophysics and Space Research, and the MIT Physics Department.

The NEWAGE collaboration wishes to acknowledge the support of
Grant-in-Aids for KAKENHI (19684005) of Young Scientist(A); JSPS
Fellows; and Global COE Program ``The Next Generation of Physics, Spun
from Universality and Emergence'' from Ministry of Education, Culture,
Sports, Science and Technology (MEXT) of Japan.

The MIMAC collaboration acknowledges ANR-07-BLAN-255 funding.

The Emulsions work was supported by the Global COE Program of Nagoya
University, ``Quest for Fundamental Principles in the Universe
(QFPU)'' from JSPS, and MEXT of Japan.

This manuscript has been authored by an author at Lawrence Berkeley
National Laboratory under Contract No. DE-AC02-05CH11231 with the U.S.
Department of Energy. The U.S. Government retains, and the publisher,
by accepting the article for publication, acknowledges, that the U.S.
Government retains a non-exclusive, paid-up, irrevocable, world-wide
license to publish or reproduce the published form of this manuscript,
or allow others to do so, for U.S. Government purposes.

%% file: cygnus2009Whitepaper.bbl
\begin{thebibliography}{10}

\bibitem{BertonePhysRept2005}
G.~Bertone, D.~Hooper and J.~Silk, {\em Phys. Rept.} {\bf 405}, 279 (2005).

\bibitem{JungmanPhysRept1996}
G.~Jungman, M.~Kamionkowski and K.~Griest, {\em Phys. Rept.} {\bf 267}, 195
  (1996).

\bibitem{KaneModPhysLett2008}
G.~Kane and S.~Watson, {\em Mod. Phys. Lett.} {\bf A23}, 2103 (2008).

\bibitem{Goodman1985}
M.~W. Goodman and E.~Witten, {\em Phys. Rev.} {\bf D31}, p. 3059 (1985).

\bibitem{PAMELA2009}
PAMELA, O.~Adriani {\em et~al.}, {\em Nature} {\bf 458}, 607 (2009).

\bibitem{ATIC2008}
J.~Chang {\em et~al.}, {\em Nature} {\bf 456}, 362 (2008).

\bibitem{FERMILAT2009}
The Fermi LAT, A.~A. Abdo {\em et~al.}, {\em Phys. Rev. Lett.} {\bf 102}, p.
  181101 (2009).

\bibitem{Gaitskell2004}
R.~J. {Gaitskell}, {\em Ann. Rev. Nucl. Part. Sci.} {\bf 54}, 315 (2004).

\bibitem{CDMS2009}
CDMS, Z.~Ahmed {\em et~al.}, {\em Phys. Rev. Lett.} {\bf 102}, 011301 (2009).

\bibitem{AnglePRL2008}
XENON, J.~Angle {\em et~al.}, {\em Phys. Rev. Lett.} {\bf 100}, p. 021303
  (2008).

\bibitem{LewinAndSmith1996}
J.~D. Lewin and P.~F. Smith, {\em Astropart. Phys.} {\bf 6}, 87 (1996).

\bibitem{MorganPhysRevD2005}
B.~Morgan, A.~M. Green and N.~J.~C. Spooner, {\em Phys. Rev.} {\bf D71}, p.
  103507 (2005).

\bibitem{Vergados2007}
J.~D. Vergados and A.~Faessler, {\em Phys. Rev.} {\bf 75}, 055007 (2007).

\bibitem{Drukier1986}
A.~K. Drukier, K.~Freese and D.~N. Spergel, {\em Phys. Rev.} {\bf D33}, 3495
  (1986).

\bibitem{Freese1988}
K.~Freese, J.~A. Frieman and A.~Gould, {\em Phys. Rev.} {\bf D37}, p. 3388
  (1988).

\bibitem{DAMALIBRA2008}
DAMA, R.~Bernabei {\em et~al.}, {\em Eur. Phys. J.} {\bf C56}, 333 (2008).

\bibitem{Spergel1988}
D.~N. Spergel, {\em Phys. Rev.} {\bf D37}, p. 1353 (1988).

\bibitem{Copi1999}
C.~J. Copi, J.~Heo and L.~M. Krauss, {\em Phys. Lett.} {\bf B461}, 43 (1999).

\bibitem{Stiff2003}
D.~Stiff and L.~M. Widrow, {\em Phys. Rev. Lett.} {\bf 90}, p. 211301 (2003).

\bibitem{Finkbeiner2009}
D.~P. Finkbeiner, T.~Lin and N.~Weiner, {\em arXiv/} {\bf 0906.0002} (2009).

\bibitem{TuckerSmith2001}
D.~Tucker-Smith and N.~Weiner, {\em Phys. Rev.} {\bf D64}, p. 043502 (2001).

\bibitem{MayetPLB2002}
F.~Mayet {\em et~al.}, {\em Phys. Lett.} {\bf B538}, p. 257 (2002).

\bibitem{MoulinPLB2005}
E.~Moulin, F.~Mayet and D.~Santos, {\em Phys. Lett.} {\bf B614}, 143 (2005).

\bibitem{GondoloJCAP2004}
P.~Gondolo {\em et~al.}, {\em JCAP} {\bf 0407}, p. 008 (2004).

\bibitem{MartoffNIMA2000}
C.~J. Martoff {\em et~al.}, {\em Nucl. Instrum. Meth.} {\bf A440}, 355 (2000).

\bibitem{OhnukiNIMA2001}
T.~Ohnuki, D.~P. Snowden-Ifft and C.~J. Martoff, {\em Nucl. Instrum. Meth.}
  {\bf A463}, 142 (2001).

\bibitem{AlnerNIMA2004}
DRIFT, G.~J. Alner {\em et~al.}, {\em Nucl. Instrum. Meth.} {\bf A535}, 644
  (2004).

\bibitem{MorganNIMA2003}
DRIFT, B.~Morgan {\em et~al.}, {\em Nucl. Instrum. Meth.} {\bf A513}, 226
  (2003).

\bibitem{AlnerNIMA2005}
DRIFT, G.~J. Alner {\em et~al.}, {\em Nucl. Instrum. Meth.} {\bf A555}, 173
  (2005).

\bibitem{BurgosAstroPart2009}
DRIFT, S.~Burgos {\em et~al.}, {\em Astropart. Phys.} {\bf 31}, 261(May 2009).

\bibitem{MartoffNIMA2005}
C.~J. {Martoff} {\em et~al.}, {\em Nucl. Instrum. Meth.} {\bf A555}, 55 (2005).

\bibitem{MiyamotoNIMA2004}
J.~{Miyamoto} {\em et~al.}, {\em Nucl. Instrum. Meth.} {\bf A526}, 409 (2004).

\bibitem{LightfootAstroPart2007}
P.~K. Lightfoot {\em et~al.}, {\em Astropart. Phys.} {\bf 27}, 490 (2007).

\bibitem{MorganAstroPart2005}
B.~Morgan {\em et~al.}, {\em Astropart. Phys.} {\bf 23}, 287 (2005).

\bibitem{BurgosAstroPart2007}
DRIFT, S.~Burgos {\em et~al.}, {\em Astropart. Phys.} {\bf 28}, 409 (2007).

\bibitem{CarsonNIMA2005}
M.~J. Carson {\em et~al.}, {\em Nucl. Instrum. Meth.} {\bf A546}, 509 (2005).

\bibitem{BurgosNIMA2009}
DRIFT, S.~Burgos {\em et~al.}, {\em Nucl. Instrum. Meth.} {\bf A600}, 417
  (2009).

\bibitem{driftNitricCleaning}
http://www.youtube.com/watch?v=G4270rjtDnY.

\bibitem{BurgosNIMA2008}
DRIFT, S.~Burgos {\em et~al.}, {\em Nucl. Instrum. Meth.} {\bf A584}, 114
  (2008).

\bibitem{munaPhD2008}
D.~Muna, {Three dimensional analysis and track reconstruction in the DRIFT-II
  dark atter detector}, PhD thesis, University of Sheffield, (Sheffield, UK,
  2008).

\bibitem{SpoonerPhysSocJap2007}
N.~J. Spooner, {\em J. Phys. Soc. Jap.} {\bf 76}, p. 111016 (2007).

\bibitem{Majewski2009}
P.~Majewski {\em et~al.}, {\em arXiv/} {\bf 0902.4430} (2009).

\bibitem{SnowdenIfftNIMA2004}
D.~P. Snowden-Ifft {\em et~al.}, {\em Nucl. Instrum. Meth.} {\bf A516}, 406
  (2004).

\bibitem{BattestiLectNotesPhys2008}
R.~Battesti {\em et~al.}, {\em Lect. Notes Phys.} {\bf 741}, 199 (2008).

\bibitem{BurgosJINST2009}
DRIFT, S.~Burgos {\em et~al.}, {\em JINST} {\bf 4}, p. P04014 (2009).

\bibitem{crane1961}
H.~R. {Crane}, {\em Rev. Sci. Inst.} {\bf 32}, 953 (1961).

\bibitem{Pushkin2009}
K.~{Pushkin} and D.~{Snowden-Ifft}, {\em Nucl. Instrum. Meth.} {\bf A606}, 569
  (2009).

\bibitem{AraujoNIMA2005}
H.~M. Araujo, V.~A. Kudryavtsev, N.~J.~C. Spooner and T.~J. Sumner, {\em Nucl.
  Instrum. Meth.} {\bf A545}, 398 (2005).

\bibitem{KudryavtsevNIMA2003}
V.~A. Kudryavtsev, N.~J.~C. Spooner and J.~E. McMillan, {\em Nucl. Instrum.
  Meth.} {\bf A505}, 688 (2003).

\bibitem{LemraniNIMA2006}
R.~Lemrani {\em et~al.}, {\em Nucl. Instrum. Meth.} {\bf A560}, 454 (2006).

\bibitem{AraujoAstroPart2008}
H.~M. Araujo {\em et~al.}, {\em Astropart. Phys.} {\bf 29}, 471 (2008).

\bibitem{RobinsonNIMA2003}
M.~Robinson {\em et~al.}, {\em Nucl. Instrum. Meth.} {\bf A511}, 347 (2003).

\bibitem{TziaferiAstroPart2007}
E.~Tziaferi {\em et~al.}, {\em Astropart. Phys.} {\bf 27}, 326 (2007).

\bibitem{dujmicTAUP2007}
DMTPC, D.~Dujmic {\em et~al.}, {\em J. Phys. Conf. Ser.} {\bf 120}, p. 042030
  (2008).

\bibitem{caldwell2009}
DMTPC, T.~Caldwell {\em et~al.}, {\em arXiv/} {\bf 0905.2549} (2009).

\bibitem{kabothCF4}
DMTPC, A.~Kaboth {\em et~al.}, {\em Nucl. Instrum. Meth.} {\bf A592}, 63
  (2008).

\bibitem{dujmicNIMA2008}
DMTPC, D.~Dujmic {\em et~al.}, {\em Nucl. Instrum. Meth.} {\bf A592}, p. 123
  (2008).

\bibitem{dujmicAstroPart2008}
DMTPC, D.~Dujmic {\em et~al.}, {\em Astropart. Phys.} {\bf 30}, 58 (2008).

\bibitem{TanimoriPLB2004}
T.~Tanimori {\em et~al.}, {\em Phys. Lett.} {\bf B578}, 241 (2004).

\bibitem{MiuchiPLB2007}
NEWAGE, K.~Miuchi {\em et~al.}, {\em Phys. Lett.} {\bf B654}, 58 (2007).

\bibitem{NishimuraAstroPart2009}
NEWAGE, H.~Nishimura {\em et~al.}, {\em Astropart. Phys.} {\bf 31}, 185 (2009).

\bibitem{TakadaNIMA2007}
A.~{Takada} {\em et~al.}, {\em Nucl. Instrum. Meth.} {\bf A573}, 195(April
  2007).

\bibitem{NishimuraPhD2009}
H.~Nishimura, {Direction-sensitive direct dark matter search experiment with a
  gaseous TPC}, PhD thesis, Kyoto University, (Kyoto, Japan, 2009).

\bibitem{MatsuzawaTIPP09}
A.~Matsuzawa {\em et~al.} Proceedings of TIPP09, (2009), Tsukuba, Japan.

\bibitem{GreenAstroPart2007}
A.~M. {Green} and B.~{Morgan}, {\em Astropart. Phys.} {\bf 27}, 142(March
  2007).

\bibitem{GuillaudinJPhys2009}
O.~{Guillaudin} {\em et~al.}, {\em J. Phys. Conf. Ser.} {\bf 179}, 012012(July
  2009).

\bibitem{MayetJPhys2009}
F.~{Mayet} {\em et~al.}, {\em J. Phys. Conf. Ser.} {\bf 179}, 012011(July
  2009).

\bibitem{Santos2008}
D.~Santos {\em et~al.}, {\em arXiv/} {\bf 0810.1137} (2008).

\bibitem{Lindhard1963}
J.~Lindhard {\em et~al.}, {\em Mat. Fys. Medd. Dan. Vid. Selsk.} {\bf 33(14)},
  1 (1963).

\bibitem{Hitachi2008}
A.~Hitachi, {\em Rad. Phys. Chem.} {\bf 77}, 1311 (2008), note: The W-value for
  CF$_4$ used in this reference was too large. The proper value of 34.3 eV
  (G.F. Reinking et al., J. Appl. Phys. 60, 499, 1986) makes the ionization
  1.57 times the values (the right axis) shown in Fig. 7 of this reference.

\bibitem{srim}
J.~Ziegler {\em et~al.}, http://www.srim.org.

\bibitem{Lamy2009}
T.~Lamy {\em et~al.}, in preparation.

\bibitem{Geller1996}
R.~Geller, {\em Electron cyclotron resonance ion sources and ECR plasmas}
  (Institute of Physics Publishing, Philadelphia, 1996).

\bibitem{GiomatarisNIMA2006}
I.~Giomataris {\em et~al.}, {\em Nucl. Instrum. Meth.} {\bf A560}, 405 (2006).

\bibitem{GiomatarisNIMA1996}
Y.~Giomataris, P.~Rebourgeard, J.~P. Robert and G.~Charpak, {\em Nucl. Instrum.
  Meth.} {\bf A376}, 29 (1996).

\bibitem{Richer2009}
J.~P. Richer {\em et~al.}, in preparation.

\bibitem{Mayet2009}
F.~Mayet {\em et~al.}, in preparation.

\bibitem{kodamaPRD2008}
DONuT, K.~Kodama {\em et~al.}, {\em Phys. Rev.} {\bf D78}, p. 052002 (2008).

\bibitem{aokiProgTheorPhys1991}
S.~Aoki {\em et~al.}, {\em Prog. Theor. Phy.} {\bf 85}, 951 (1991).

\bibitem{aokiNIMB1990}
S.~Aoki {\em et~al.}, {\em Nucl. Instrum. Meth.} {\bf B51}, 466 (1990).

\bibitem{AcquafreddaJINST2009}
R.~Acquafredda {\em et~al.}, {\em JINST} {\bf 4} (2009).

\bibitem{NatsumeNIMA2007}
M.~Natsume {\em et~al.}, {\em Nucl. Instrum. Meth.} {\bf A575}, 439 (2007).

\bibitem{NakaNIMA2007}
Emulsions, T.~Naka {\em et~al.}, {\em Nucl. Instrum. Meth.} {\bf A581}, 761
  (2007).

\bibitem{KugeJIST2009}
K.~Kuge {\em et~al.}, {\em J. Imaging Science and Tech.} {\bf 53(1)}, 010507
  (2009).

\bibitem{KimNIMA2008}
T.~Kim {\em et~al.}, {\em Nucl. Instrum. Meth.} {\bf A589}, 173 (2008).

\bibitem{OedNIMA1988}
A.~{Oed}, {\em Nucl. Instrum. Meth.} {\bf A263}, 351 (1988).

\bibitem{Derre2002}
J.~{Derr{\'e}} and I.~{Giomataris}, {\em Nucl. Instrum. Meth.} {\bf A477}, 23
  (2002).

\bibitem{DafniNIMA2009}
T.~{Dafni} {\em et~al.}, {\em Nucl. Instrum. Meth.} {\bf A608}, 259 (2009).

\bibitem{AnvarNIMA2009}
S.~{Anvar} {\em et~al.}, {\em Nucl. Instrum. Meth.} {\bf A602}, 415 (2009).

\bibitem{AbbonNJP2007}
P.~{Abbon} {\em et~al.}, {\em New Journal of Physics} {\bf 9}, 170 (2007).

\bibitem{AuneJPCS2009}
S.~{Aune} {\em et~al.}, {\em J. Phys. Conf. Ser.} {\bf 179}, 012015 (2009).

\end{thebibliography}


@STRING{NIM="Nucl. Instrum. Meth."}
@STRING{ASTROPART="Astropart. Phys."}
@STRING{NEWJOFPHYS="New J. Phys."}
@STRING{PRL="Phys. Rev. Lett."}
@STRING{JINST="JINST"}
@STRING{NATURE="Nature"}
@STRING{PTP="Prog. Theor. Phy."}
@STRING{JPCS="J. Phys. Conf. Ser."}
@STRING{LNP="Lect. Notes Phys."}
@STRING{EPJ="Eur. Phys. J."}
@STRING{PR="Phys. Rept."}
@STRING{PL="Phys. Lett."}
@STRING{PHYSREV="Phys. Rev."}
@STRING{ARNP="Ann. Rev. Nucl. Part. Sci."}
@STRING{JCAP="JCAP"}
@STRING{RPC="Rad. Phys. Chem."}
@STRING{MPL="Mod. Phys. Lett."}
@STRING{JIST="J. Imag. Sci. Tech."}
@STRING{MFMDVS="Mat. Fys. Medd. Dan. Vid. Selsk."}
@STRING{JPSJ="J. Phys. Soc. Jap."}
@STRING{REVSCIINST="Rev. Sci. Inst."}
